\documentclass[]{aa}
\usepackage[T1]{fontenc}
\usepackage{graphicx}
\usepackage{color}
\usepackage{txfonts}
\usepackage{microtype}
\usepackage{ellipsis}
\usepackage{psfrag}
\usepackage{natbib}
\bibpunct{(}{)}{;}{a}{}{,}
\usepackage[british]{babel}
\definecolor{green4}{rgb}{0,0.6,0}
\begin{document}
\title{The Earliest Phases of Star formation observed with
\textit{Herschel}\thanks{\textit{Herschel} is an ESA space observatory with
science instruments provided by European-led Principal Investigator consortia
and with important participation from NASA.}\\(EPoS): The dust temperature and
density distributions of B68\thanks{Partially based on observations collected at
the European Organisation for Astronomical Research in the Southern Hemisphere,
Chile, as part of the observing programme 78.F-9012.}}

\author{M.~Nielbock\inst{1}
        \and
        R.~Launhardt\inst{1}
        \and
        J.~Steinacker\inst{2,1}
        \and
        A.~M.~Stutz\inst{1}
        \and
        Z.~Balog\inst{1}
        \and
        H.~Beuther\inst{1}
        \and
        J.~Bouwman\inst{1}
        \and
        Th.~Henning\inst{1}
        \and
        P.~Hily-Blant\inst{2}
        \and
        J.~Kainulainen\inst{1}
        \and
        O.~Krause\inst{1}
        \and
        H.~Linz\inst{1}
        \and
        N.~Lippok\inst{1}
        \and
        S.~Ragan\inst{1}
        \and
        C.~Risacher\inst{3,4}
        \and
        A.~Schmiedeke\inst{1,5}
        }

\offprints{M. Nielbock, \email{nielbock@mpia.de}}

\institute{Max-Planck-Institut f\"ur Astronomie,
           K\"onigstuhl 17, D-69117 Heidelberg, Germany
           \and
           Institut de Plan\'etologie et d'Astrophysique de Grenoble, 
           Universit\'e de Grenoble, BP 53, F-38041 Grenoble C\'edex 9, France
           \and
           SRON Netherlands Institute for Space Research, PO Box 800,
           9700 AV Groningen, The Netherlands
           \and
           Max-Planck-Institut f\"ur Radioastronomie, Auf dem H\"ugel 69,
           D-53121 Bonn, Germany
           \and
           Universit\"at zu K\"oln, Z\"ulpicher Stra{\ss}e 77, D-50937 K\"oln,
           Germany
           }

\date{Received: 29 February 2012 / Accepted: 21 August 2012}

\titlerunning{EPoS: The dust temperature and density distributions of B68}

\abstract
{Isolated starless cores within molecular clouds can be used as a testbed to
investigate the conditions prior to the onset of fragmentation and gravitational
proto-stellar collapse.}
{We aim to determine the distribution of the dust temperature and
the density of the starless core B68.}
{In the framework of the \textit{Herschel} Guaranteed-Time Key Programme ``The
Earliest Phases of Star formation'' (EPoS), we have imaged B68 between 100
and $500~\mu$m. Ancillary data at (sub)millimetre
wavelengths, spectral line maps of the \element[][12]{CO}~(2--1), and
\element[][13]{CO}~(2--1) transitions as well as an NIR extinction map
were added to the analysis. We employed a ray-tracing algorithm to derive the 2D
mid-plane dust temperature and volume density distribution without suffering
from the line-of-sight averaging effects of simple SED fitting procedures.
Additional 3D radiative transfer calculations were employed to investigate the
connection between the external irradiation and the peculiar crescent-shaped
morphology found in the FIR maps.}
{For the first time, we spatially resolve the dust temperature and
density distribution of B68, convolved to a beam size of 36\farcs4. We find a
temperature gradient dropping from $(16.7 {+1.3 \above0pt -1.0})$~K at the edge
to $(8.2 {+2.1 \above0pt -0.7})$~K in the centre, which is about $4$~K lower
than the result of the simple SED fitting approach. The column density peaks at
$N_\mathrm{H} = (4.3 {+1.4 \above0pt -2.8})~\times 10^{22}$~cm$^{-2}$, and the
central volume density was determined to $n_\mathrm{H} = (3.4 {+0.9 \above0pt
-2.5})~\times 10^{5}$~cm$^{-3}$. B68 has a mass of $3.1~M_{\sun}$ of material
with $A_K > 0.2$~mag for an assumed distance of 150~pc.
We detect a compact source in the southeastern trunk, which is also seen in
extinction and CO. At 100 and $160~\mu$m, we observe a crescent of enhanced emission to the south.}
{The dust temperature profile of B68 agrees well with previous estimates. We
find the radial density distribution from the edge of the inner plateau outward
to be $n_\mathrm{H} \propto r^{-3.5}$. Such a steep profile can arise from
either or both of the following: external irradiation with a significant UV
contribution or the fragmentation of filamentary structures. Our 3D radiative
transfer model of an externally irradiated core by an anisotropic ISRF
reproduces the crescent morphology seen at 100 and 160~$\mu$m. Our CO
observations show that B68 is part of a chain of globules in both space and
velocity, which may indicate that it was once part of a filament that
dispersed. We also resolve a new compact source in the southeastern trunk and
find that it is slightly shifted in centroid velocity from B68, lending
qualitative support to core collision scenarios.}
\maketitle
%

\section{Introduction}
There is a general consensus about the formation of low and intermediate-mass
stars that they form via gravitational collapse of cold pre-stellar cores.
A reasonably robust evolutionary sequence has been established
over the past years from gravitationally bound pre-stellar cores
\citep{ward-thompson02} over collapsing Class~0 and Class~I protostars to
Class~II and Class~III pre-main sequence stars \citep{andre93,shu87}.
The scenario of inside-out collapse of a singular isothermal sphere
\citep[e.g.][]{shu77,young05} seems to be consistent with many observations of
cores that already host a first hydrostatically stable protostellar object
\citep[e.g.][]{motte01}. There are also disagreeing results in the
case of starless cores, where inward motions exist before the formation of a
central luminosity source \citep{difrancesco07}, and infall can be
observed throughout the cloud, not only in the centre \citep[e.g.][]{tafalla98}.

\begin{table*}[t]
\caption{Details of the \textit{Herschel} observations of B68.}
\label{t:obs}
\begin{tabular}{ccccccccc}
\hline\hline
OD       & OBSID
         & Instrument
         & Wavelength
         & Map size
         & Repetitions
         & Duration
         & Scan speed
         & Orientation \\
         &
         &
         & ($\mu$m)
         &
         &
         & (s)
         & (\arcsec~s$^{-1}$)
         & (\degr) \\
\hline
287      & 1342191191
         & SPIRE
         & 250, 350, 500 
         & $9\arcmin\times9\arcmin$
         & 1
         & 555
         & 30 
         & $\pm$ 42\\
\raisebox{-5pt}[0pt][-5pt]{320} & 1342193055
         & \raisebox{-5pt}[0pt][-5pt]{PACS}
         & \raisebox{-5pt}[0pt][-5pt]{100, 160}
         & \raisebox{-5pt}[0pt][-5pt]{$7\arcmin\times7\arcmin$}
         & 30
         & 4689
         & 20
         & 45 \\ 
         & 1342193056
         &
         &
         &
         & 30
         & 4568
         & 20
         & 135 \\ 
\hline
\end{tabular}
\tablefoot{The orientation angles of the scan direction listed in the last
column are defined relative to the instrument reference frame. The PACS
maps were obtained using the \textit{homogeneous coverage} and \textit{square map}
options of the scan map AOT. The SPIRE maps were obtained separately.}
\end{table*}

\citet{evans01} and \citet{andre04} consider positive temperature gradients,
which are caused by external heating and internal shielding, for their density
profile models of pre-stellar cores. Although their results differed
quantitatively, both come to the conclusion that the slopes of the density
profiles in the centre of pre-stellar cores are significantly smaller than
previously assumed.  However, these temperature profiles were derived from
radiative transfer models with certain assumptions about the local interstellar
radiation field, but lack an observational confirmation.

To improve this situation and measure the dust temperature and density
distribution in a realistic way, we used the imaging capabilities of the
Photodetector Array Camera and Spectrograph (PACS\footnote{PACS has been
developed by a consortium of institutes led by MPE (Germany) and including UVIE
(Austria); KU Leuven, CSL, IMEC (Belgium); CEA, LAM (France); MPIA (Germany);
INAF-IFSI/OAA/OAP/OAT, LENS, SISSA (Italy); IAC (Spain). This development has
been supported by the funding agencies BMVIT (Austria), ESA-PRODEX (Belgium),
CEA/CNES (France), DLR (Germany), ASI/INAF (Italy), and CICYT/MCYT (Spain).})
\citep{pacs} and the Spectral and Photometric Imaging Receiver (SPIRE)
\citep{spire} instruments on board the \textit{Herschel} Space Observatory
\citep{herschel} to observe the Bok globule \object{Barnard~68} (B68, LDN~57,
CB~82) \citep{barnard27} as part of the \textit{Herschel} Guaranteed Time Key
Programme ``The Earliest Phases of Star formation'' \citep[EPoS; P.I.~O.~Krause;
e.g.][]{beuther10,beuther12,henning10,linz10,stutz10,ragan12} of the PACS
consortium. With these observations, we close the important gap that covers the
peak of the spectral energy distribution (SED) that determines the temperature
of the dust. These observations are complemented by data that cover the SED at
longer wavelengths reaching into the millimetre range.

Bok globules \citep{bok47} like B68 are nearby, isolated, and largely spherical
molecular clouds that in general show a relatively simple structure with one or
two cores \citep[e.g.~][]{larson72,keene83,chen07,stutz10}. A number of globules
appear to be gravitationally stable and show no star formation activity at all.
Whether or not this is only a transitional stage depends on the conditions
inside and around the individual object \citep[e.g.~][]{launhardt10} like its
temperature and density structure, as well as external pressure and heating.
Irrespective of the final fate of the cores and the globules that host them,
such sources are ideal laboratories in which to study physical properties and
processes that exist prior to the onset of star formation.

B68 is regarded as an example of a prototypical starless core. It is located at
a distance of about 150~pc \citep{degeus89,hotzel02co,lombardi06,alves07} within
the constellation of Ophiuchus and on the outskirts of the \object{Pipe nebula}.
The mass of B68 was estimated to lie between 0.7 \citep{hotzel02co} and
$2.1~M_{\sun}$ \citep{alves01nature}, which also relies on different assumptions
for the distance. Previous temperature estimates were mainly based on molecular
line observations that trace the gas and resulted in mean values of the entire
cloud between 8~K and 16~K
\citep{bourke95,hotzel02temp,bianchi03,lada03,bergin06}. Dust continuum
temperature estimates lie in the range of 10~K \citep{kirk07}. However, they
usually lack the necessary spatial resolution and spectral coverage for a
precise assessment of the dust temperature. Typically, the Rayleigh-Jeans tail
is well constrained by submillimetre observations. However, the peak of the
SED, which is crucial for determining the
temperature, is measured either with little precision or too coarse a spatial
resolution that is unable to sample the interior temperature structure.

The radial density profile of B68 is often represented by fitting a Bonnor-Ebert
sphere (BES) \citep{ebert55,bonnor56}, where self gravity and external and
internal pressure are in equilibrium. Based on deep near-infrared (NIR)
extinction mapping, \citet{alves01nature} show by using this scheme that B68
appears to be on the verge of collapse. However, this interpretation is based on
several simplifying assumptions, e.g.~spherical symmetry and isothermality that
is supposed to be caused by an isotropic external radiation field.
\citet{pavlyuchenkov07} have shown that even small deviations from the
isothermal assumption can significantly affect the interpretation of the
chemistry derived from spectral line observations, especially for chemically
almost pristine prestellar cores like B68, and therefore also affect the balance
of cooling and heating.

We demonstrate in this paper that none of these three conditions are met, and we
conclude that even though the radial column density distribution can be fitted
by a Bonnor-Ebert (BE) profile, any interpretation of the nature of B68 that is
based on this assumption should be treated with caution.

\section{Observations, archival data, and data reduction}
 
\subsection{Herschel}
\begin{figure}
\centering
\resizebox{0.95\hsize}{!}{\includegraphics{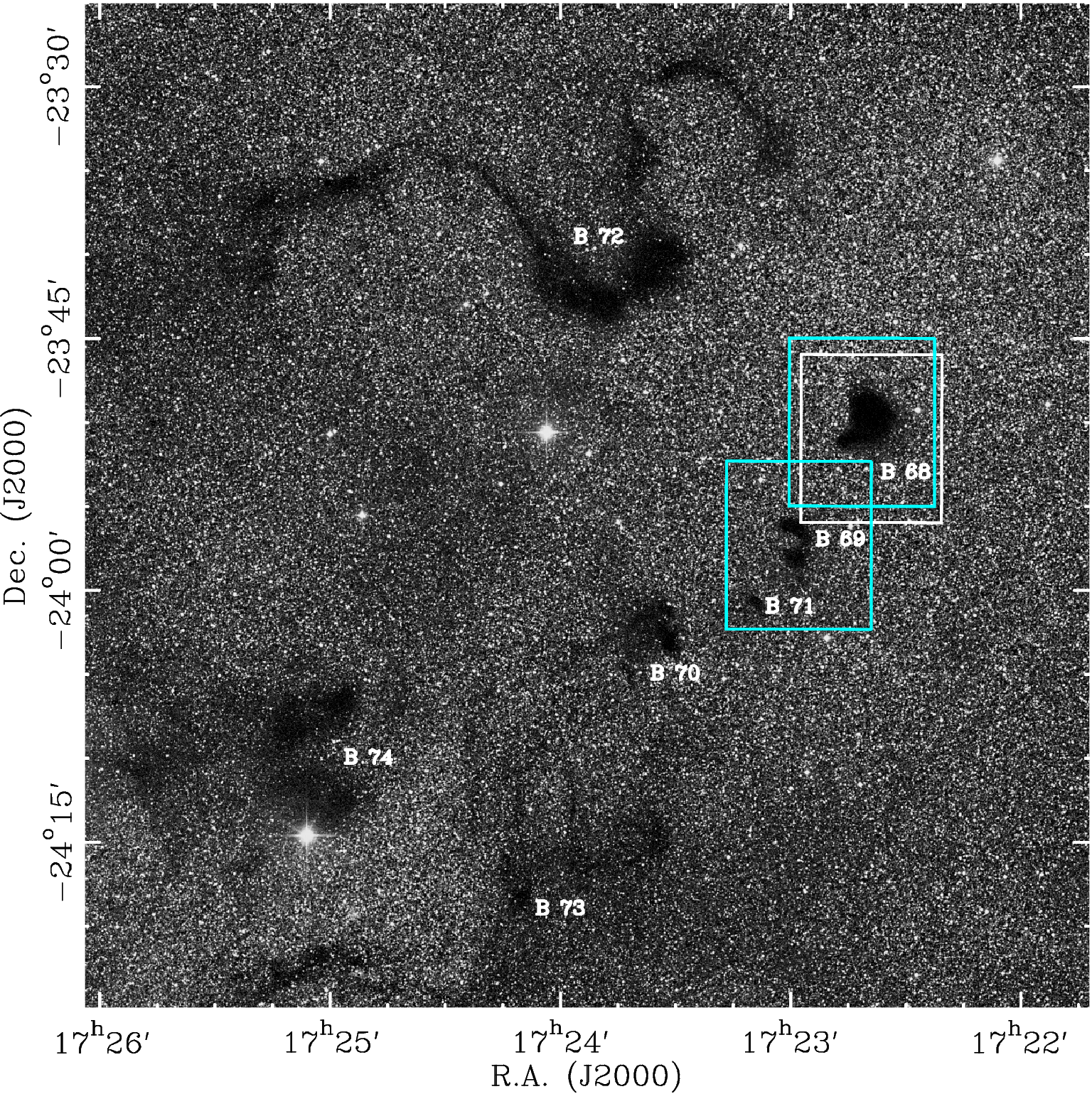}}
\caption{\label{f:b68obs}Digitized Sky Survey (red) image of the area around
B68. The field-of-view (FoV) is $1\degr\times 1\degr$. The Barnard dark clouds
68 to 74 are indicated. The white frame represents the approximate FoV of the
combined \textit{Herschel} observations, while the blue frames denote the areas
mapped in the \element[][12]{CO} and \element[][13]{CO} lines. North is up and
east is to the left.}
\end{figure}

The details of the observational design of the scan map Astronomical Observing
Templates (AOT) are given in Table~\ref{t:obs}. We observed B68 at 100, 160,
250, 350, and 500~$\mu$m using both the imagers of the \textit{Herschel} Space
Telescope, i.e. PACS and SPIRE in prime mode. We chose not to employ
the SPIRE/PACS parallel mode in order to avoid the inherent
disadvantageous side effects, such as beam distortion, reduced spatial sampling,
and lower effective sensitivity, in particular introduced by an increased bit
rounding in the digitisation of the data that especially affects areas with low
surface brightness. In this way, we were able to benefit from the best
resolution and sensitivity available. The latter advantage is especially crucial
for the $100~\mu$m data that are needed to constrain the temperature well. The
approximate area covered by these observations is indicated in
Fig.~\ref{f:b68obs}.

The PACS data reduction comprised a processing to the Level~1 stage using the
\textit{Herschel} Interactive Processing Environment \citep[HIPE, V6.0 user
release,][]{hipe} of the \textit{Herschel} Common Science System
\citep[HCSS,][]{hcss}. For the refinement to the Level~2 stage, we decided to
employ the
Scanamorphos\footnote{\texttt{http://www2.iap.fr/users/roussel/herschel/}}
software \citep{scanamorphos_arxiv}, which provides the best compromise to date
of a good recovery of the extended emission and, at the same time, a proper
handling of compact sources. Still, the resulting background offset derived from
all non-cooled telescopes (e.g.~\textit{Spitzer} and \textit{Herschel}) remain
arbitrary and are mainly influenced by the emission of the warm telescope and
the data reduction techniques. For processing the SPIRE data, we used HIPE, V5.0
(software build 1892). For details of the reduction of the \textit{Herschel}
data, we refer to Launhardt et al.~(in prep.).

\subsection{APEX}
Observations with the Large APEX Bolometer Camera (LABOCA)
\citep{laboca} attached to the Atacama Pathfinder Experiment (APEX)
telescope \citep{apex} at a wavelength of $870~\mu$m were carried out on
17~July~2007 with $2\times2$ spiral patterns that were separated by
$27\arcsec$. These observations were part of a programme aimed to cover a larger
area, which will be presented in Risacher~\&~Hily-Blant~(in prep.). 
The original full width at half maximum ($FWHM$) of the main beam
is 19\farcs2, but the maps were smoothed with a 2D Gaussian to reduce
the background noise, resulting in an effective beam size of $27\arcsec$
matching the separation between individual observations. The absolute
flux-calibration accuracy is assumed to be of the order of 15\%. The root mean
square (rms) noise in the map amounts to 5~mJy per beam.

The LABOCA data were reduced with the BoA software mainly following 
the procedures described in section 3.1 of \citet{schuller09}. It comprised
flagging bad and noisy channels, despiking, and removing correlated noise.
In addition, low-frequency filtering was performed to reduce the effects of the
slow instrumental drifts that are uncorrelated between bolometers. This
filtering is performed in the Fourier domain, and applies to frequencies below
0.2 Hz, which, given the scanning speed of 3\arcmin\,s$^{-1}$, translates to
spatial scales of above 15\arcmin, i.e.~larger than the LABOCA field of
view.

The entire reduction was done in an iterative fashion. No source model was
included for the first instance. Therefore, a first-order baseline removal was
applied to the entire time stream. The signal above a threshold of $4\sigma$
obtained from result of the first reduction was then used as a source model in
the second iteration to blank the regions. This iterative process was done five
times in order to recover most of the extended emission.

The final map was created using a natural weighting algorithm in which each data
point has a weight of $1/rms^2$, where $rms$ is the standard deviation assigned
to each pixel.

\subsection{Spitzer}
B68 was observed with the Infrared Array Camera \citep[IRAC,][]{spitzer-irac}
on board the \textit{Spitzer} Space Telescope
\citep{spitzer} on 4~September~2004 in the framework of Prog~94 (PI:
Lawrence). We used the archived data processed by the \textit{Spitzer} Science
Center (SSC) IRAC pipeline S18.7.0. Mosaics were created from the basic
calibrated data (BCD) frames using a custom IDL program that is described in
\citet{gutermuth08}.

\begin{figure}
\centering
\resizebox{\hsize}{!}{\includegraphics{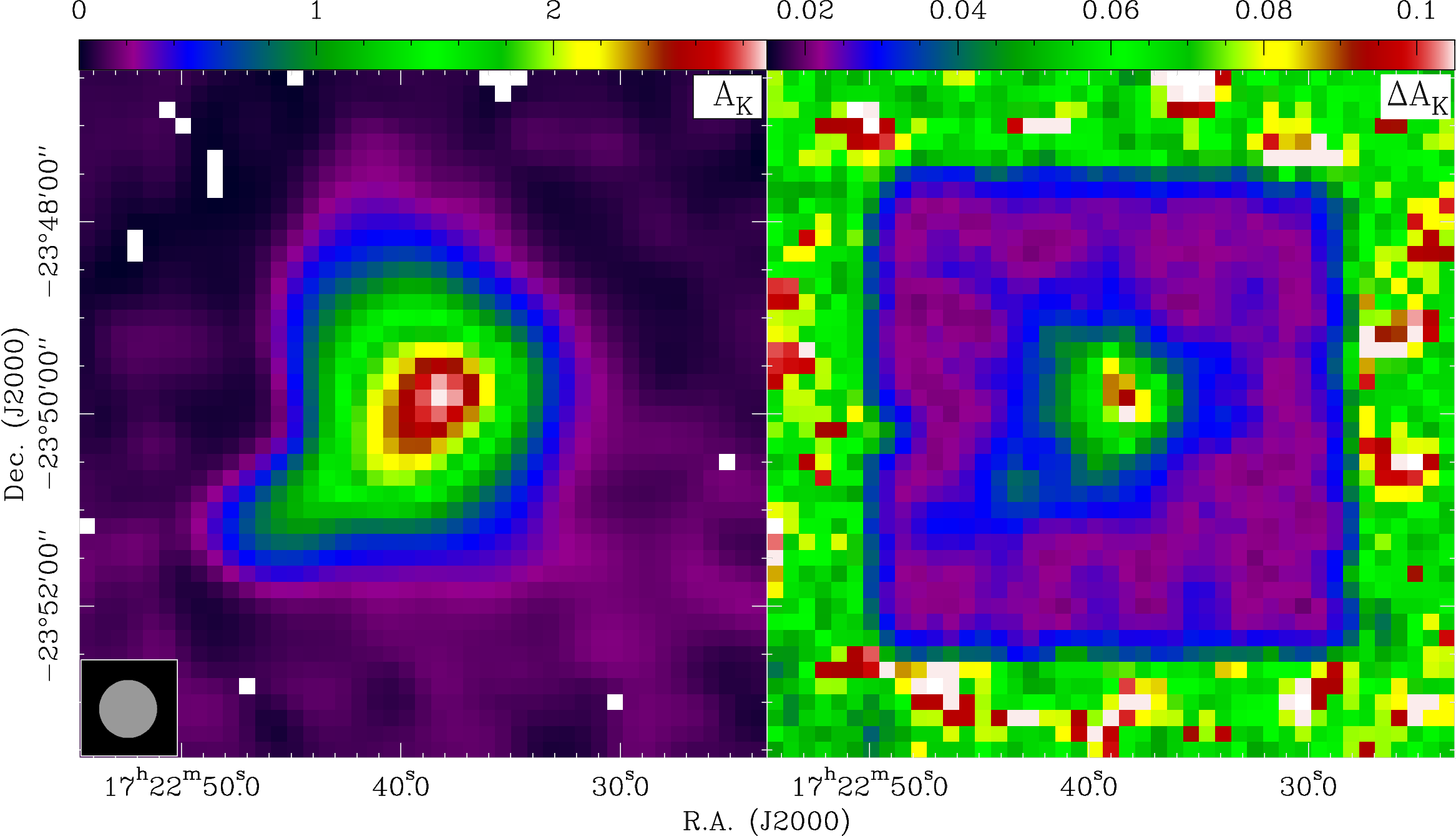}}
\caption{\label{f:b68ext}Distribution of $K$-band extinction $A_K$
recalculated from the data presented in \citet{alves01nature} and 
complementary 2MASS data according to Sect.~\ref{s:nirextobs}.
Left: extinction map; right: corresponding error map. The original spatial
resolution was convolved to match the SPIRE $500~\mu$m data. The spatial
scale is the same as in Fig.~\ref{f:b68cont}.}
\end{figure}

In addition, we included 24~$\mu$m observations obtained with the Multiband
Imaging Photometer for \textit{Spitzer} \citep[MIPS,][]{spitzer-mips}. The
observations were carried out in scan map mode during September 2004 and are
part of programme 53 (PI: Rieke). Data reduction was done with the Data
Analysis Tool \citep[DAT;][]{gordon05}. The data were processed according to
\citet{stutz07}. These images are presented here for the first time.

\begin{figure*}[t]
\centering
\includegraphics[width=\textwidth]{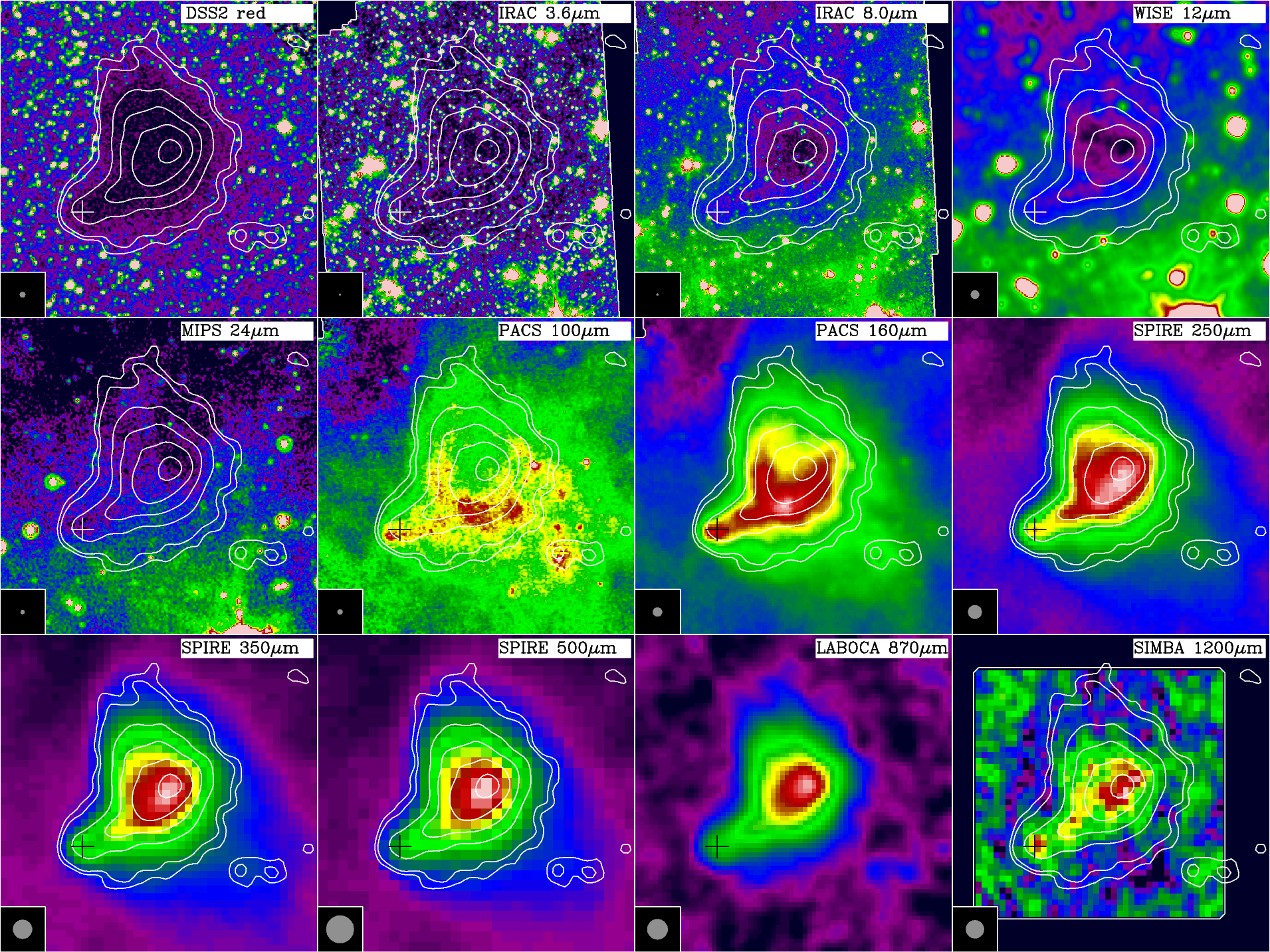}
\caption{\label{f:b68cont}
Image gallery of B68 \textit{Herschel} and archival data
covering a wide range of wavelengths. The instruments and wavelengths at which
the images were obtained are indicated in the white annotation boxes. Each image
has an arbitrary flux density scale, whose settings are given in
Table~\ref{t:scale}. They were chosen to facilitate a comparison of the
morphology. The maps are centred on RA~$=17^\mathrm{h}22^\mathrm{m}39^\mathrm{s}$,
Dec~$=23\degr50\arcmin00\arcsec$ and have a field of view of
$7\arcmin\times7\arcmin$. All coordinates in this manuscript are based on the
J2000.0 reference frame. The corresponding beam sizes are indicated in the
inserts. Contours of the LABOCA data obtained at $\lambda=870~\mu$m show
flux densities of 30, 50, 150, 250, 350, and 450~mJy/beam and are superimposed
for comparison. The black and white crosses indicate the position of the point
source at the tip of the trunk.}
\end{figure*}

B68 was also observed with the Infrared Spectrograph
\citep[IRS,][]{spitzer-irs} on board the \textit{Spitzer} Space Telescope.
With this
instrument, low-resolution ($R\approx60-120$) spectra were obtained on 23~March,
20 and 21~April~2005 in the framework of programmes~3290 (PI: Langer) and 3320
(PI: Pendleton). The spectra used in this paper are based on the {\tt droopres}
intermediate data product processed through the SSC pipeline S18.8.0. Partially
based on the {\tt SMART} software package \citep[][for details on this tool and
extraction methods]{higdon04smart}, our data are further processed using
spectral extraction tools developed for the FEPS \textit{Spitzer} science legacy
programme \citep[for further details on our data reduction, see
also][]{bouwman08,swain08,juhasz10}. The spectra are extracted using a
five-pixel fixed-width aperture in the spatial dimension for the observations in
the first spectral order between 7.5 and 14.2~$\mu$m. Absolute flux calibration
was achieved in a self-consistent manner using an ensemble of observations on
the calibration stars \object{HR~2194}, \object{HR~6606} and \object{HR~7891}
observed within the standard calibration programme for the IRS. The data
were used for a different purpose than extracting the background emission
spectra in \citet{chiar07}.

\subsection{Heinrich Hertz Telescope}
Molecular line maps of the \element[][12]{CO} and \element[][13]{CO} species at
the $J=2-1$ line transition (220.399 and 230.538~GHz) were obtained with the
\textit{Heinrich Hertz} Telescope (HHT) on Mt.~Graham, AZ, USA. The field-of-view (FoV)
of these observations is indicated in Fig.~\ref{f:b68obs}. The spectral
resolution was $\approx 0.3$~km\,s$^{-1}$ and the angular resolution of the
telescope was $32\arcsec$ ($FWHM$). For more details on the observations and
data reduction, we refer to \citet{stutz09}.

\begin{table}
\caption[]{\label{t:scale}Intensity cuts of the 12 images in
Fig.~\ref{f:b68cont}.}
\begin{tabular}{lcc}
\hline
\hline
            & \multicolumn{1}{c}{Low cut} & \multicolumn{1}{c}{High cut} \\ 
Image label & \multicolumn{1}{c}{(MJy~sr$^{-1}$)} &
              \multicolumn{1}{c}{(MJy~sr$^{-1}$)} \\ 
\hline
DSS2 red          & \multicolumn{1}{c}{4900$^\ast$} &
\multicolumn{1}{c}{10000$^\ast$}  \\
IRAC 3.6~$\mu$m   &  0.4 &  2.1 \\
IRAC 8.0~$\mu$m   & 15.3 & 17.9 \\
WISE 12~$\mu$m    & 13.6 & 14.2 \\
MIPS 24~$\mu$m    &  0.0 &  0.4 \\
PACS 100~$\mu$m   & 20.3 & 30.5 \\
PACS 160~$\mu$m   &  7.4 & 42.0 \\
SPIRE 250~$\mu$m  &  4.1 & 69.7 \\
SPIRE 350~$\mu$m  &  2.0 & 50.4 \\
SPIRE 500~$\mu$m  &  0.5 & 25.1 \\
LABOCA 870~$\mu$m &  0.0 &  6.7 \\
SIMBA 1200~$\mu$m &  0.0 &  4.1 \\
\hline
\multicolumn{3}{l}{$^\ast$ counts per pixel}\\
\end{tabular}
\end{table}

\subsection{Near-infrared dust extinction mapping}
\label{s:nirextobs}
We used NIR data from the literature to derive the dust extinction
through the B68 globule. Photometric observations of the sources towards the
globule were performed in $JHK_\mathrm{S}$ bands by
\citet[][]{alves01nature} and \citet[][]{roman10} using the Son of ISAAC
(SOFI) instrument \citep[][]{sofi,sofi2} that is attached to the New
Technology Telescope (NTT) at La Silla, Chile. The observed field of
approximately $4\arcmin\times 4\arcmin$ only covers the innermost part of the
globule. To estimate the extinction outside this area, we included photometric
data covering a wider field from the Two Micron All Sky Survey (2MASS)
archive \citep[][]{2mass}. For sources with detections in both 2MASS and
SOFI data, the SOFI photometry was used \citep[SOFI data were
calibrated using 2MASS data, see][]{roman10}.

We used the $JHK_\mathrm{S}$ band photometric data in conjunction with the
\textsf{NICEST} colour-excess mapping technique \citep[][]{lombardi09} to derive
the dust extinction. In general, the observed colours of stars shining through
the dust cloud are related to the dust extinction via the equation
\begin{equation}
\label{e:colourexcess}
E_{i-j} = \left(m_i - m_j\right) - \left\langle m_i -m_j\right\rangle_0 =
A_j\left(\frac{\tau_i}{\tau_j}-1\right),
\end{equation}
where $E_{i-j}$ is the colour-excess, $A_j$ extinction, and $\langle m_i
-m_j\rangle_0$ the mean intrinsic colour of stars, i.e.~the mean colour of stars
in a control field that can be assumed to be free from extinction. This control
field was chosen as a field of negligible extinction from the large-scale dust
extinction map of the \object{Pipe nebula} \citep{kainulainen09}. When applying
the \textsf{NICEST} colour-excess mapping technique, colour-excess measurements
towards individual stars (observations in $JHK_\mathrm{S}$ bands yield two
colour-excess measurements towards each star) are first combined using a maximum
likelihood technique, yielding an estimate of extinction $A_K$ towards each
detected source. Then, these measurements are used to produce a regularly
sampled map of $A_K$ by computing the mean extinction in regular intervals (map
pixels) using as weights the variance of each extinction measurement and a
Gaussian-like\footnote{As the spatial weighting function, we used the function
approximating the SPIRE PSF, provided in
http://dirty.as.arizona.edu/$\sim$kgordon/mips/conv$\_$psfs/conv$\_$psfs.html .}
spatial weighting function with $FWHM\approx 36\farcs4$ to match the resolution
of the SPIRE $500~\mu$m observations. In deriving the extinction from
Eq.~\ref{e:colourexcess}, we adopted the coefficients for the reddening law
$\tau_i$ from \citet{cardelli89}:
\begin{equation}
\label{e:tauK}
\tau_K = 0.600 \cdot \tau_H = 0.404 \cdot \tau_J \quad .
\end{equation}
The relation holds for the NIR extinction produced by the diffuse ISM and denser
regions \citep[e.g.][]{chapman09a,chapman09b}. The resulting $A_K$ extinction
map is shown in Fig.~\ref{f:b68ext}.

The changes to the NIR extinction law introduced by varying $R_V = A_V/E(B-V)$
are insignificant. However, when extending the extinction law into the visual
range, modifying $R_V$ between 3.1 and 5.0 changes $\tau_J/\tau_V$,
$\tau_H/\tau_V$ and $\tau_K/\tau_V$ by approximately 16\%. Since the values and
the distribution of $R_V$ are a priori unknown, we use $A_K$ instead of $A_V$ in
the subsequent ray-tracing modelling.

We select an empirical extinction law in the NIR instead of taking the opacities
of the dust model of \citet{ossenkopf94} as used for the subsequent modelling of
the FIR to mm data. The reason is that by design this model does not account for
scattering, which is an important contribution to the overall extinction in the
NIR and difficult to predict, but which turns out to be negligible for FIR
properties. Furthermore, \citet{fritz11} have shown that none of the currently
available dust models can actually reproduce the observed NIR extinction
properties well.

\begin{figure}
\centering
\resizebox{\hsize}{!}{\includegraphics{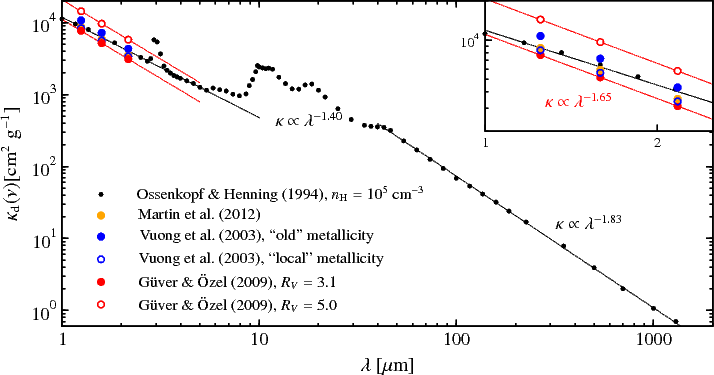}}
\caption{\label{f:opacities}Dust opacities taken from \citet{ossenkopf94}
(coagulated, with thin ice mantles, $n_\mathrm{H} = 10^5 \mathrm{cm}^{-3}$) and
from \citet{cardelli89}, derived by using a gas-to-dust ratio of 150
\citep{sodroski97}. Different ratios of $N_\mathrm{H}/A_\lambda$ have been used
\citep{vuong03,guever09,martin12} when applying Eqs~\ref{e:kappa} and
\ref{e:kappa_R}. The insert shows an enlargement of the range covered by the
$JHK$ bands.}
\end{figure}

To visualise the different behaviours in the NIR range in
Fig.~\ref{f:opacities}, we compare the mass absorption coefficients of the dust
model we use for the modelling and the ones obtained from the adopted empirical
NIR extinction law via Eqs.~\ref{e:kappa} and \ref{e:kappa_R}:
\begin{eqnarray}
\label{e:kappa}
\kappa_\mathrm{d}(\nu)
&=& 6.02 \times 10^{25} \mathrm{g}^{-1} \left(\frac{N_\mathrm{H}}{A_\lambda}\right)^{-1} \\ 
\label{e:kappa_R}
&=& 8800~\mathrm{cm}^{2}\mathrm{g}^{-1} \cdot R_V \cdot
\left(\frac{A_\lambda}{A_V}\right)(R_V) \quad .
\end{eqnarray}
They were derived by assuming a gas-to-dust ratio of 150 \citep[][]{sodroski97}
and a conversion of the total gas to hydrogen mass of 1.36 (see
Sect.~\ref{s:kappa} of the online appendix). In the NIR,
$N_\mathrm{H}/A_\lambda$ can be determined using the estimates of
\citet{vuong03} and \citet{martin12} along with the extinction law of
\citet{cardelli89}. For a discussion on the implicit assumptions and inherent
uncertainties, we refer to Sect.~\ref{s:discussion:n}.

Equation~\ref{e:kappa_R} provides a description that depends on $R_V$. Here, we
adopted the relation $N_\mathrm{H}/A_V$ for the diffuse ISM ($R_V=3.1$)
according to \citet[][]{guever09} and $A_\lambda/A_V$ as a function of $R_V$,
following the extinction law of \citet[][Eq.~1, Tab.~3]{cardelli89}. The
numerical value for the ratio as established by \citet{bohlin78} is 10400
instead of 8800. As a result, $\kappa_\mathrm{d}(\nu)$ increases by a factor of
1.6 in the NIR when raising $R_V$ from 3.1 to 5.0. At the same time, the
corresponding $N_\mathrm{H}$ value is reduced by $1-3.1/5.0=0.38$. This means
that in dense cores, where $R_V$ is assumed to exceed the value of 3.1, the
column density in the densest regions (i.e. the core centre) may be
overestimated by this ratio when derived from $A_V$.

For further details of the \textsf{NICEST} colour-excess mapping technique and
for a discussion of its properties, we refer to
\citet[][]{lombardi05,lombardi09}.

\subsection{Ancillary data}
We included $450$ and $850~\mu$m data obtained with the Submillimetre
Common-User Bolometer Array (SCUBA) at the \textit{James Clerk Maxwell}
Telescope (JCMT) at Mauna Kea, Hawaii, USA, in our analysis. These data
were retrieved from the SCUBA Legacy Catalogues. The maps are not shown
here, but have already been presented in \citet{difrancesco08}.

Observations at a wavelength of 1.2~mm were obtained with the SEST Imaging
Bolometer Array (SIMBA) mounted at the Swedish-ESO Submillimetre Telescope (SEST) at La Silla, Chile. Standard reduction methods were applied as
illustrated in \citet{chini03}. For a detailed description of the data
reduction, we refer to \citet{bianchi03}. The map was also presented by
\citet{nyman01}.

Mid-infrared data obtained at $\lambda = 12~\mu$m were taken from the data archive of the Wide-field Infrared Survey Explorer \citep[WISE,][]{wise}.

\section{Modelling}
\label{s:modelling}

\subsection{Optical-depth-averaged dust temperature and column density maps from SED fitting}
\label{s:modelling:SED}
In a first step, we use the calibrated dust emission maps at the various
wavelengths to extract for each image pixel an SED with up to eight data points
between $100~\mu$m and 1.2~mm.
For this purpose, they were prepared in the
following way:
\begin{enumerate}
\item registered to a common coordinate system
\item regridded to the same pixel scale
\item converted to the same physical surface brightness
\item background levels subtracted
\item convolved to the SPIRE 500~$\mu$m beam
\end{enumerate}
All these steps are described in detail in Launhardt et al.~(in prep.).
Nevertheless, we explain a few crucial procedures in more detail here. For all
maps, the flux density scale was converted to a common surface brightness using
convolution kernels of \citet{aniano11}. In addition, we have applied a extended
emission calibration correction to the SPIRE maps as recommended by the SPIRE
Observers Manual.

The residual background offset levels to be subtracted from the individual maps
were determined by an iterative scheme that uses Gaussian fitting and $\sigma$
clipping to the noise spectrum inside a common map area that was identified as
being ``dark'' relative to the dust emission and relatively free of
\element[][12]{CO}~(2--1) emission.  During each iteration, the flux range was
extended until the distribution deviated from the Gaussian profile. This was
only feasible because of the selection criterion of an isolated prestellar core.

The remaining emission seen by a given map pixel is then given by
\begin{equation}
\label{e:flux}
S_\nu(\nu) = \Omega\left(1-e^{-\tau(\nu)}\right)
             \left(B_\nu(\nu,T_\mathrm{d})-I_\mathrm{bg}(\nu)\right)
\end{equation}
with
\begin{equation}
\tau(\nu) = N_\mathrm{H} m_\mathrm{H} \frac{M_\mathrm{d}}{M_\mathrm{H}}
\kappa_\mathrm{d}(\nu)
\end{equation}
where $S_\nu(\nu)$ is the observed flux density at frequency $\nu$, $\Omega$ the
solid angle from which the flux arises, $\tau(\nu)$ the optical depth in the
cloud, $B_\nu(\nu, T_\mathrm{d})$ the Planck function, $T_\mathrm{d}$ the dust
temperature, $I_\mathrm{bg}(\nu)$ the background intensity,
$N_\mathrm{H} = 2 \times N(\mathrm{H}_2)+N(\mathrm{H})$ the total hydrogen
atom column density, $m_\mathrm{H}$ the hydrogen mass,
$M_\mathrm{d}/M_\mathrm{H}$ the dust-to-hydrogen mass ratio, and
$\kappa_\mathrm{d}(\nu)$ the dust opacity. For the purpose of this paper, we
assume $T_\mathrm{d}$ and $\kappa_\mathrm{d}(\nu)$ to be constant along the LoS
and discuss the uncertainties and limitations of this approach in
Launhardt et al.~(in prep.).

The fluxes of the radiation background $I_\mathrm{bg}(\nu)$ and the instrumental
contributions from the warm telescope mirrors were removed during the background
subtraction described above. Its exact values are unknown, but
$I_\mathrm{bg}(\nu)$ is dominated by contributions from the cosmic infrared and
microwave backgrounds, as well as the diffuse galactic background. A detailed
investigation shows that $I_\mathrm{bg} \ll B_\nu(10~\textrm{K})$, so that
neglecting this contribution introduces a maximum uncertainty of 0.2~K to the
final dust temperature estimate.

\begin{figure}
\centering
\resizebox{\hsize}{!}{\includegraphics{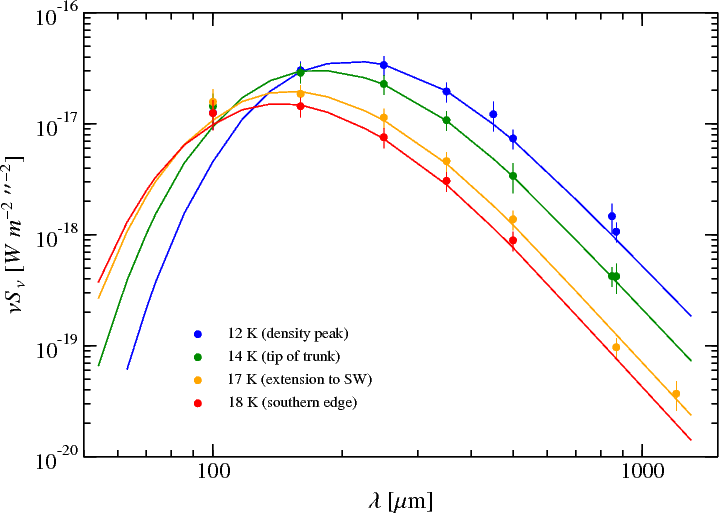}}
\caption{\label{f:b68sed}Spectral energy distributions at four selected
positions across B68. The dots represent $10\arcsec\times 10\arcsec$ pixel
values extracted from the maps that are used for the modified black body
fitting. The solid curves are the SEDs fitted for a given dust temperature and column density.}
\end{figure}

For the dust opacity, $\kappa_\mathrm{d}(\nu)$, we use the tabulated values
listed by
\citet{ossenkopf94}\footnote{\texttt{ftp://cdsarc.u-strasbg.fr/pub/cats/J/A+A/291/943}}
for mildly coagulated ($10^5$ yrs coagulation time at gas density
$10^5$~cm$^{-3}$) composite dust grains with thin ice mantles. The resulting
optical depth averaged dust temperature and column density maps are presented in
Sect.~\ref{s:results:SEDfit}. They provide an accurate estimate to the actual
dust temperature and column density of the outer envelope, where the emission is
optically thin at all wavelengths and LoS temperature gradients are negligible.
Towards the core centre, where cooling and shielding can produce significant LoS
temperature gradients and the observed SEDs are therefore broader than
single-temperature SEDs, the central dust temperatures are overestimated and the
column density is underestimated. Since this effect gradually increases from the
edges to the centre of the cloud, the resulting radial column density profile
mimics a slope that is shallower than the actual underlying radial distribution.

\begin{figure*}
\centering
\includegraphics[width=0.40\textwidth]{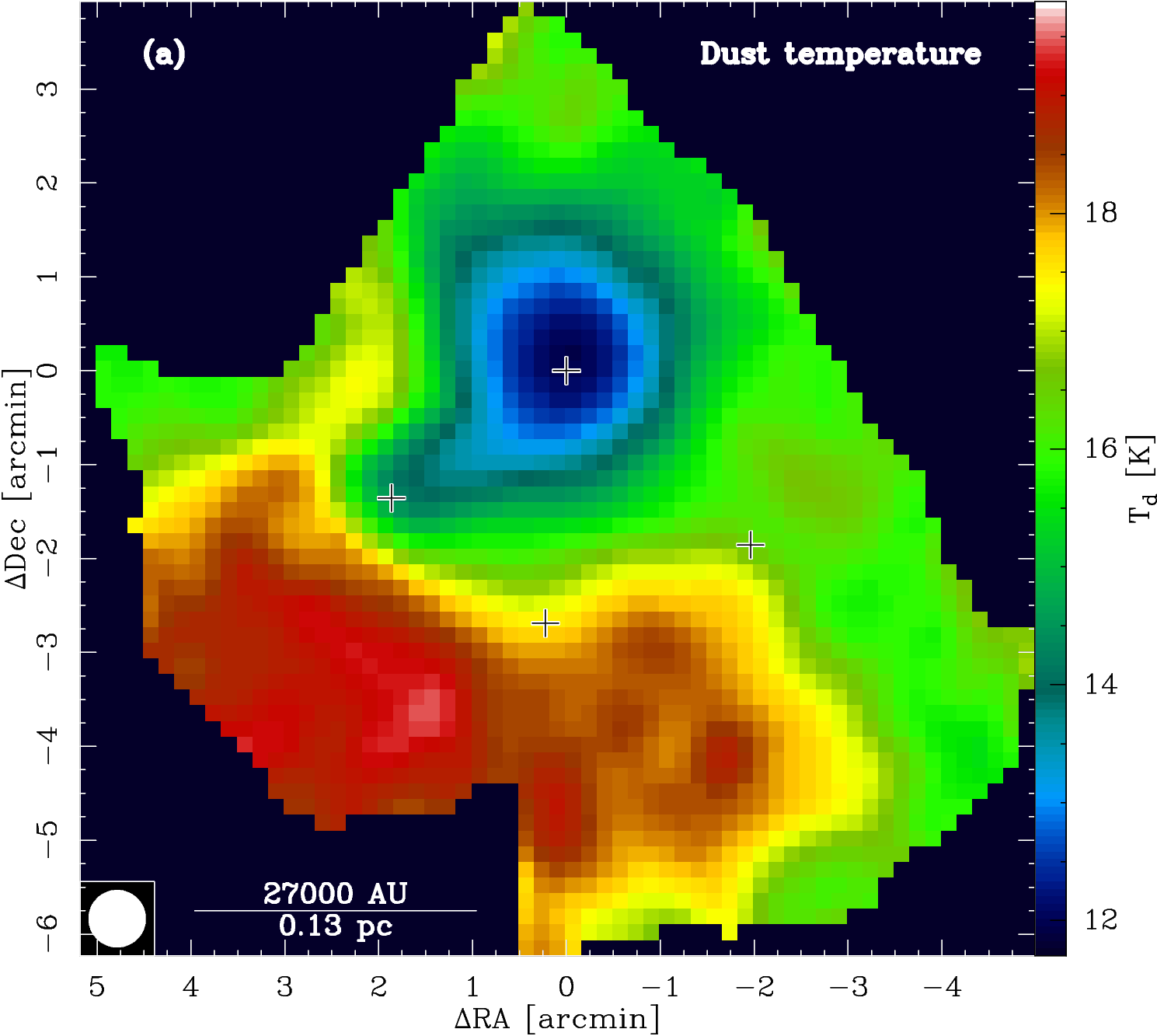}
\hfill
\includegraphics[width=0.40\textwidth]{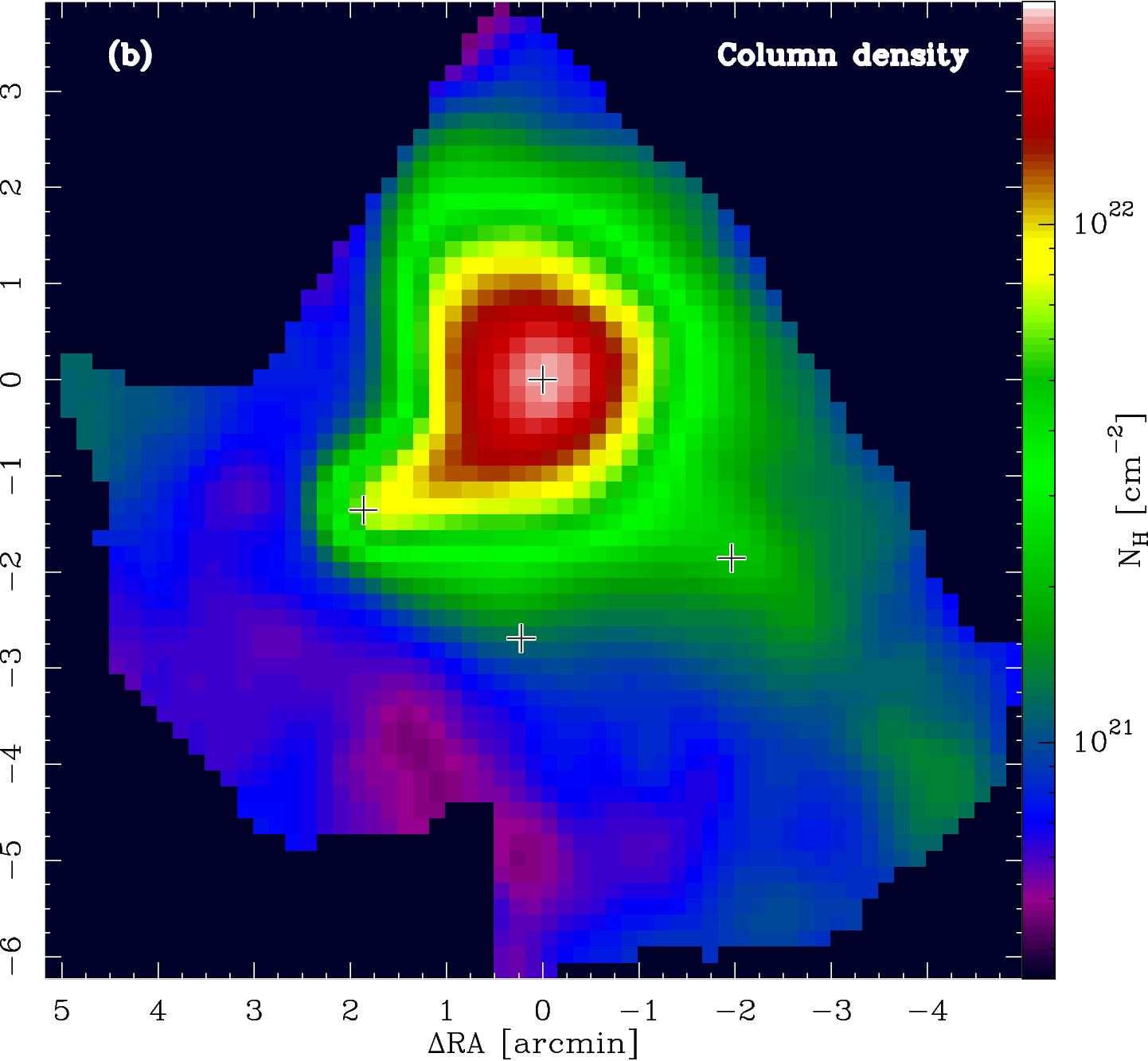}
\caption{\label{f:b68gbmap} Dust temperature (a) and column density (b) maps of
B68 derived by line-of-sight SED fitting as described in
Sect.~\ref{s:modelling:SED}. The temperature minimum attains 11.9~K and is
offset by $5\arcsec$ from the column density peak. The four positions from which
we extracted the individual SEDs shown in Fig.~\ref{f:b68sed} are indicated with
$+$. The central column density is $N_\mathrm{H}=2.7\times10^{22}$~cm$^{-2}$.
The coordinates are given in arcminutes relative to the centre of the column
density distribution, i.e.~RA~$=17^\mathrm{h}22^\mathrm{m}38\fs6$,
Dec~$=-23\degr49\arcmin51\arcsec$.}
\end{figure*}

This procedure is illustrated in Fig~\ref{f:b68sed}, where the SEDs based on the
map pixel fluxes at four positions in the B68 maps are plotted and fitted. This
process includes an iterative colour correction for each of the derived
temperatures. We chose four exemplary locations that cover the full temperature
range, i.e.~at the peak of the column density, at the tip of the southeastern
extension (the trunk) within the faint extension to the southwest where the
$870~\mu$m map has a faint local maximum, and towards the southern edge of B68
(see Fig.~\ref{f:b68gbmap}). The fits to the SEDs demonstrate that the flux
levels are very consistent with a flux distribution of a modified Planck
function. Only the $100~\mu$m data are generally too high for the coldest
temperatures, which we attribute to contributions from higher temperature
components along the LoS. We have assigned a reduced weight in the single-temperature SED fitting to them. The reason for them being at a similar level
irrespective of the map position is related to the small flux density contrast
in the $100~\mu$m map (Tab.~\ref{t:scale}). Also physical effects like
stochastic heating of very small grains (VSGs) may play a role, but dealing with
these issues is beyond the scope of this paper.

\subsection{Dust temperature and density structure from ray-tracing models}
\label{s:modelling:RT}
In a second step, we attempt to reconstruct an estimate of the full temperature
and density structure of the cloud by accounting for LoS temperature and density
gradients in the SED modelling. We make the simplifying starting assumption of
spherical geometry, i.e.~we assume that the LoS profiles are equal to the mean
profiles in the plane of the sky (PoS), but then allow for local deviations. For
this purpose, we construct a cube with the $x$ and $y$ dimension and pixel size
equal to the flux maps and 100 pixels in the $z$ direction. Each cell in this
cube represents a volume element of the size \mbox{$(10\,D_{\mathrm
pc}\,\mathrm{AU})^3$}, where 10 is the chosen angular pixel size in arcseconds
and $D_{\mathrm pc}$ is the distance towards the source in pc, resulting in a
resolution of $1.13\times10^{49}$~cm$^3$ per cell. Each cell is assigned a
density and a temperature. We start with a spherical model centred on the column
density peak derived by the LoS SED fitting (Sect.~\ref{s:modelling:SED}) and at
the cube centre in $z$ direction. For the density profile, we assume a
``Plummer-like'' function \citep[][]{plummer11}
\begin{equation}
\label{e:n_profile}
n_\mathrm{H}(r) =  
n_0 \left(\frac{r_0}{\left(r_0^2+r^2\right)^{1/2}}\right)^\eta=
\frac{n_0}{\left(1+\left(\frac{r}{r_0}\right)^2\right)^{\eta/2}}\quad,
\end{equation}
introduced by \citet{whitworth01} to characterise the radial density
distribution of prestellar cores on the verge of gravitational collapse. It is
analytically identical to the Moffat profile \citep{moffat69} and can also mimic
a BES density profile very closely for a given choice of parameters. To account
for the observed outer density ``plateau'', we modify it by adding a constant
term:
\begin{equation}
\label{e:n_profile_const}
n_\mathrm{H}(r) =  
\left\{ \begin{array}{ll}
\frac{\Delta n}{\left(1+\left(\frac{r}{r_\mathrm{0}}\right)^2\right)^{\eta/2}} + n_\mathrm{out} &\mbox{ if $r\le r_\mathrm{out}$} \\
0            &\mbox{ if $r > r_\mathrm{out}$} \quad . 
       \end{array}
\right.
\end{equation}
The radius $r_\mathrm{out}$ sets the outer boundary of the modelling. Anything
beyond this radius is set to zero. The peak density is then
\begin{equation} 
\label{e:n_peak}
n_0 = n_\mathrm{out} + \Delta n \quad .
\end{equation}
Adding a constant term to Eq.~\ref{e:n_profile} slightly changes the behaviour
of $r_0$ and $\eta$. Although this complicates comparing the results with
previous findings, we apply Eq.~\ref{e:n_profile_const} to analytically describe
the radial distribution for the ray-tracing fitting, because it simplifies the
modelling and fits better to the radial profiles than Eq.~\ref{e:n_profile}.
This profile (i) accounts for an inner flat density core inside $r_0$ with a
peak density $n_0$, (ii) approaches a modified power-law behaviour with an
exponent $\eta$\ at $r\gg r_0$, (iii) turns over into a flat-density halo
outside
\begin{equation}                      
\label{e:n_rout}
r_1 = r_\mathrm{0}\,\sqrt{\left(\frac{\Delta n}{n_\mathrm{out}}\right)^{2/\eta}-1} \quad ,
\end{equation}
which probably belongs to the ambient medium around and along the LoS of B68,
and (iv) is manually cut off at $r_\mathrm{out}$. The outer flat-density halo
and the sharp outer boundary are artificial assumptions to simplify the
modelling and account for the observed finite outer halo. This outer tenuous
envelope of material is neither azimuthally symmetric nor fully spatially
covered by our observations. Therefore, its real size remains unconstrained.
Furthermore, its temperature and density remain very uncertain because of the
low flux levels and uncertainties introduced by the background and its attempted
subtraction. Therefore, the value of $r_\mathrm{out}$ should not be considered
as an actual source property.

\begin{table*}
\caption{\label{t:rtfit}
Parameters and results of the ray-tracing modelling for the radial
distributions using Eq.~\ref{e:n_profile_const}.}
\begin{tabular}{lccccccccccccccccccc}
\hline
\hline
 & &
\multicolumn{2}{c}{$T$} & &
\multicolumn{2}{c}{$N_\mathrm{H}$} & &
\multicolumn{2}{c}{$n_\mathrm{H}$} & &
\multicolumn{3}{c}{} & &
\multicolumn{5}{c}{} \\
 & &
\multicolumn{2}{c}{(K)} & &
\multicolumn{2}{c}{$(10^{20}~\mathrm{cm}^{-2})$} & &
\multicolumn{2}{c}{$(10^{2}~\mathrm{cm}^{-3})$} & &
\multicolumn{3}{c}{LoS} & &
\multicolumn{5}{c}{PoS} \\
Solution & & 
0 & out & &
0 & out & &
0 & out & &
$\tau_0$ &
$r_{0,n}~(\arcsec)$ &
$\eta_{n}$ & &
$\tau_0$ &
$r_{0,N}~(\arcsec)$ &
$\eta_{N}$ &
$r_{0,n}~(\arcsec)$ &
$\eta_{n}$ \\
\hline
best  & &
8.2 & 16.7 & &
430 & 8 & &
3400 & 4 & &
1.5 & 35 & 4.0 & &
2.1 & 76 & 4.6 & 56 & 4.9 \\
steep & &
9.7 & 16.7 & &
450 & 9 & &
2600 & 5 & &
3.0 & 40 & 3.5 & &
3.3 & 99 & 6.7 & 77 & 6.6 \\
flat & &
8.2 & 17.0 & &
350 & 8 & &
2200 & 1 & &
0.5 & 40 & 3.7 & &
1.8 & 85 & 4.9 & 48 & 3.7 \\
\hline
\end{tabular}
\tablefoot{The corresponding radial distributions are shown in
Fig.~\ref{f:rtfit}.}
\end{table*}

For the temperature profile of an externally heated cloud, 
we adopt the following empirical prescription. It resembles the radiation
transfer equations in coupling the local temperature to the effective optical
depth towards the outer ``rim'', where the ISRF impacts:
\begin{equation}
\label{e:T_profile}
T(r) = T_\mathrm{out}-\Delta T \left(1-e^{-\tau(r)}\right)
\end{equation}
with
\begin{equation}
\label{e:T_in}
\Delta T =  T_\mathrm{out}-T_0\quad,
\end{equation}
the frequency-averaged effective optical depth
\begin{equation}
\label{e:profile_tau}
\tau(r) = \tau_0 \int_r^{r_\mathrm{out}} \frac{n_\mathrm{H}(x)}{n_0}\,dx\quad,
\end{equation}
and $\tau_0$\ being an empirical (i.e. free) scaling parameter that accounts for
the a priori unknown mean dust opacity and the UV shape of the ISRF. The
constant term $T_0$, which is also free in the modelling, accounts for the
additional IR heating (external and internal). The functional shape of this
simple model was compared to results of full radiative transfer simulations over
a range of parameters relevant for this study \citep[e.g.][]{zucconi01} and
found to agree reasonably well.

This description has a total of eight free parameters, of which $r_\mathrm{out}$
and $T_\mathrm{out}$ are derived from the azimuthally averaged profile of the
column density map resulting from the SED fitting (Sect.~\ref{s:modelling:SED})
and are then kept fixed. The parameters $r_0$, $\eta$, and $n_\mathrm{out}$ are
estimated from the same source, but then are iterated via the azimuthally
averaged profile of the resulting volume density map. $\tau_0$ is initially set
to 1 and is then also iterated via the azimuthally averaged profiles of the
resulting volume density and mid-plane temperature maps. Finally, $n_0$ and
$\Delta T$ are derived independently for each image pixel by ray-tracing through
the model cloud and least squares fitting to the measured SEDs. An iterative
colour correction algorithm was applied for each of the derived temperatures. At
$r > r_1$, the cloud is assumed to be isothermal, and we fit for a single $T$
instead of a temperature profile with $\Delta T$. The best-fit solutions for all
image pixels are checked to be unique in all cases, albeit with non-negligible
uncertainty ranges, i.e.~the fit never converges to local (secondary) minima.

This procedure yields cubes of density and temperature values, from which we
extract the mid-plane (in the PoS) density and temperature maps, as well as the
total column density map (by integrating the density cube along the LoS). The
resulting mid-plane density and temperature maps are then azimuthally averaged
and fitted for $n_0$, $r_0$, $\eta$, $n_\mathrm{out}$, and $\tau_0$. The
resulting best-fit values are then used to update the LoS profiles of the model
for the next ray-tracing iteration. Approximate convergence between the input
LoS profile parameters and the PoS profiles derived from the mid-plane density
and temperature maps of the independently fitted image pixels is usually reached
after two to three iterations. To account for beam convolution effects
in the PoS, we use a slightly smaller $r_0$ along the LoS than derived in the
PoS. This procedure allows us to fit for local deviations from the spherical
profile in the plane of the sky, while preserving the on average symmetric
profile along the LoS.

Since this iteration between PoS and LoS profiles does not lead to a
well-defined exact solution, we derive the uncertainties on the density and
temperature values not simply from the formal statistical fitting errors, but
proceed as follows. Once we have a parameter set for which PoS and LoS profiles
converge, we decrease and increase $\tau_0$ by a factor of two and adjust the
other parameters for the LoS profile such that we obtain the smoothest possible
PoS profile; convergence between PoS and LoS is no longer reached with these
$\tau_0$ values. A larger $\tau_0$ representing a better shielding leads to a
steeper drop of $T$ with a wide inner plateau of nearly constant $T_0$ at a
higher value than in the best-fit case. A smaller $\tau_0$ simulating a weaker
shielding leads to a shallower drop of $T$ with a small inner region of very low
$T_0$. To this range of profiles we add the statistical fitting uncertainties to
define the total uncertainty, as illustrated in Fig.~\ref{f:rtfit}. We have
listed the corresponding parameters used for the LoS and the PoS fitting in
Table~\ref{t:rtfit}. The resulting temperature and density profiles
calculated along the LoS provide a significantly better approximation to the
true underlying source structure than the results of the single SED fitting; in
the isothermal SED analysis, the individual pixel SEDs represent average LoS
quantities projected to the PoS.

\section{Results}
\subsection{From absorption to emission}
As shown by Fig.~\ref{f:b68cont}, B68 causes extinction from the optical to
the mid-IR (MIR) range. Various authors have used this property to derive
extinction maps \citep[e.g.~][]{alves01nature,lombardi06,racca09} and determine
its density distribution. We have augmented the data of \citet{alves01nature}
with the 2MASS catalogue \citep{2mass} to extend the resulting size of the
extinction map (Fig.~\ref{f:b68ext}) so that it matches the extents and
resolution of the \textit{Herschel} maps (see Sect.~\ref{s:nirextobs}).

The absorption that is seen at the \textit{Spitzer} IRAC 8~$\mu$m and the
WISE 12~$\mu$m bands is most likely caused by the attenuation of background
emission originating in polycyclic aromatic hydrocarbons (PAHs) by the dust in
B68. Such spectral features are visible in \textit{Spitzer} IRS background spectra
(available as online material, Fig.~\ref{f:irs}) at prominent wavelengths
(8.6, 11.3, 12.7~$\mu$m) without any spatial variation across B68. This
indicates that the emission is dominated by contributions from the ambient
diffuse interstellar medium (ISM) and is not related to B68 itself.

There is no evidence for coreshine \citep{steinacker10} in the \textit{Spitzer}
IRAC 3.6 and 4.5~$\mu$m images. Since coreshine is interstellar radiation
scattered by dust grains larger than the majority of grains in the ISM,
it could indicate that grain growth is or has been inefficient in B68. Only
about half of the cores investigated so far show the effect
\citep{stutz09,pagani10}.

In the $24~\mu$m maps, B68 appears as neither a clear absorption nor an emission
feature. The low dust temperatures cause the thermal emission to remain well
below the MIPS $24~\mu$m detection limit. Since the extinction law is
very similar at 8, 12, and 24~$\mu$m \citep[][]{flaherty07}, the lack of the
absorption feature may be attributed to the missing strong background emission
that is present as PAH emission in the 8 and 12~$\mu$m images. The inherent
contrast between emission and absorption at 
$24~\mu$m could be below the detection limit of the MIPS observations.
It is not until $100~\mu$m that B68 appears faintly in emission above the
diffuse signal from the surrounding ISM, mainly in the southern part of the
globule (see next Sect.). As seen in Fig.~\ref{f:b68sed},
the maximum of the SEDs is between 100 and 250~$\mu$m.

\subsection{Morphology}
\label{s:morphology}
Based on observations at visual, near-infrared (NIR), and millimetre (mm)
wavelengths, the morphology of B68 is usually described in terms of a nearly
spherical globule with a small extension to the southeast
\citep[e.g.~][]{bok77,nyman01}. This is also confirmed by extinction maps
that are based on NIR imaging as shown in Fig.~\ref{f:b68ext}
\citep[][]{alves01nature,racca09}. However, the spatial distribution of the
far-infrared (FIR) emission as measured by \textit{Herschel} depends on the
observed wavelength and does not show the symmetry that one would expect by
extrapolating from previous observations.

At 100~$\mu$m (Fig.~\ref{f:b68cont}, second row, second column), B68 is
dominated by a rather faint extended emission that almost blends in with the
ambient background emission. The brightest region forms a crescent that covers
the southern edge of the globule and resembles the distribution of C$^{18}$O gas
as shown by \citet{bergin02}. The peak of the $100~\mu$m emission is not located
at the position of the highest extinction. As shown by the superimposed
LABOCA map, there is a gap of emission instead. A handful of background
stars also detected at shorter wavelengths is present in this map. The
southeastern extension is also visible.

The morphology is still present at 160~$\mu$m (Fig.~\ref{f:b68cont}, second row,
third column), where the crescent-shaped extended emission is even more
pronounced. As the observed wavelength increases, the crescent-shaped feature
converges to the more familiar morphology observed in, e.g., near-IR extinction
and long wavelength continuum data (c.f.~Fig.~\ref{f:b68cont}, third row, third
column). In Sect.~\ref{s:aniso}, we demonstrate that this changing morphology
can be explained by anisotropic irradiation effects. It is interesting that the
MIR images, in particular at 8 and $12~\mu$m, show a brightness gradient from
the north to the south. Drawing a connection to the anisotropic irradiation also
in these wavebands appears tempting, because they contain prominent spectral
bands of PAH emission from VSGs that might be induced by an incident UV field.
However, the actual nature of the excess emission south of B68 is uncertain (PAH
or continuum emission) and there is no evidence of a physical connection of that
emission to B68 (see also the discussion on PAH emission in the previous
section). Therefore, we leave a more detailed analysis of this phenomenon to
later studies.

At the tip of the southeastern extension, there appears to be a compact object
whose detection is confirmed by the \textit{Herschel} 160~$\mu$m and 250~$\mu$m
maps, indicated in Fig.~\ref{f:b68cont}, and by our \element[][13]{CO}~(2--1)
mean intensity map (Fig.~\ref{f:b68co}). It is also present in the extinction
map of \citet{alves01eso}. There is a point source $15\arcsec$ east of that
position in the 100~$\mu$m map, but this one corresponds to a background source
that is also visible in the \textit{Spitzer} images. The newly discovered source
disappears and blends in with the more extended emission at longer wavelengths,
but reappears at 1.2~mm. We successfully performed PSF photometry of this
feature in the two PACS bands, while we used the background star visible in the
100~$\mu$m map as an upper limit for the true flux at this position. A
calibration observation of the asteroid Vesta obtained during the operational
day (OD) 160 of the \textit{Herschel} mission was selected to serve as a
template for the instrumental point spread function (PSF). With a PSF fitting
correlation coefficient of $> 80\%$, we find a point source at
RA~$=17^\mathrm{h}22^\mathrm{m}46\fs3$, Dec~$=-23\degr51\arcmin15\arcsec$ (see
Fig.~\ref{f:b68cont}), where the visual extinction attains $A_V=11.3$~mag. The
photometry yields colour-corrected flux densities of $(20.0\pm0.5)$~mJy at
100~$\mu$m and $(178.5\pm1.4)$~mJy at 160~$\mu$m. By fitting a Planck function,
we derive a formal colour temperature of 15~K, i.e.~the true colour temperature
is even lower, since the 100~$\mu$m flux is taken as an upper limit. A mass
estimate can be attempted by using the relation
\begin{equation}
\label{e:mass}
M_\mathrm{H} = m_\mathrm{H} \sum_\mathrm{\#pix} N_\mathrm{H} A_\mathrm{pix}
\quad . 
\end{equation}
We applied it to the extinction map of \citet{alves01nature}, which has a
better resolution than the \textit{Herschel}-based density maps. The pixel scale is
$5\arcsec \times 5\arcsec$, resulting in a pixel area of
$A_\mathrm{pix}=1.26\times10^{32}$~cm$^2$ at an assumed distance of 150~pc. The
conversion factor of $N_\mathrm{H}/A_V=2.2\times 10^{21}$~cm$^{-2}$~mag$^{-1}$
\citep{ryter96,guever09,watson11} was applied. Then we extracted the source from
this map via a 2D Gaussian fit after an unsharp masking procedure by convolving
the original extinction map to three resolutions, $30\arcsec$, $40\arcsec$ and
$50\arcsec$. The resulting extracted mass is of the order of $M_\mathrm{tot} =
1.36~M_\mathrm{H} = 8\times 10^{-2}~M_{\sun}$.

\begin{figure}
\centering
\resizebox{0.8\hsize}{!}{\includegraphics{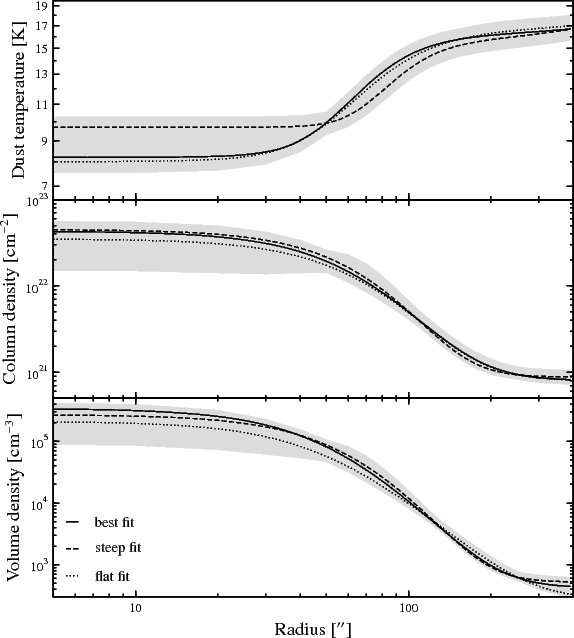}}
\caption{\label{f:rtfit}Visualisation of the ray-tracing fitting to the B68 data
according to Table~\ref{t:rtfit}. We show the functional relations we used for
fitting the LoS distributions of the dust temperature and the volume density
that cover the range of valid solutions for each of the modelled quantities. The
column density is the result of the integration of modelled volume densities
along the LoS. The shaded areas represent the modelled range of values including
the uncertainties of the fitting algorithm. Only the best fits (solid lines) are
used to create the maps that show the dust temperature, the column density, and
the volume density distributions.}
\end{figure}

\begin{figure*}
\centering
\includegraphics[width=0.3\textwidth]{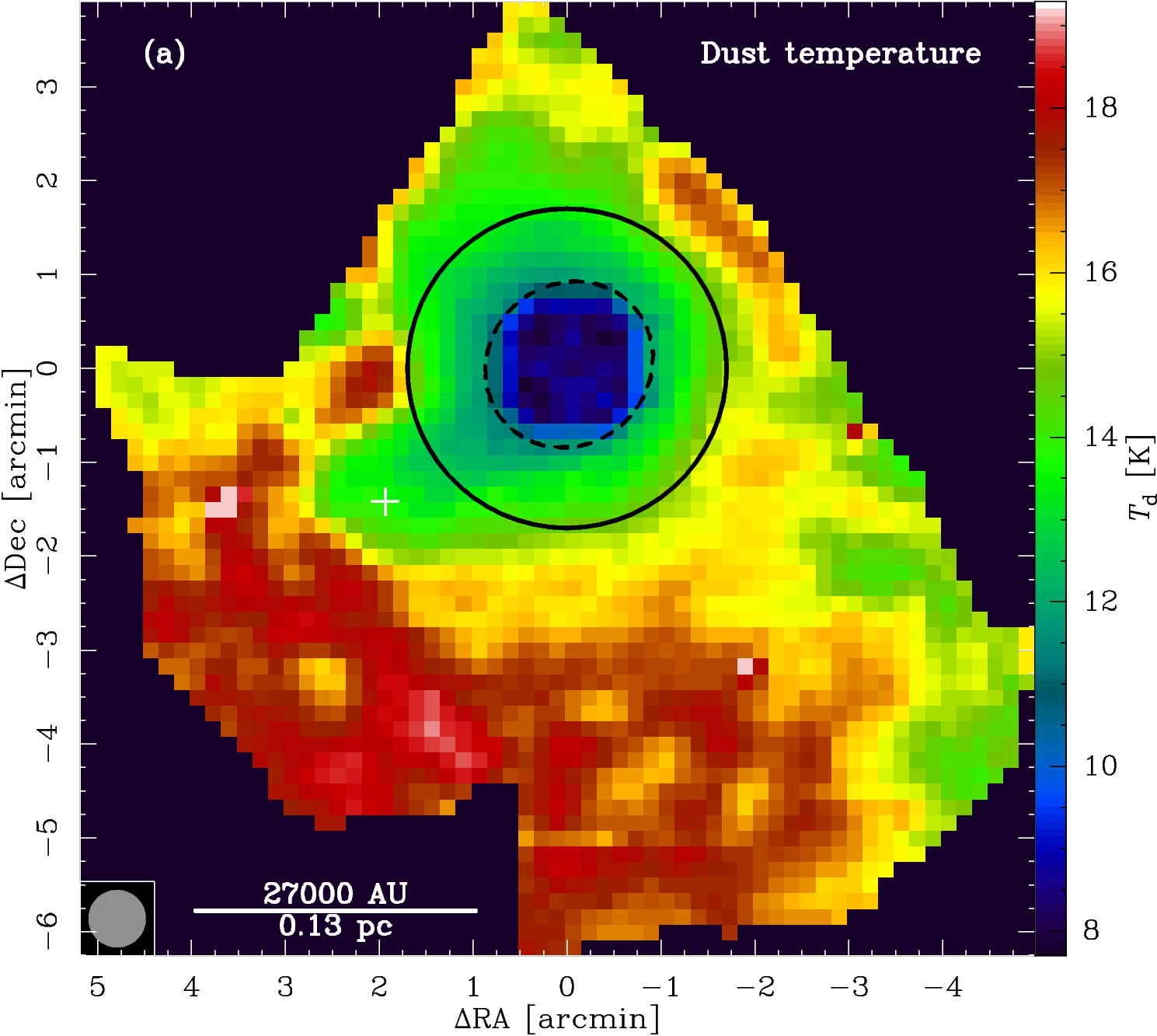}
\hfill
\includegraphics[width=0.3\textwidth]{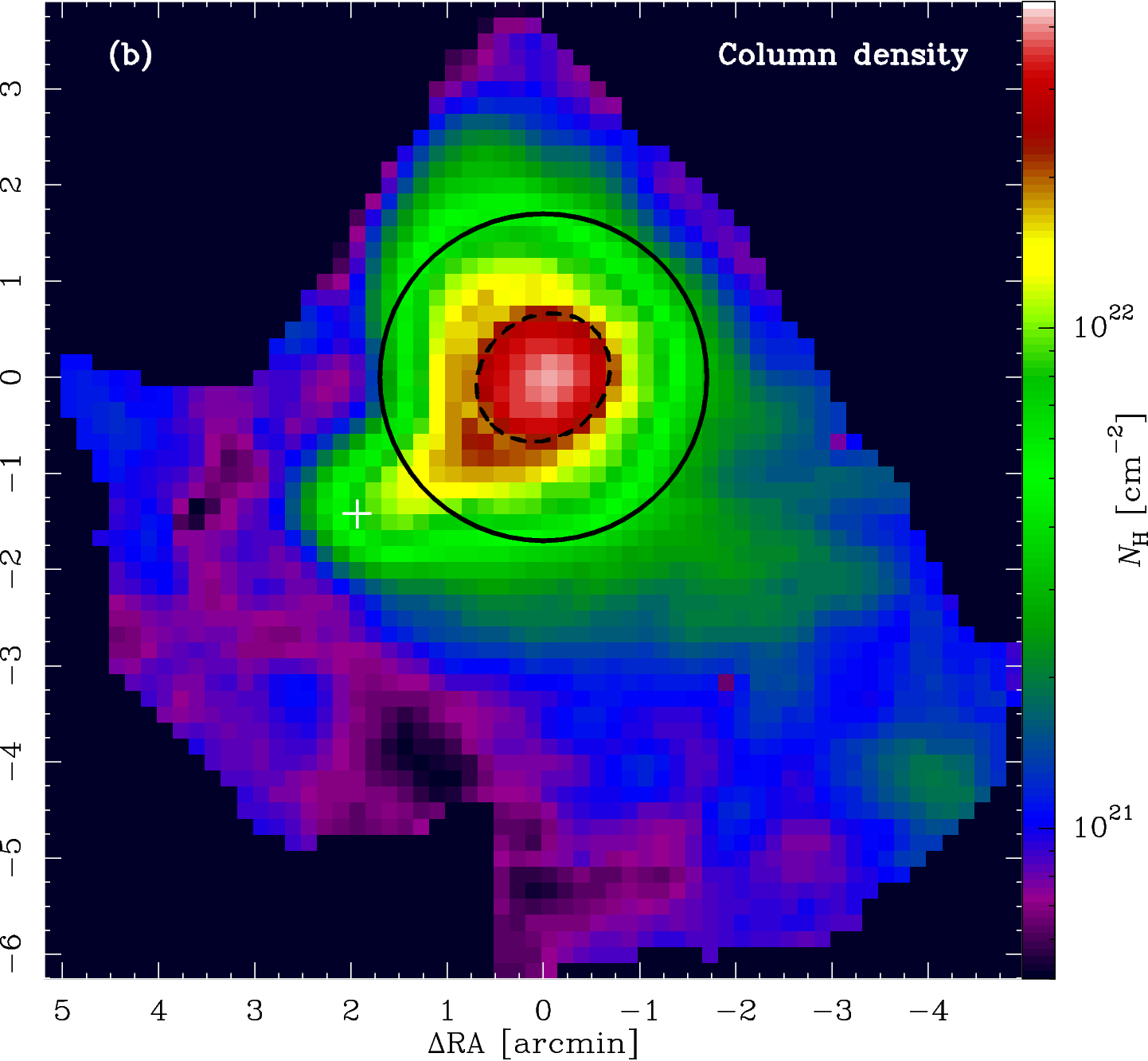}
\hfill
\includegraphics[width=0.3\textwidth]{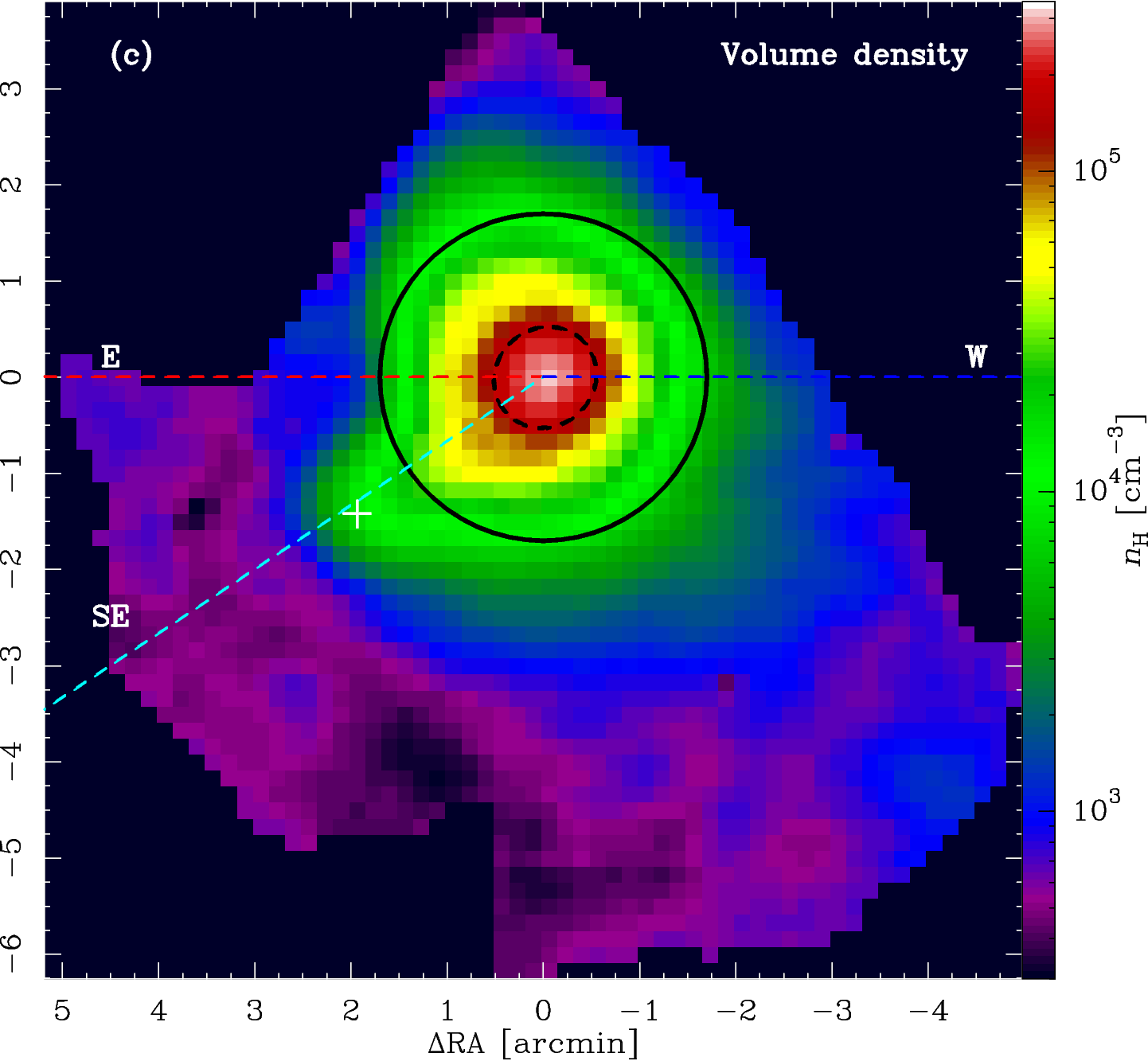}
\caption{\label{f:b68rtmap} Dust temperature (a), column density (b), and volume
density (c) maps of B68 derived by the ray-tracing algorithm as described in
Sect.~\ref{s:modelling:RT}. The maps showing the dust temperature and the volume
density are the ones calculated for the mid-plane across the plane of the sky.
The global temperature minimum coincides with the density peak and attains
a value of 8~K. The central column density is $N_{\rm
H}=4.3\times10^{22}$~cm$^{-2}$. The central particle density is $n_{\rm
H}=3.4\times10^5$~cm$^{-3}$. The black solid circles represent the
size of the BES fitted by \citet{alves01nature}. The major
and the minor axes of the black dashed ellipses denote the $FWHM$ of the 2D
Gaussians fitted to temperature and density peaks (see text). We have plotted
the three cuts through the maps as dashed lines, whose colours match the ones in
the radial profile plots. The white crosses indicate the location of the point
source we have identified and described in Sect.~\ref{s:morphology}. The
coordinates are given in arcminutes relative to the centre of the density
distributions, i.e.~RA~$=17^\mathrm{h}22^\mathrm{m}38\fs5$,
Dec~$=-23\degr49\arcmin50\arcsec$.}
\end{figure*}

\subsection{Line-of-sight-averaged SED fitting: Temperature and column density distribution}
\label{s:results:SEDfit}
To illustrate the differences between the two modelling approaches, we briefly
summarise the results of the modified black body fitting in this section. The
procedure is illustrated in Fig.~\ref{f:b68sed}, where the SEDs based on the map
pixel fluxes at four positions in the B68 maps (see Fig.~\ref{f:b68gbmap}) are
plotted and fitted. We chose four exemplary locations that cover the full
temperature range, i.e.~at the peak of the column density, at the tip of the
southeastern extension (the trunk), within the faint extension to the southwest
where the $870~\mu$m map has a faint local maximum, and towards the southern
edge of B68. The fits to the SEDs demonstrate that the flux levels are very
consistent with a flux distribution of a modified Planck function. Only the
$100~\mu$m data are generally too high for the coldest temperatures, which we
attribute to contributions from higher temperature components along the LoS. The
single-temperature SED fit is not a good approximation for including also the
warm dust components at the surface.

Figure~\ref{f:b68gbmap} (a) shows the dust temperature map of B68 measured by
LoS optical depth averaged SED fitting as described in
Sect.~\ref{s:modelling:SED}. The dust temperature in the map ranges between
11.9~K and 19.6~K, with an uncertainty from the modified black body fitting that
is $\leq 5$\% throughout the map. The minimum of the dust temperature is at
RA~$=17^\mathrm{h}22^\mathrm{m}38\fs4$, Dec~$=-23\degr49\arcmin42\arcsec$. The
location of the column density peak is offset by $5\arcsec$.

At 3\farcm2 (0.14~pc at a distance of 150~pc) to the south and southeast, there
is a sudden rise in the temperature to $\approx 20$~K that is not seen anywhere else around B68. This temperature is derived from SEDs based on
low flux densities that suffer more from uncertainties in determining the
subtracted global background emission than the brighter interiors of B68.
Therefore, the absolute value of the dust temperature is less certain, too. If
this temperature increase is real, this would indicate effects of an anisotropic
irradiation, perhaps due to the distance to the galactic plane ($\approx
40$~pc) or the nearby B2IV star \object{$\theta$~Oph}. The aspect of external
irradiation is discussed in more detail in Sect.~\ref{s:aniso}.

Figure~\ref{f:b68gbmap} (b) shows the column density map of B68 derived from
LoS optical depth averaged SED fitting. It resembles the NIR extinction
map (Fig.~\ref{f:b68ext}). The column density ranges from $N_\mathrm{H} =
2.7\times 10^{22}$~cm$^{-2}$ in the centre at
RA~$=17^\mathrm{h}22^\mathrm{m}38\fs6$,
Dec~$=-23\degr49\arcmin51\arcsec$ to approximately
$4.5\times 10^{20}$~cm$^{-2}$ in the southeastern region, where the dust
temperature rises to nearly 20~K. The uncertainty of the measured column density
introduced by the SED fitting is better than 20\% for all map pixels.

\subsection{Modelling via ray-tracing}
\label{s:results:RTfit}
Figure~\ref{f:b68rtmap} shows the dust temperature, column density and volume
density distributions in the mid-plane, i.e.~half way along the LoS through B68
as determined by the ray-tracing algorithm described in
Sect.~\ref{s:modelling:RT}. These maps were constructed by applying the best
LoS fit parametrised by
$\tau_0 = 1.5$, $r_0 = 35\arcsec$, $\eta = 4.0$,
$n_0=3.4\times10^5$~cm$^{-3}$, $n_\mathrm{out}=4\times10^2$~cm$^{-3}$,
$T_0=8.2$~K, and $T_\mathrm{out}=16.7$~K.
The statistical uncertainties for a given solution of the modelling are
generally better than 5\% for the dust temperature and 20\% for the density. The
systematic uncertainties, i.e. the range of valid solutions of the modelling,
dominate the error budget (see Table~\ref{t:rtfit}, Fig.~\ref{f:rtfit}).
They are given in the individual results sections.

\subsubsection{Dust temperature distribution}
\label{s:results:RTfit:T}
The global temperature minimum of the dust temperature map in
Fig.~\ref{f:b68rtmap} (a) at RA~$=17^\mathrm{h}22^\mathrm{m}38\fs7$,
Dec~$=-23\degr49\arcmin46\arcsec$ can be fitted by a 2D Gaussian with an
$FWHM$ of $(111\pm1)\arcsec \times (103\pm1)\arcsec$, i.e.~more than three times
the beam size. A second, local temperature minimum of $\approx
14$~K remains at RA~$=17^\mathrm{h}22^\mathrm{m}44\fs0$,
Dec~$=-23\degr50\arcmin50\arcsec$ after subtracting the fitted Gaussian
from the map.

\begin{figure*}
\centering
\resizebox{0.750\hsize}{!}{\includegraphics{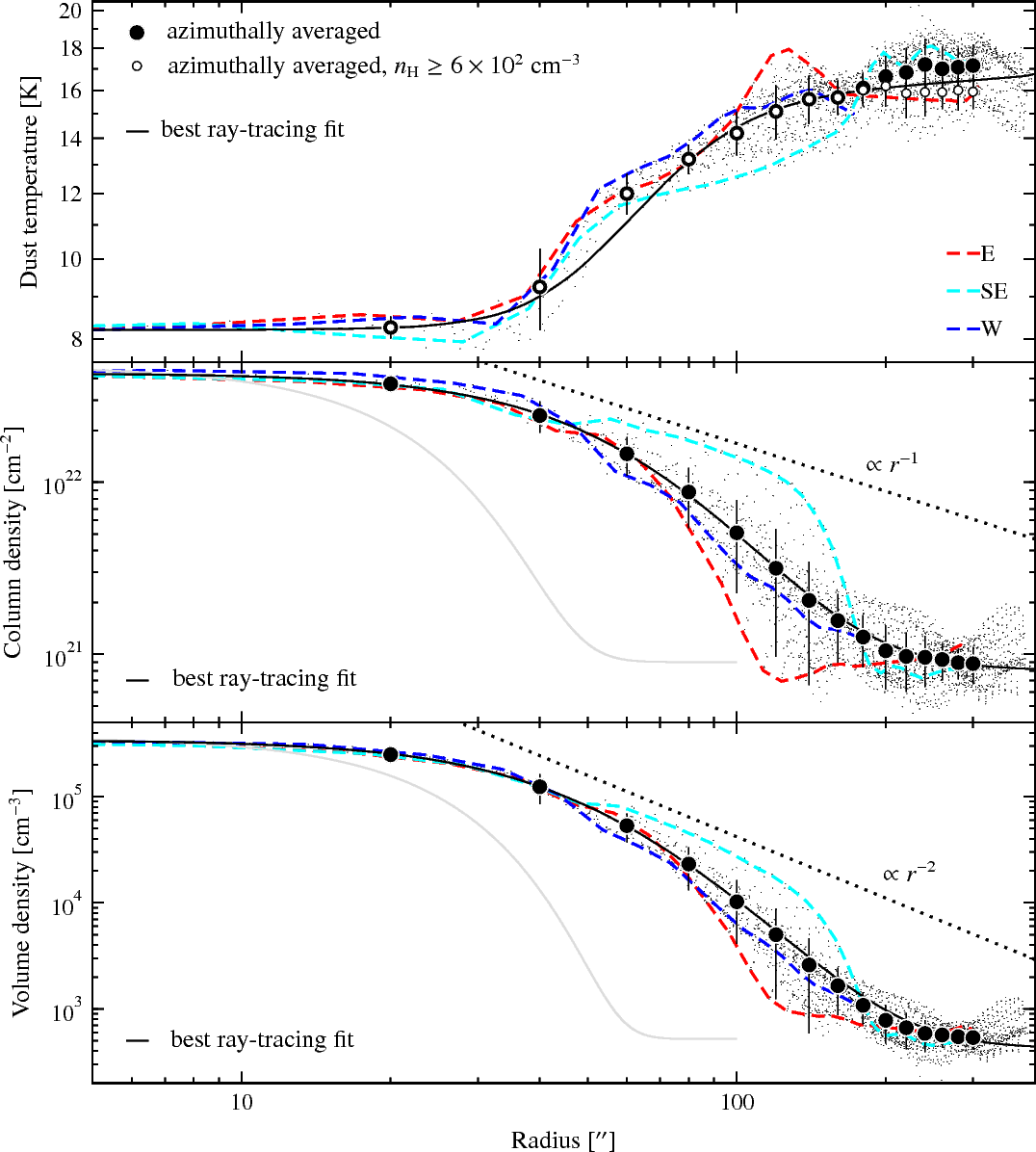}}
\caption{\label{f:b68radP_RT}Radial profiles of the dust temperature, the column
density and the volume density of B68 based on the distributions shown in
Fig~\ref{f:b68rtmap}. All values of the three maps are presented by the small
dots. The big filled dots represent azimuthally averaged
values inside $20\arcsec$ wide annuli. The open white dots in the upper panel
show the same for those dust temperatures that are attained at densities above
$6\times10^{2}$~cm$^{-2}$. The error bars reflect the $1\sigma$ r.m.s. scatter of the azimuthal averaging and hence indicate the deviation from the
spheroid assumption. We show the best ray-tracing fits to the data that were
used to calculate the three maps, indicated by the solid 
lines. The coloured dashed lines correspond to the radial distributions along
three selected directions, as outlined in Fig~\ref{f:b68rtmap} (c). The dotted
lines show the canonical density power-laws for self-gravitating isothermal
spheres, and the grey curve depicts a Gaussian profile with a $FWHM$
equal to the spatial resolution of the maps.}
\end{figure*}

Figure~\ref{f:b68radP_RT} (upper panel) summarises the radial distribution of
the dust temperature. An azimuthal averaging of the data in annuli of $20\arcsec$ is also
shown, both for the entire data and restricted to the a region, where the volume
density is $n_\mathrm{H} \geq 6\times10^2$~cm$^{-2}$. The error bars represent
the $1\sigma$ r.m.s. scatter of the azimuthal averaging. Based on these
measurements, we find a central dust temperature of $T_0=(8.2\pm0.1)$~K. The
azimuthally averaged dust temperature at the edge of B68 amounts to
$T_\mathrm{out} = (17.6 \pm 1.0)$~K for the entire data and $T_\mathrm{out} =
(16.0 \pm 1.0)$~K when omitting the tenuous material where $n_\mathrm{H} <
6\times10^2$~cm$^{-2}$. We have added a corresponding radial profile fit
according to Eq.~\ref{e:T_profile}.

When considering also the systematic uncertainties of the ray-tracing modelling,
the central dust temperature was determined to
$T_\mathrm{d}=(8.2^{+2.1}_{-0.7})$~K, with the errors reflecting both the range
of the valid solutions and statistical errors. We emphasise that this is about
$4$~K below the dust temperature that we derived from the LoS averaged SED
fitting. As a result, the simple SED fitting approach cannot be regarded as a
good approximation for the central dust temperature.

Although the largest scatter in the radial dust temperature distribution of the
best model is only of the order of $1.2$~K ($1\sigma$), the peak-to-peak
variation at the edge of B68 amounts to about 5~K. To illustrate the
contribution of various radial paths to the overall temperature distribution, we
have added three diagnostic cuts through the maps from the centre to the
eastern, the southeastern, and the western directions, indicated in
Figs.~\ref{f:b68rtmap} (c) and \ref{f:b68radP_RT}.

\subsubsection{Column density distribution}
\label{s:results:RTfit:N}
Figure~\ref{f:b68rtmap} (b) shows the column density map of B68 derived from
the ray-tracing algorithm as described in Sect.~\ref{s:modelling:RT}.
It was constructed by LoS integration of the volume density
(Sect.~\ref{s:results:RTfit:n}). The peak of the column density distribution at
RA~$=17^\mathrm{h}22^\mathrm{m}38\fs9$,
Dec~$=-23\degr49\arcmin52\arcsec$ can be fitted by a 2D Gaussian with an
$FWHM$ of $(87\pm1)\arcsec \times (76\pm1)\arcsec$, i.e.~about twice the beam
size.

The radial column density distribution is presented in Fig.~\ref{f:b68radP_RT}
(central panel). It visualises the map values transformed into a radial
projection relative to the density peak as well as, azimuthal averages within
radial bins of $20\arcsec$. When considering the overall modelling uncertainties
as shown in Fig.~\ref{f:rtfit}, the central column density amounts to
$N_\mathrm{H}= (4.3^{+1.4}_{-2.8})\times 10^{22}$~cm$^{-2}$, while we find
$N_\mathrm{H}=(8.2^{+2.3}_{-1.0})\times 10^{20}$~cm$^{-2}$ at the edge of the
distribution.

The radial profile is also consistent with a relation according to
\begin{equation}
\label{e:N_powerlaw}
N_\mathrm{H}(r) = \frac{N_{\mathrm{H},0}}{1+\left(\frac{r}{r_0}\right)^\alpha}
\quad ,
\end{equation}
which has the same properties as Eq.~\ref{e:n_profile}, but with a different
parametrisation (see also Eq.~\ref{e:model_n}). When restricting the fit to $r
\leq 200\arcsec$, i.e. before the column density flattens out into the
background value, we obtain $r_0=45\arcsec$ and $\alpha = 2.5$.

Using the method proposed by \citet{tassis11}, the derived column
density corresponds to a central hydrogen particle density of $n_\mathrm{H}=4.7
\times 10^5$~cm$^{-3}$ with $r_{90} = 2505$~AU, which is about 40\% above the
value we derive from the modelling (see Sect.~\ref{s:results:RTfit:n}).

As already described in Sect.~\ref{s:results:RTfit:T}, we added three radial
cuts through the maps to the profile plots in Fig.~\ref{f:b68radP_RT}. They
include the most extreme column density variations for the most part of the
radial profile, although they are all closely confined to the mean distribution
up to an angular radius of $40\arcsec$, whose seemingly very good approximation
to a spherical symmetry is owed to a poor sampling of the distribution and beam
convolution effects. From here, the eastern cut through the column density drops
more quickly than the globule average with $r^{-5.4}$ between radii of
$75\arcsec$ and $120\arcsec$, before it reaches the common background value. At
approximately $r=50\arcsec$, the southeastern trunk attains a column density
plateau at $N_\mathrm{H} = 2.2\times 10^{22}$~cm$^{-2}$, from which it declines
with $r^{-0.9}$ to a distance of $90\arcsec$ from the density peak. Finally, it
steeply drops with $r^{-9.3}$ between radii of $150\arcsec$ and $190\arcsec$,
where it levels out with the background column density.

\begin{figure*}
\centering
\includegraphics[scale=0.56]{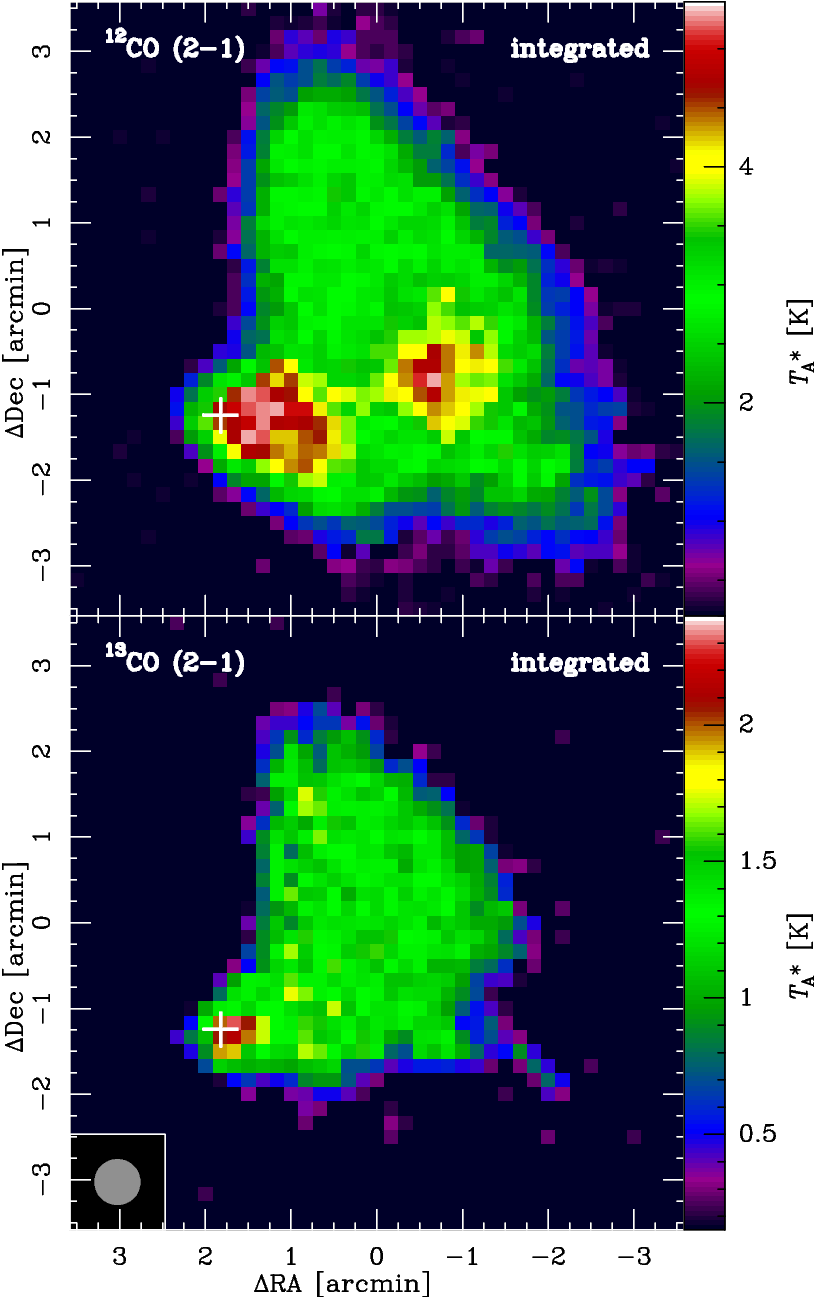}
\hfill
\includegraphics[scale=0.56]{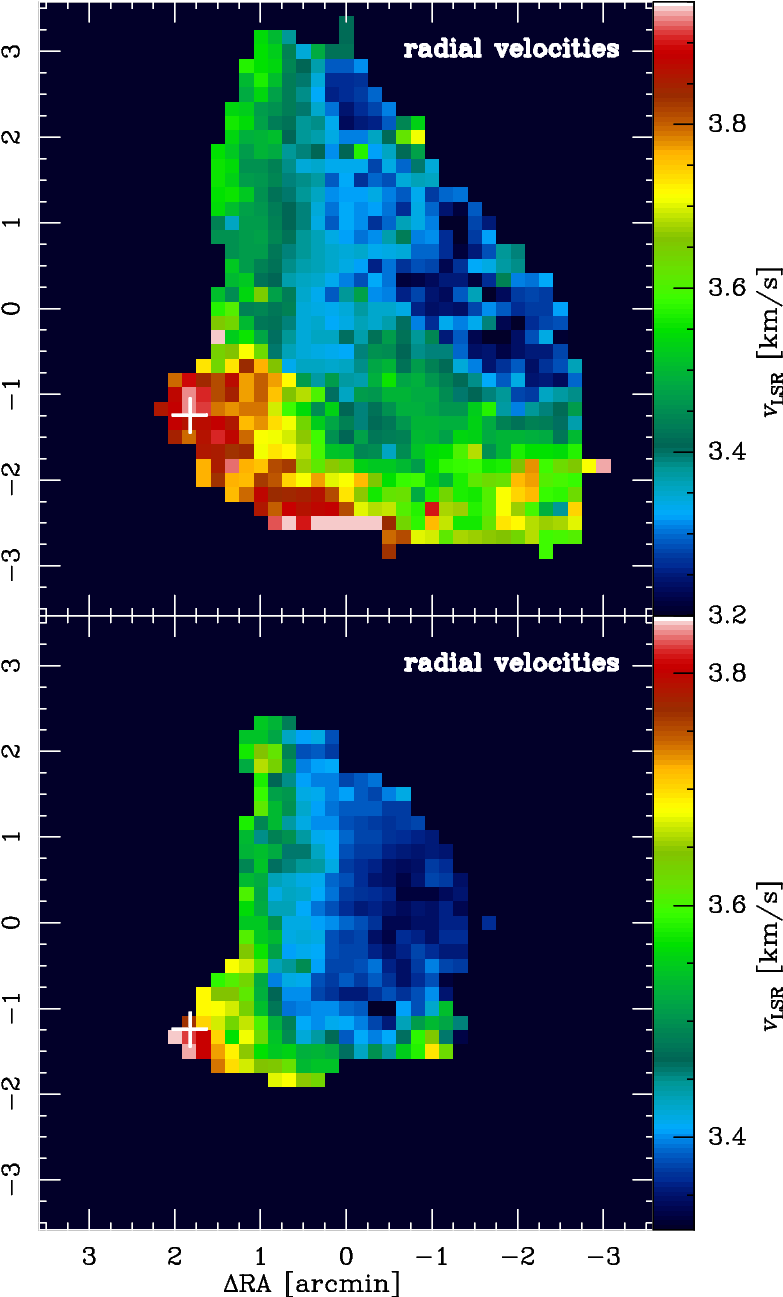}
\hfill
\includegraphics[scale=0.56]{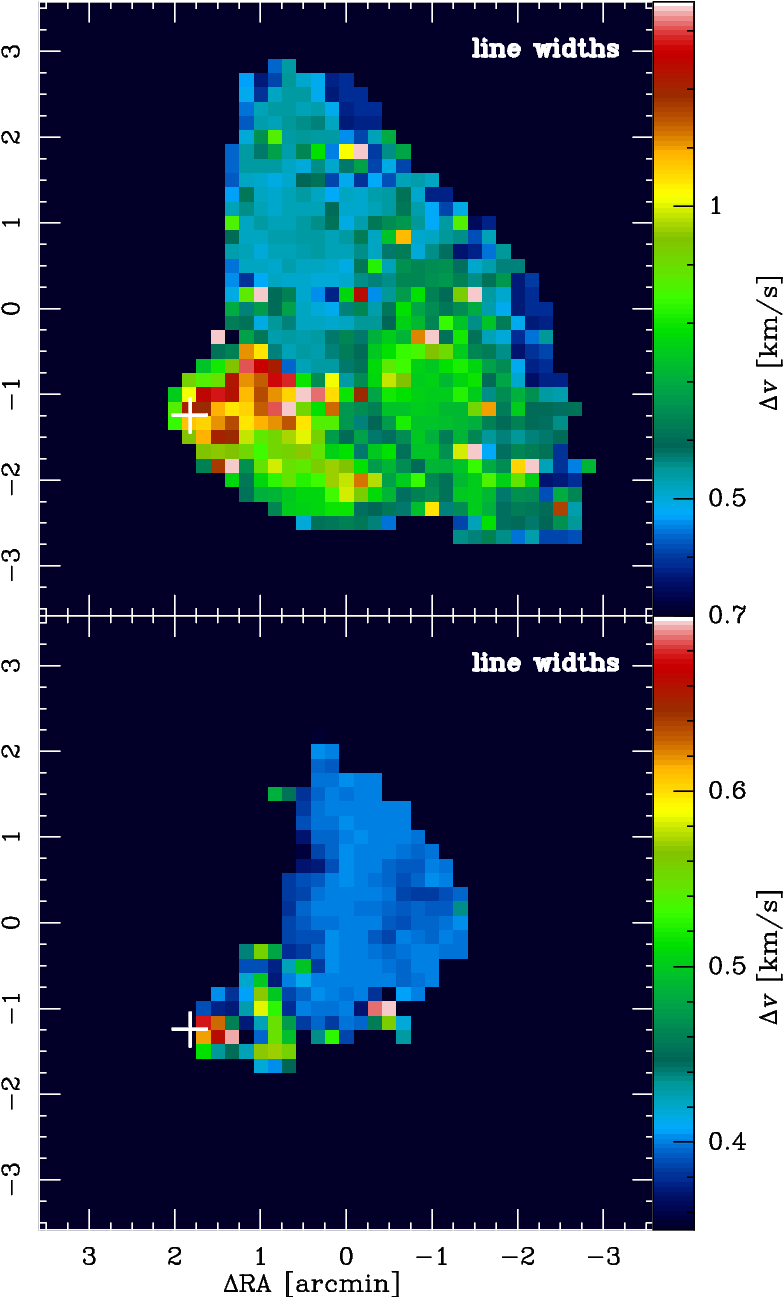}
\caption{\label{f:b68co}Maps of B68 generated from the CO data. The upper row
contains the \element[][12]{CO}~(2--1) line data, while the lower row shows the
\element[][13]{CO}~(2--1) observations. The columns from left to right
indicate the line flux integrated over all spectrometer channels, the radial
velocity distribution, and the corresponding line widths. The white crosses
indicate the position of the point source at the tip of the trunk.}
\end{figure*}

The column density map can be used to derive a mass estimate for B68. When
including all the data in the column density map, which cover an area of
$60\sq\arcmin$, we derive a total mass of $4.2~M_{\sun}$. For the material that
represents $A_K \geq 0.2$~mag, we get a total mass of $3.1~M_{\sun}$. This value
is in good agreement with the one given in \citet{alves01nature} after
correcting for the different assumed distance. In general, the uncertainty on
the mass estimate from the modelling alone is rather small ($\pm 0.5~M_{\sun}$),
because the various solutions more or less only redistribute the material
instead of changing the total amount of mass significantly. We are aware that
the uncertainties due to the dust properties are much greater, but they do not
change the consistency between the different solutions. That the mass derived
from the best ray-tracing fit to the data actually coincides well with the one
extracted independently from the extinction mapping is reassuring and
demonstrates that the quoted uncertainties are conservative.

\subsubsection{Volume density distribution}
\label{s:results:RTfit:n}

Figure~\ref{f:b68rtmap} (c) shows the volume density distribution across the PoS
and in the mid-plane along the LoS through of B68 derived from the ray-tracing
algorithm described in Sect.~\ref{s:modelling:RT}, so it is only the
mid-plane slice of the entire ray-tracing cube-fitting result. The peak of the
volume density distribution RA~$=17^\mathrm{h}22^\mathrm{m}38\fs5$,
Dec~$=-23\degr49\arcmin50\arcsec$ can be fitted by a 2D Gaussian with an $FWHM$
of $(66\pm1)\arcsec \times (61\pm1)\arcsec$, i.e.~just below twice the beam
size.

We show the resulting radial volume density distribution in
Fig.~\ref{f:b68radP_RT} (lower panel). Azimuthal averages are added, whose error
bars indicate the rms noise of the averaging. The best fit to the data according
to Eq.~\ref{e:n_profile_const} is listed in Table~\ref{t:rtfit}. With
Eq.~\ref{e:n_rout}, we obtain $r_1 = 180\arcsec$ for the LoS profile. The fit is
parametrised by $n_0 = 3.4\times 10^5$~cm$^{-3}$, $r_0 = 35\arcsec$, and $\eta =
4.0$. Including the overall conservative ray-tracing uncertainty
(Fig.~\ref{f:rtfit}), the central volume density amounts to $n_\mathrm{H} =
(3.4^{+0.9}_{-2.5})\times 10^5$~cm$^{-3}$. The volume density of the constant
background level is modelled as $n_\mathrm{H} = (4.4^{+1.8}_{-1.5})\times
10^2$~cm$^{-3}$.
 
Similar to Eqs.~\ref{e:N_powerlaw} and \ref{e:model_n}, the radial distribution
can also be fitted with a combination of a flat centre and a power-law at large
radii. The resulting fit is not unique and covers a range of fit parameters with
$r_0=35\arcsec - 40\arcsec$ and $\alpha = 3.4 - 3.7$, when restricting the fit
to $r \leq 190\arcsec$.

The resulting mid-plane volume density distribution for the selected three
directions are also shown. Up to a distance of $1\arcmin$ from the density peak,
all profiles agree very well, but deviate in the transition between the flat
central profile and the decline towards the edge of the globule. As for the
column density distribution, the reduced data sampling and beam convolution
effects can contribute to the seemingly good agreement with a spherical geometry
in the centre of B68. While the average density profile drops, as shown,
approximately with $r^{-3.5}$, the distribution to the eastern edge falls off
more sharply, i.e.~with $r^{-6.5}$ between $r=75\arcsec$ and $120\arcsec$. The
previously reported behaviour of the southeastern trail along the trunk is also
reflected in the volume density that reaches a plateau of $n_\mathrm{H} =
9\times 10^{4}$~cm$^{-3}$ between $50\arcsec$ and $70\arcsec$ away from the
peak. Then it declines up to $r=90\arcsec$ with $r^{-1.8}$, from which it
steeply drops with $r^{-2.7}$ up to a radius of $130\arcsec$, and then with
$r^{-10.2}$ between radii of $150$ and $190\arcsec$.

This analysis shows that, although the mean volume density profile is consistent
with a spherical 1D symmetry in the centre of the globule, where the density
profile is very shallow, that symmetry is broken below densities of
$n_\mathrm{H} = 10^{5}$~cm$^{-3}$ and beyond $r\approx 40\arcsec$ (6000~AU at a
distance of 150~pc). The largest deviations are seen between the western and
southeastern directions, where the volume densities differ by a factor of four
at a distance of $90\arcsec$ from the density peak and attain a maximum of
factor 20 at $r=120\arcsec$. But even if the peculiar southeastern trunk is not
included, we find density deviations between different radial vectors of a
factor of 2 at $90\arcsec$ to a factor of 5 at $120\arcsec$.
 
\subsection{CO data}
\label{s:co}
We have mapped the larger area around B68 in the
\mbox{\element[][12]{CO}~(2--1)} and the \element[][13]{CO}~(2--1) lines. The
resulting integrated line, radial velocity, and line width maps are shown in
Fig.~\ref{f:b68co}. Keeping in mind that particularly the centre of B68 is
strongly affected by CO gas depletion \citep[e.g.~][]{bergin02}, the global
shape resembles the known morphology quite well. We see two maxima in the
\element[][12]{CO}~(2--1) integrated line maps, while only one peak is visible
in the optically thinner \element[][13]{CO}~(2--1) line maps. Interestingly, it
coincides with the local maximum at the tip of the southeastern extension of the
extinction map of \citep{alves01eso} and the source we discovered at 160 and
250~$\mu$m.

We confirm a global radial velocity gradient with slightly increasing values
from the northwest to the southeast that have been reported by others
\citep[e.g.~][]{lada03,redman06} and may indicate a slow global rotation or
shear movements. The line widths are rather small and in most cases cannot be
resolved by our observations. Strikingly, the highest redshifts and largest
apparent line widths are observed towards the aforementioned peak at the tip of
the extension, identified as a point source in Sect.~\ref{s:morphology}. To
investigate this phenomenon closer, we have extracted
\mbox{\element[][13]{CO}~(2--1)} spectra at that position and, for comparison,
one at the western edge of B68 (available as online material,
Fig.~\ref{f:13coline}).

The spectrum taken at the position of the point source is shifted by
0.4~km\,s$^{-1}$, which corresponds to the spectral resolution of the data, and
broadened by approximately the same amount. A significant local increase in the
temperature can be ruled out by the dust temperature maps and the low colour
temperature extracted from the photometry in the \textit{Herschel} maps, so that a
line broadening due to thermal effects seems unlikely, at least at the spectral
resolution obtained with our data. Instead, it can be explained by a
superposition of two neighbouring radial velocities ($\sim 3.3$ and $\sim
3.7$~km~s$^{-1}$). This interpretation is supported by already published
spectral line observations at higher spectral resolution that reveal a
double-peaked line shape \citep[e.g.~][]{lada03,redman06}.

\section{Discussions}
\subsection{Temperature and density}
\label{s:tnd}

\subsubsection{The temperature distribution}
\label{s:discussion:t}
In Sect.~\ref{s:results:RTfit:T}, we have presented the results of the dust
temperature distribution of B68. At the resolution of the SPIRE 500~$\mu$m
data ($36\farcs4$), the central dust temperature was determined to $(8.2 {+2.1
\above0pt -0.7})$~K, considering the uncertainties of the ray-tracing modelling
algorithm (see Sect.~\ref{s:modelling:RT}). The temperature at the edge of B68
including volume densities below $6\times10^{2}$~cm$^{-3}$ was measured to
16.7~K with an uncertainty of $\pm1$~K. When considering only material that is
denser than this value, the outer dust temperature is $0.7$~K lower, which
results in an edge-to-core temperature ratio of $2.0$.

Model calculations using radiative transfer towards idealised isolated 1D BES of
similar densities show central temperatures of the order of $7.5$~K and
higher \citep{zucconi01} which confirms our result. Modifying the BE approach
to a non-isothermal configuration does not change the situation significantly
\citep{sipilae11}. The coldest prestellar cores found to date populate a
temperature range around 6~K, but also possess densities and masses higher
than B68 \citep[e.g.~][]{pagani04,doty05,crapsi07,pagani07,stutz09}.

Previous estimates of the dust temperature in B68 lacked either the necessary
spectral or spatial coverage and quote mean values around 10~K for the entire
core \citep[e.g.~][]{kirk07}. In addition, \citet{bianchi03} were able to deduce
a deviation from the isothermal temperature distribution by analysing (sub)mm
data. They derived an external dust temperature of $14\pm2$~K, which, in the
absence of higher frequency data, tends to underestimate the real value.
Nevertheless, it is consistent with our results when discarding the warmer
low-density regime.

Determining the central gas temperature and its radial profile turned out be
more successful \citep[e.g.~][]{hotzel02temp,bergin06}, and generally one would
expect both the gas and the dust temperature to be tightly coupled in the core
centre, where they experience densities of $n_\mathrm{H} \geq 10^5$~cm$^{-3}$
\citep[e.g.~][]{lesaffre05,zhilkin09}, with the strength of the coupling
gradually increasing already from $n_\mathrm{H} \approx 10^4$~cm$^{-3}$
\citep{goldsmith01}. However, \citet{bergin06} found evidence that this might
not be necessarily the case. They have reconstructed the dust temperature
profile by chemical and radiative transfer modelling of CO observations (their
Fig.~2a) and also predict a central value of $\approx7.5$~K, which is
significantly below what they find for the central gas temperature. Their
results even revealed a gas temperature decrease to larger radii, which is
confirmed by \citet{juvela11} in a more generalised modelling programme. They
demonstrate that the gas temperature in dense cores can be affected by many
contributions and thus alter the central temperature by up to 2~K, depending on
the composition of the various physical conditions. This can lead to diverging
gas and dust temperatures in the centre of dense cores
\citep[e.g.~][]{zhilkin09}. According to \citet{bergin06}, this can only be
explained by a reduced gas-to-dust coupling in the centre of B68, especially
when -- as in this case -- the gas is strongly depleted.

\begin{figure}
\centering
\resizebox{\hsize}{!}{\includegraphics{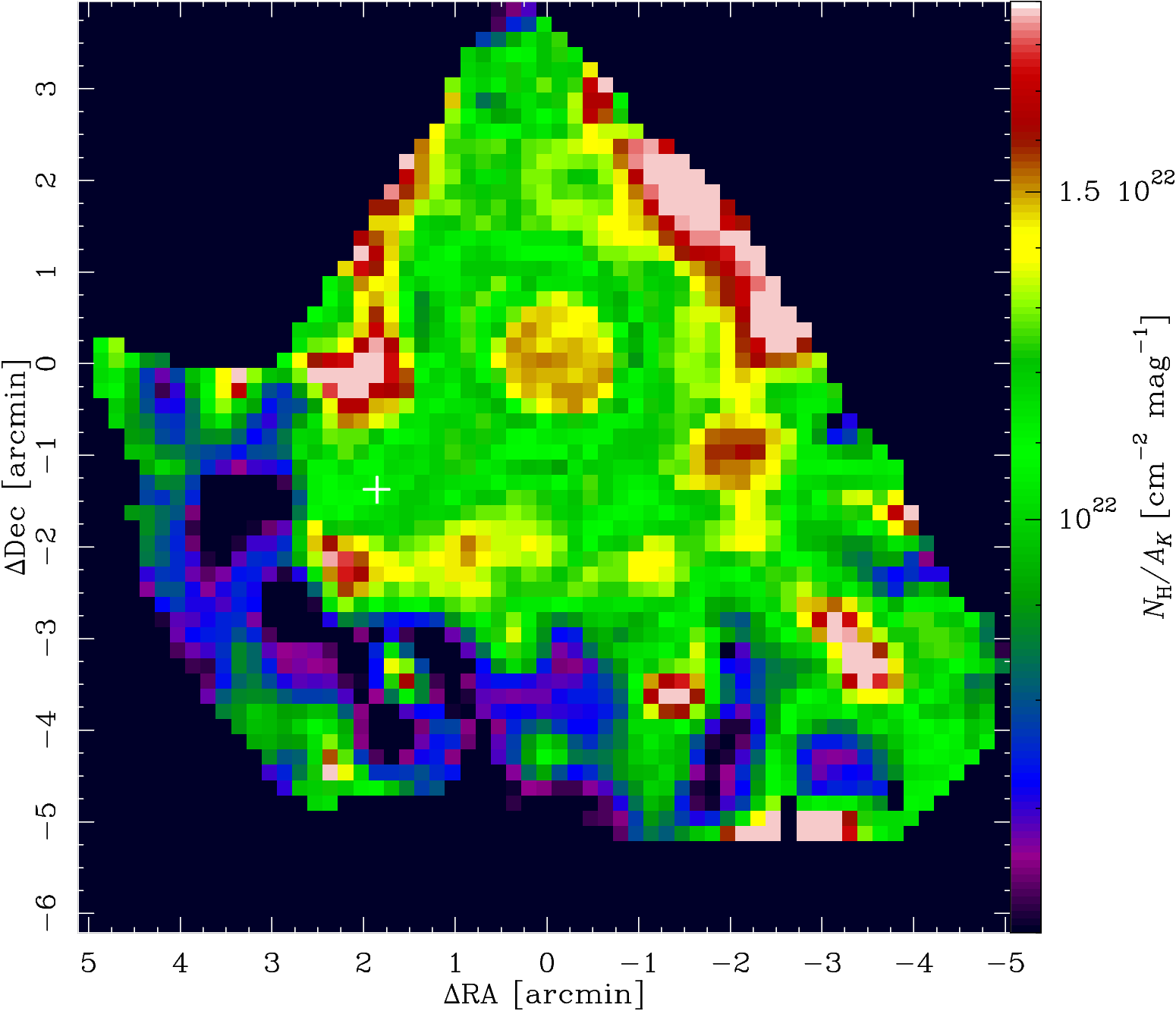}}
\caption[]{\label{f:AKvsNmap}$N_\mathrm{H}$ vs. $A_K$ ratio map. The
uncertainties increase strongly beyond a radius of $180''$, because a) the flux
densities and extinction values are low towards the southeast, b) the
extinction map has a higher uncertainty there, c) edge effects in the modelled
flux distribution. The position of the embedded point source is indicated with a
white cross.}
\end{figure}

\subsubsection{The density distribution}
\label{s:discussion:n}
In Sect.~\ref{s:results:RTfit:N}, we have described and analysed the column
density distribution that we derived from the ray-tracing modelling. It peaks at
$N_\mathrm{H}= (4.3 {+1.4 \above0pt -2.8})\times 10^{22}$~cm$^{-2}$ towards the
centre of B68 and blends in with the ambient medium of $N_\mathrm{H} =
(8.2^{+2.3}_{-1.0})\times 10^{20}$~cm$^{-2}$ at $\approx 200\arcsec$. The
central value is in good agreement with the results of \citet{hotzel02temp}.
Further investigations led them to the conclusion that the gas-to-dust ratio in
B68 is below average, i.e.~also less than our assumption made for the modelling.
The actual column density may thus be lower.

\begin{figure}
\centering
\resizebox{\hsize}{!}{\includegraphics{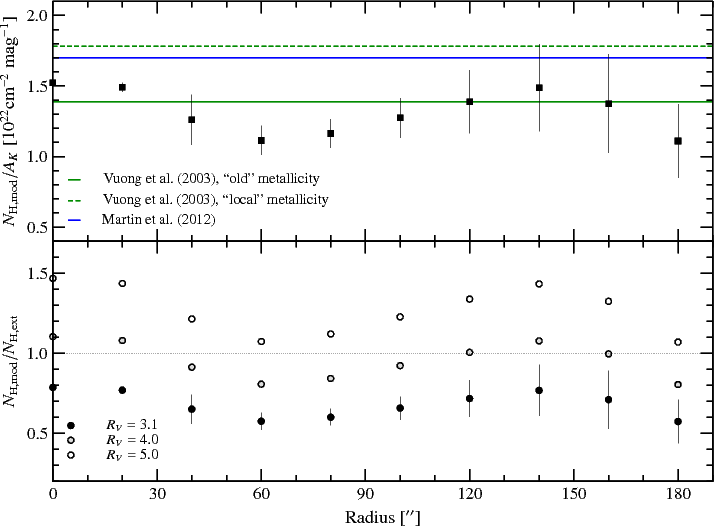}}
\caption[]{\label{f:AKvsN}Radial profiles of the $N_\mathrm{H,mod}/A_K$ ratio
(upper panel) and the $N_\mathrm{H}$ ratio between the modelled column density
and the one derived from the NIR extinction for three values of $R_V$ (lower
panel).  The $N_\mathrm{H,mod}/A_K$ ratio attains a mean value of $1.3 \times
10^{22}$~cm$^{-2}$~mag$^{-1}$ with a scatter of 12\%.  This corroborates the
estimate of \citet{vuong03} using the standard metallicity of the ISM.  A good
agreement between the modelled column density and the one based on the
extinction is achieved for a slightly increased $R_V$ value of about 4.}
\end{figure}

Since we have two independent measurements of the column density, the extinction
and the emission data, we can compare both and test how consistent they are. For
this purpose, we first analysed a possible variation in the ratio between the
$N_\mathrm{H}$ and $A_K$ maps. A ratio map is shown in Fig.~\ref{f:AKvsNmap}.
The strong drop towards the southeast is very uncertain due to higher
uncertainties in the extinction and the modelled column density maps. Therefore,
we only consider the area up to $r = 180''$. Within this range, the mean value
is $N_\mathrm{H}/A_K = 1.3 \times 10^{22}$~cm$^{-2}$~mag$^{-1}$. The ratios at
the centre of B68 appear slightly increased when compared to the rest of the
globule. An azimuthally averaged radial profile up to $r=180''$ is presented in
Fig.~\ref{f:AKvsN} (upper panel). The variation amounts to 12\% around the mean
value. We also plot the relationships found by \citet[][]{vuong03} and
\citet[][]{martin12}. In fact, the latter work presents a conversion between
$N_\mathrm{H}$ and $E_{J-K}$, which is based on a linear fit to data points with
a significant amount of scatter. In addition, those data only covered
extinctions up to $A_V \approx 3$~mag. Nevertheless, when applying the
extinction law of \citet{cardelli89}, we find $N_\mathrm{H}/A_K = 1.7\times
10^{22}$cm$^{-2}$mag$^{-1}$.

\citet{vuong03} derive values in the $\rho$~Oph cloud using two different
assumptions of the metallicity that sensitively modify the extinction determined
from their X-ray measurements. The corresponding values for the $N_{\rm H}/A_K$
ratio differ by almost 30\%. They range between
$1.4\times10^{22}$cm$^{-2}$mag$^{-1}$ for standard ISM abundances and
$1.8\times10^{22}$cm$^{-2}$mag$^{-1}$ for a revised metallicity of the solar
neighbourhood \citep{holweger01,allende01,allende02}, which is apparently closer
to the mean value of the Galaxy, if $R_V = 3.1$ is assumed. However, one should
probably not even expect a good agreement considering the different LoS towards
the Ophiuchus region and the Galactic centre. In Fig.~\ref{f:AKvsN}, the value
for the standard ISM abundances is in remarkably good agreement with our
analysis, although the alternative ratios still fit reasonably well, considering
the large uncertainties.

For a suitable conversion from $A_K$ to $N_\mathrm{H}$, we can also compare
ratios of the column densities derived from the modelling and the extinction map
directly. \citet{vuong03} have already discussed the possibility that their
apparent mismatch of the $N_\mathrm{H}/A_J$ ratio with the galactic mean value
may not be related to a difference in metallicities, but more a variation of the
$R_V$ value from the one of the diffuse ISM. They refer to a previous estimate
of $R_V = 4.1$ in the outskirts of the $\rho$~Oph cloud \citep{vrba93}. In
general, increased $R_V$ values are common in dense regions
\citep[e.g.][]{chini83,whittet92,kruegel03,chapman09a,olofsson10}. Therefore, we
chose to apply a conversion via $A_V$ by assuming the extinction law of
\citet{cardelli89} and a given value of $R_V$.

Recent conversions between $A_V$ and $N_\mathrm{H}$ are determined for the
diffuse ISM that on average attains $R_V =3.1$. For such an assumption, we can
use $2.2\times 10^{21}$~cm$^{-2}$~mag$^{-1}$ \citep{ryter96,guever09,watson11}.
We can then derive modified conversions for varying values of $R_V$, which in
our analysis was iterated between 3.1, 4.0, and 5.0. The corresponding column
density ratios between the modelled distribution and the one derived from the
extinction are shown in Fig.~\ref{f:AKvsN} (lower panel). A reasonable agreement
is only reached for a slightly increased $R_V$ value around 4, which is in
excellent agreement with the value obtained by \citet{vrba93}. A change in $R_V$
is usually attributed to an anomalous grain size distribution
\citep[e.g.~][]{roman07,mazzei08,fitzpatrick09}, which would be consistent with
the interpretation of \citet{bergin06} for a reduced gas-to-dust coupling rate.
While the result of an increased $R_V$ relative to the canonical $R_V = 3.1$ in
the diffuse ISM may point to the possible presence of grain growth processes in
B68, systematic uncertainties in the modelling could affect the derived $R_V =
4$ value. Therefore this result should be interpreted with caution.

The radial distribution of column densities or extinction values of prestellar
cores is often modelled to represent a BES. For B68, this was done by
\citet{alves01nature}, who were able to fit a BE profile with an outer radius of
$100''$ (see Fig.~\ref{f:extprofiles}, available as online material). A quick
investigation demonstrates that power-law profiles according to
Eqs.~\ref{e:n_profile} and \ref{e:N_powerlaw} with a wide variety of parameters
fit the radial distribution up to $r=60''$ equally well. However, the derived BE
profile fails to match the column density distribution between radii of $100''$
and $200''$. This does not mean that no BES model fits the full column density
profile of B68, but at least the one derived by \citet{alves01nature} cannot
account for the extended distribution shown in Figs.~\ref{f:b68radP_RT} and
\ref{f:extprofiles}. In contrast, the profiles we use to characterise the radial
distributions fulfil this requirement. They are even very similar for the two
radial extinction profiles we show in Fig.~\ref{f:extprofiles}, which indicates
that, within the mutually covered area, the revised extinction map
(Fig.~\ref{f:b68ext}) agrees well with the one of \citet{alves01eso}.

Section~\ref{s:results:RTfit:n} presents our ray-tracing modelling results of
the volume density distribution in the mid-plane of B68, i.e.~traced across the
PoS at a position half way through the globule. We find a central particle
density of $n_\mathrm{H} = (3.4 {+0.9 \above0pt -2.5})~\times 10^{5}$~cm$^{-3}$,
while it drops to $n_\mathrm{H} = (4.4 {+1.8 \above0pt -1.5})~\times
10^{2}$~cm$^{-3}$ at the background level. That the resulting mass range of $3.1
- 4.2~M_{\sun}$ for an assumed distance of 150~pc is in good agreement with the
previous independent estimates puts the valid density range into a different
perspective and demonstrates that assumed uncertainties are rather conservative.

The central density is a factor of $0.4$ below what we estimate by applying
the simplified method of \citet{tassis11}, but it is consistent with the
result of \citet{dapp09}. \citet{tassis11} expect the deviations in the results
derived with their method to be within a factor of two of the real value, and
\citet{dapp09} assume a temperature of $11$~K for the entire core. Thus,
considering the uncertainties, their results are in reasonably good agreement
with our analysis.

The theoretical model of inside-out collapse of prestellar cores predicts a
density profile of $n_\mathrm{H} \propto r^{-2}$ \citep[e.g.~][]{shu77,shu87},
which appears to be present in most of the cores observed to date
\citep[e.g.~][]{andre96}. Nevertheless, steeper profiles have been reported as
well \citep[e.g.~][]{abergel96,bacmann00,whitworth01,langer01,johnstone03g11}.
As mentioned in Sect.~\ref{s:results:RTfit:n}, we also find a rather steep
power-law relation for the mean profile and most of the radial cuts through the
density map of the order of $n_\mathrm{H} \propto r^{-3.5}$. It is consistent
with the relation we find for the column density, whose slope follows a power
law of $N_\mathrm{H} \propto r^{-2.5}$. \citet{bacmann00} show that the simple
relation $N_\mathrm{H} \propto r^{-m} \Leftrightarrow n_\mathrm{H} \propto
r^{-(m+1)}$ \citep[e.g.~][]{adams91,yun91} does not hold for a more realistic
description by piecewise power-laws. For large radii, the difference of the two
slopes is reduced to $\approx 0.6$. That the revised extinction map
(Fig.~\ref{f:b68ext}) shows similar power-law behaviour of the form $A_K \propto
r^{-3.1}$ for $60\arcsec \leq r \leq 180\arcsec$ (available as online material,
Fig.~\ref{f:extprofiles}) independently confirms our result. In addition, we
confirm this finding with the better resolved NIR extinction map of
\citet{alves01eso}, although the area covered with these data is considerably
smaller (available as online material, Fig.~\ref{f:extprofiles}). This
demonstrates that resolution effects are negligible in this aspect.

To be able to evaluate possible physical reasons for such steep slopes, we first
have to discuss contributions that result from the observing strategy and the
data handling. Our ray-tracing approach produces a volume density map that
relies on the profile fitting of the first estimate of the column density
distribution we get from the SED fitting. Therefore, any modification introduced
by technical and procedural contributions affects the column density
distribution first and then propagates through to the modelled volume density.
Since the profiles derived across the PoS are used to parametrise the behaviour
along the LoS, they have a direct impact on the ray-tracing results for the
volume density distribution as well. To find non-physical reasons for
the steep volume density profiles, we must therefore take one step backward and
have a look at the equally steep column density profiles.

One contribution comes from resolution effects. To test their influence on the
radial profiles, we performed 1D radiative transfer calculations of a model core
with a flat inner density distribution that turns over into $n_\mathrm{H}\propto
r^{-2}$ behaviour towards the edge. At large radii, the resulting column density
profile possesses a relation $N_\mathrm{H}\propto r^{-1.4}$ due to LoS
projection effects, as already shown by \citet{bacmann00}. After convolving this
configuration with a beam of $37\arcsec$, the $r^{-1.4}$ relation has migrated
to smaller radii, simulating a steeper slope in the inner parts of the core. In
combination with the flattening of the core centre, this produces the impression
of a relative strong turnaround from a rather flat to a steep slope at
intermediate core radii. However, this effect cannot change the steepness of the
slopes. As a result, it cannot explain the steep $N_\mathrm{H} \propto r^{-2.5}$
behaviour beyond $r\approx 60\arcsec$, unless the actual profile reaches such a
slope anywhere along the observed radii. We emphasise that this effect is not
unique to the data used in this investigation, and to some degree all
observations suffer from this effect, depending on the spatial resolution. If it
were responsible for the observed steepness, we should expect to see a relation
of the observed slope with resolving power. Already the moderately convolved
($10\arcsec$) extinction data of \citet{alves01nature} produce a mean radial
profile whose slope is nearly indistinguishable from the one of the column
density profile shown in Fig.~\ref{f:b68radP_RT}
(cf.~Sect.~\ref{s:discussion:n}), so, the more than three times worse resolution
does not change the situation significantly.

Another reason for the steepness of the mean radial profile is the asymmetry of
B68 that can be best judged by the profiles of the three cuts through the
density map. In particular, the path towards the east drops much more quickly
than others, which points to a smaller extent in this direction. Such parts
reduce the azimuthal averages, which leads to a faster decline of the mean
radial profile in relation to one that belongs to a symmetric core. This also
means that for asymmetric cores -- and most cores are like that -- an
azimuthally averaged radial profile is of limited value, especially at large
distances from the core centre. Instead, a proper 3D treatment is needed to
actually infer realistic physical core properties. This will be attempted in a
forthcoming paper.

Nevertheless, Fig.~\ref{f:b68radP_RT} demonstrates that most data
follow the same power-law as the mean profile which means that for the
majority of the core radii, the density does behave like $n_\mathrm{H} \propto
r^{-3}$. Interestingly, only the cut following the southeastern trunk is close
to the theoretically predicted $N_\mathrm{H}\propto r^{-1}$ and $n_\mathrm{H}
\propto r^{-2}$ relations.

An artificial offset in the density can also modify the slope of a power
law. If, for instance, we add a constant density of $10^4$~cm$^{-3}$, the mean
radial profile attains a $n_\mathrm{H} \propto r^{-2}$ power-law for
$60\arcsec\leq r \leq 120\arcsec$. However, it is difficult to imagine how such
an offset could be introduced by the SED fitting and the modelling via the
ray-tracing algorithm.

\begin{figure}
\centering
\resizebox{\hsize}{!}{\includegraphics{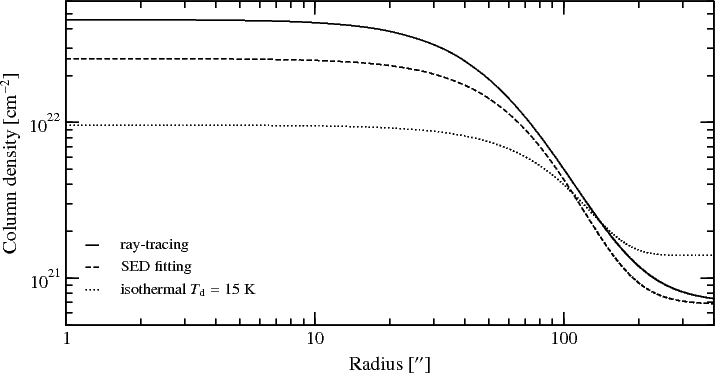}}
\caption{\label{f:NHcomp}Comparison of radial column density profiles for three
different approaches of confining the dust temperature.}
\end{figure}

In Fig.~\ref{f:NHcomp}, we demonstrate the influence of three different
modelling approaches on the resulting column density estimate for B68. The
simplest way to constrain the temperature is to assume a constant value for the
entire core, as has been done many times before. This unrealistic assumption of
isothermality apparently tends to produce an extremely broad, flat central
density distribution at a relatively low column density estimate. In the current
example of $T_\mathrm{d} = 15$~K, the central column density estimate is only
half of the ray-tracing result. The modified black body fit approach mitigates
this disadvantage to some extent, but still underestimates the central density
owing to the already discussed LoS averaging effect. Another clear difference is
the slope of the intermediate region between the core centre and its edge.
Fitting the transition region around $r=100\arcsec$ by $N_\mathrm{H} = N_0
r^{-\alpha}$, we find $\alpha = 1.6$ for the isothermal case, but $\alpha
\approx 2.4$ for the SED fitting and the ray-tracing approaches. However,
fitting Eq.~\ref{e:N_powerlaw} yields $\alpha \approx 2.5$ for all three radial
distributions, with considerable differences of the turnaround radius $r_0$
ranging from $87\arcsec$ for the isothermal case via $55\arcsec$ for the
modified black body fit to $45\arcsec$ for the ray-tracing. This demonstrates
that apparently a simple description via $N_\mathrm{H} = N_0 r^{-\alpha}$ can be
misleading, if the ratio between the calculated width of the inner core and the
total core size is too large, as it is obviously the case for the isothermal
assumption. In other words, the turnaround has not fully progressed yet, before
the profile slowly fades out into the tenuous density of the background or halo.
In such situations, it is difficult if not impossible to actually identify a
range of constant decline.

Since the ray-tracing modelling produces a range of valid power-law slopes
(Fig.~\ref{f:rtfit}) between $\alpha = 3.1$ and $4.5$, the actual value may be
uncertain. Nevertheless, it appears justified to conclude that $\alpha > 3$ and
therefore the slope we find in the ray-tracing data is clearly steeper than
$\alpha = 2$. For this reason, it is legitimate to look for physical reasons for
this phenomenon. \citet{andre00} point out that isolated prestellar cores appear
to have sharp edges with density gradients steeper than $n_\mathrm{H} \propto
r^{-3}$ for $r \ga 15\,000$~AU, i.e.~in this case $100\arcsec$, that could be
explained by a core in thermal and mechanical equilibrium interacting with an
external UV radiation field, if the models of \citet{falgarone85} and
\citet{chieze87} are considered. \citet{whitworth01} have picked up that
statement and derived an empirical, non-magnetic model for prestellar cores that
follows a Plummer-like density profile (Eq.~\ref{e:n_profile}) with $\eta = 4$.
Strikingly, we find exactly the same value in our LoS profile fit.

Interestingly, Eq.~\ref{e:n_profile} with $\eta = 4$ is analytically almost
identical to the theoretical density profile of an isothermal, non-magnetic
cylinder \citep{ostriker64}. This concept was observationally confirmed by
\citet{johnstone03g11}, who find density profiles steeper than $r^{-3}$ for
several sections in the filament of \object{GAL~011.11-0.12}. The relevance to
the case B68 is the conjecture that it belongs to a chain of individual globules
that once formed as fragments of a filament that has meanwhile been dispersed.
Figure~\ref{f:b68obs} illustrates this view, where several cores appear aligned
along a curved line. In addition, there is B72 to the north of the image that is
still embedded in a tenuous filament. It is conceivable that the remaining
fragments still contain an imprint of the density profiles of the former
filament. In that case, the other globules of that chain should also possess
steep profiles, just like B68. If this scenario is true, B68 might not be the
prototypical isolated prestellar core as often suggested. This idea plays a
central role in the concept of colliding cores we discuss in
Sect.~\ref{s:bullet}.

\begin{figure}
\centering
\resizebox{\hsize}{!}{\includegraphics{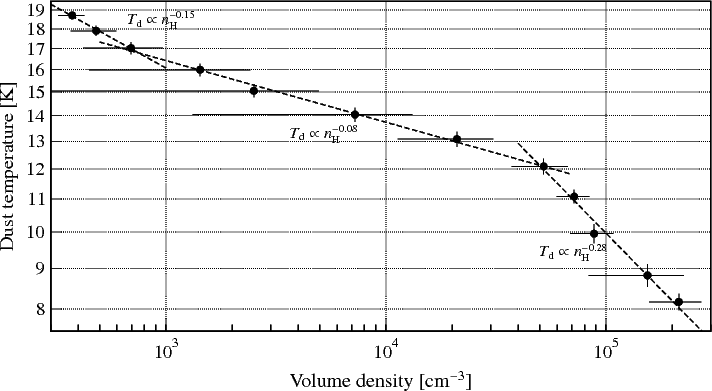}}
\caption{\label{f:Tofn}Dust temperature profile as a function of volume density
$n_\mathrm{H}$. The error bars indicate the r.m.s. scatter of the averaging
inside the individual bins. The relation can be distinguished into three
regions.}
\end{figure}

To visualise the relation between the resulting dust temperature and
volume density distributions, we have produced a temperature averaged density
profile (Fig.~\ref{f:Tofn}). The relation can be distinguished into three
sections with $T_\mathrm{d}\propto n_\mathrm{H}^\beta$, where
\begin{displaymath}
\beta = \left\{
\begin{array}{ll}
-0.15 & \quad \textrm{for \quad $n_\mathrm{H} < 10^3$~cm$^{-3}$}\\
-0.08 & \quad \textrm{for \quad $3\times 10^4$~cm$^{-3} > n_\mathrm{H} \geq 10^3$~cm$^{-3}$}\\
-0.28 & \quad \textrm{for \quad $n_\mathrm{H} > 6\times 10^4$~cm$^{-3}$} \quad .
\end{array}
\right.
\end{displaymath}
We interpret this behaviour as a result of the different heating and cooling
effects that are influencing the temperature and density structure of B68. Even
for relatively dense cores, the heating by the external ISRF is the dominating
contribution \citep{evans01}. Only the effective spectral composition changes
with the density of the core material. At the tenuous surface, heating by UV
radiation can be non-negligible, which leads to a higher heating efficiency. At
densities above $10^3$~cm$^{-3}$, heating by UV radiation is shut off. The IR
and mm radiation of the ISRF as well as cosmic rays with their secondary
effects, remain as the only heat sources. This leads to a shallower relation
between the dust temperature and the density. At densities $10^4$~cm$^{-3} <
n_\mathrm{H} < 10^5$~cm$^{-3}$, gas and dust begin to couple, which coincides
with the turnaround to a steeper relationship. This range corresponds to small
radii, so that beam smearing effects contribute to the slope measured in the
data. The actual behaviour may be different.

\subsection{Anisotropic interstellar radiation field}
\label{s:aniso}
\begin{figure}
\centering
\resizebox{0.8\hsize}{!}{\includegraphics{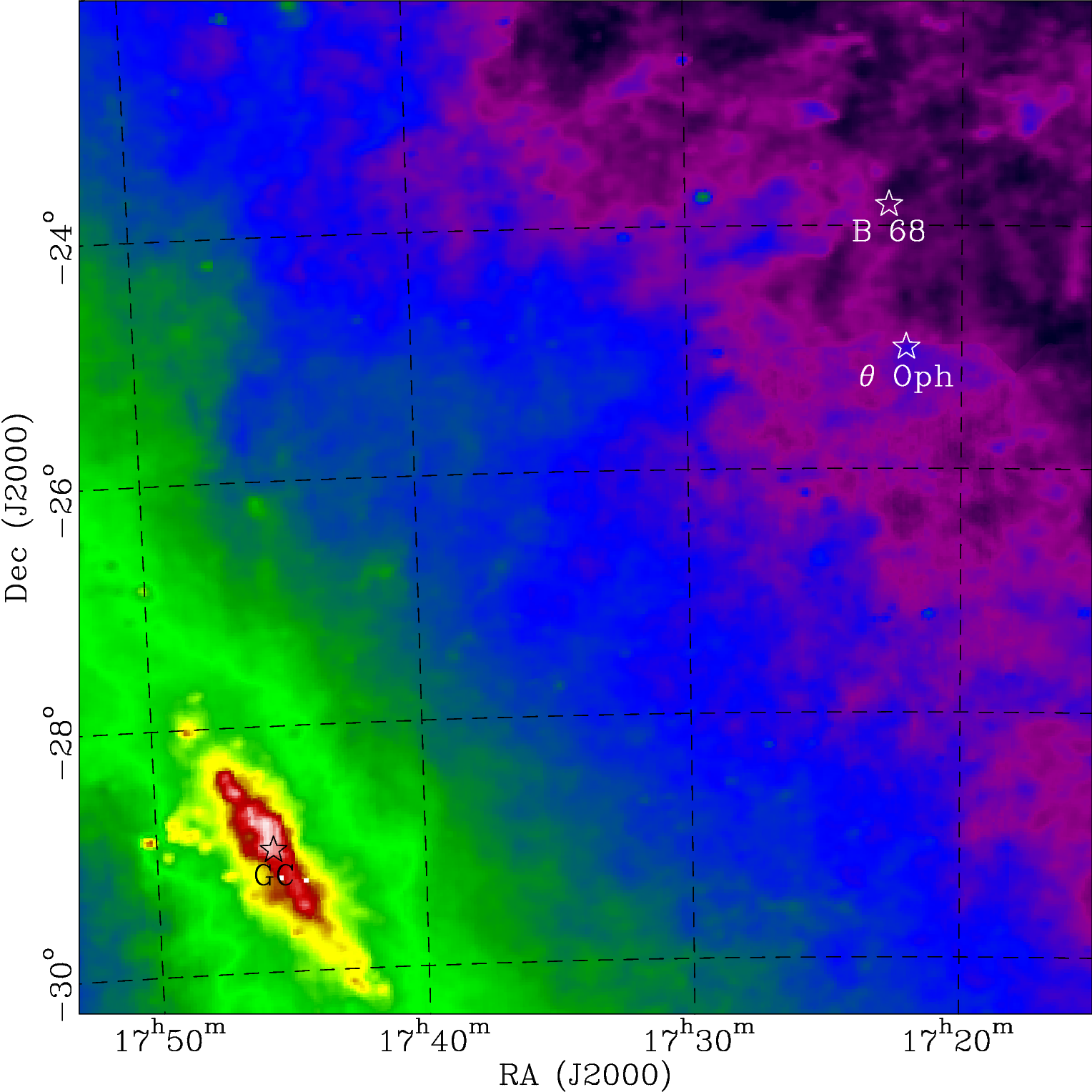}}
\caption{\label{f:b68gal}IRAS map at $\lambda = 60~\mu$m covering the
galactic plane and B68. North is up and east is left. B68 is located 7\fdg1
above the galactic plane. At a distance of 150~pc, this translates to a galactic
altitude of 40~pc. The position of the B2IV star \object{$\theta$~Oph} is
also indicated. Its parallax of $\pi=7.48\pm0.17$~mas \citep{hipparcos07} puts
it at a distance of 134~pc. Its distance to B68 is approximately 20~pc.}
\end{figure}

The ISRF at the location of B68 heating the dust particles is unknown. We
expect it to be strongly anisotropic and wavelength-dependent as revealed
e.g.~by the all-sky surveys like the Diffuse InfraRed Background Experiment
(DIRBE) at solar location. Such an anisotropic irradiation can cause
surface effects on isolated globules like B68 and may explain the crescent
structure discovered with the \textit{Herschel} observations at 100 and 160~$\mu$m.
Since this also implies an anisotropy in the heating, the discovery of a sudden
rise of the dust temperature at the southeastern rim, as described in
Sect.~\ref{s:results:SEDfit}, may be another result of the ISRF geometry. As
shown by Fig.~\ref{f:b68gal}, this side faces the galactic plane at a latitude
of about 7\fdg1. For an altitude of the sun of 20.5~pc above the galactic plane
\citep{humphreys95}, the galactic altitude is approximately 40~pc for B68. This
suggests that the local radiation field of the Milky Way may heat B68 from
behind, but at some degree also from below. With respect to dust heating, its
contribution can be divided into two parts.

At longer wavelengths, the SED of the Milky Way peaks around 100~$\mu$m that
varies only with galacto-centric distance \citep[e.g.~][]{mathis83}, aside from
the 2.7~K cosmic background contribution. The dust opacities are low enough at
FIR wavelengths to keep low-mass cores like B68 transparent. For grain heating
calculations, the anisotropy in the FIR ISRF can therefore be ignored.

At shorter wavelengths, several effects make the prediction of the local B68
ISRF and its heating more difficult. The stellar radiation sources create an
SED, which requires a superposition of Planck curves when being modelled peaking
around $1~\mu$m, all showing spatial variations. Since the optical depth at
these wavelengths becomes comparable to or greater than one for many
lines-of-sights in the Milky Way, the absorbed energy can strongly vary even for
neighbouring cores, depending on their relative positions with respect, e.g., to
the Galactic centre. Furthermore, scattering becomes important and adds a
diffuse component that heats shielded regions. Finally, outer diffuse cloud
material surrounding most molecular cloud cores can pile up enough dust mass to
effectively shield the core from the NIR and MIR radiation.

There are just a few core models that make use of an anisotropic background
radiation field. In their study of the core \object{$\rho$~Oph~D},
\citet{steinacker05} have used a combination of a mean ISRF given in
\cite{black94}, an MIR power-law distribution as suggested in \cite{zucconi01}
to account for the nearby photo-dissociation region (PDR), and the radiation
field of a nearby B2V star. A similar approach was applied by \citet{nutter09}
to model \object{LDN 1155C}. The anisotropic DIRBE all-sky map at
3.6~$\mu$m was used in \cite{steinacker10} to model the core \object{LDN~183} in
absorption and scattering.

\begin{figure}
\centering
\resizebox{\hsize}{!}{\includegraphics{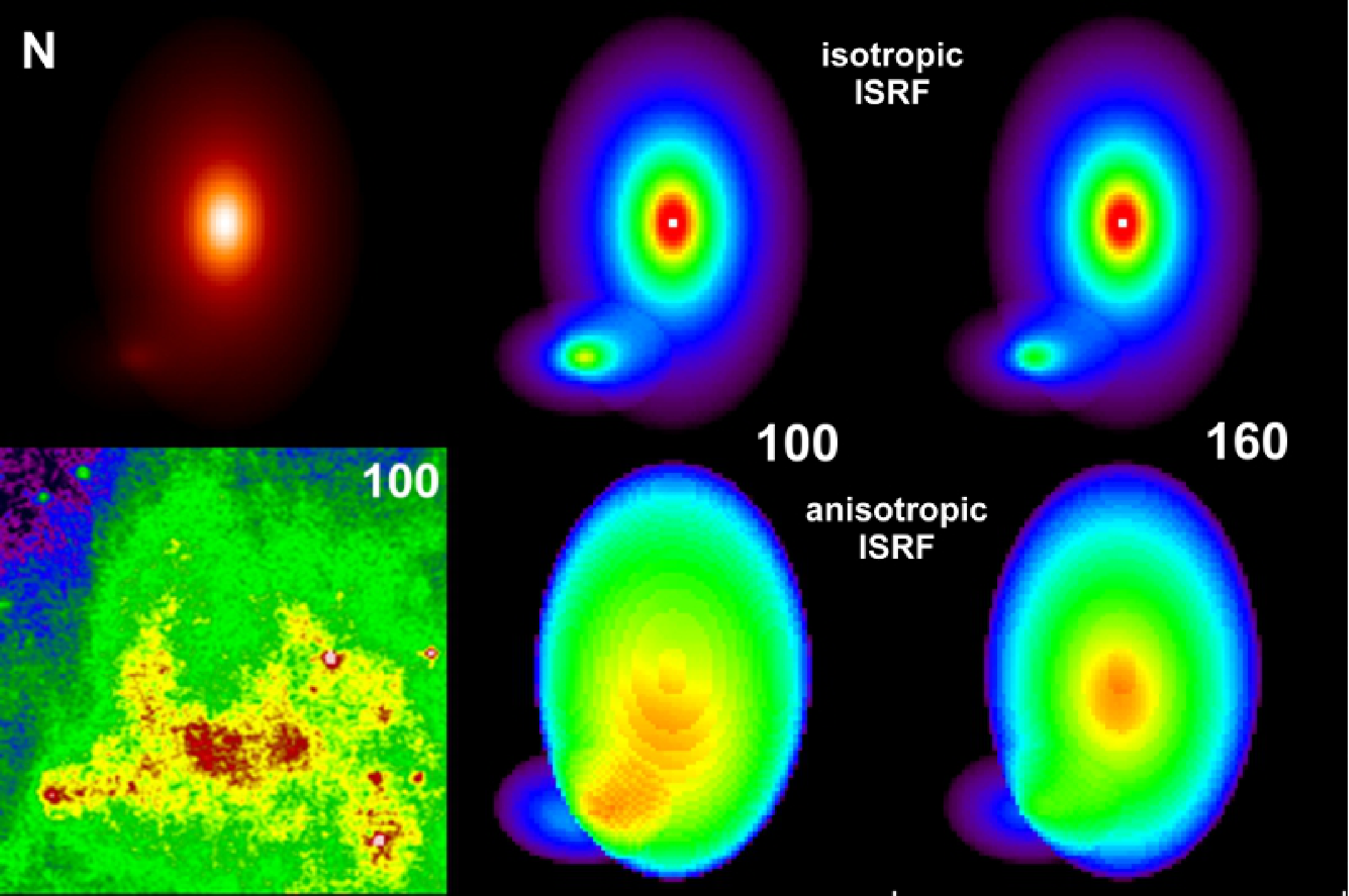}}
\caption{\label{f:b68aniso}Results of the 3D modelling showing the influence of
heating from an anisotropic ISRF on the FIR emission of a B68-like globule. The
upper left panel reflects the column density of the model globule, the lower
left panel shows the 100~$\mu$m emission as observed with \textit{Herschel}. The
remaining four images show the modelled FIR maps at 100 (central column) and
160~$\mu$m (right column) for an isotropic (top) and an anisotropic (bottom)
ISRF.}
\end{figure}

In a concept study to investigate, if the strong FIR emission in the part facing
the galactic plane and the Galactic centre is caused by an anisotropic
illumination of B68, we have constructed a simple model that mimics its basic
structural features. The main core body is approximated by a 3D clump with the
radial vector R defined by $R^2=(x/A)^2+(y/B)^2+(z/C)^2$ and a model dust number
density dropping to zero outside $r_\mathrm{out}$ following
\begin{equation}
\label{e:model_n}
n_M(\vec x)=\frac{n_{M,0}}{1+\left(\frac{R(\vec x)}{R_0}\right)^\alpha}\quad.
\end{equation}
A smaller clump with the same density structure was placed southeast of the main
clump to represent the southeastern extension of B68. The clumps do not overlap
to avoid density enhancement at the border of the main clump. The assumed masses
are 2.5 and $0.25~M_{\sun}$, respectively. With its low mass and off-centre
position, it should be heated up easily by the ISRF showing a stronger emission
at shorter wavelengths compared to shielded regions near the core centre. Two
ISRF illuminations were considered: isotropic illumination by the SED described
in \citet{zucconi01} and anisotropic illumination approximated by the same SED
through a cone with an opening angle of $20\degr$ that is oriented towards the
direction of the Galactic centre, where the NIR contribution attains its
maximum. Since B68 is located about $7\degr$ above the Galactic centre seen from
Earth, this causes a slightly stronger illumination from below. Test
calculations with a ray-tracing version of the code described in
\citet{steinacker03} yielded that considering scattering as a source term in the
radiative transfer will only lead to marginal temperature changes in the outer
parts of the model core. Since we use only an approximate ISRF and did not
consider images at shorter wavelength, we neglected it. We performed radiative
transfer calculations including extinction on a grid with $10^6$ cells, 600
equally-spaced direction nodes per cell \citep{steinacker96}, and 61 wavelengths
with simple opacities from \citet{draine84} to compare the results during the
code testing with former results obtained with the same opacities.

\begin{figure}
\centering
\resizebox{\hsize}{!}{\includegraphics[]{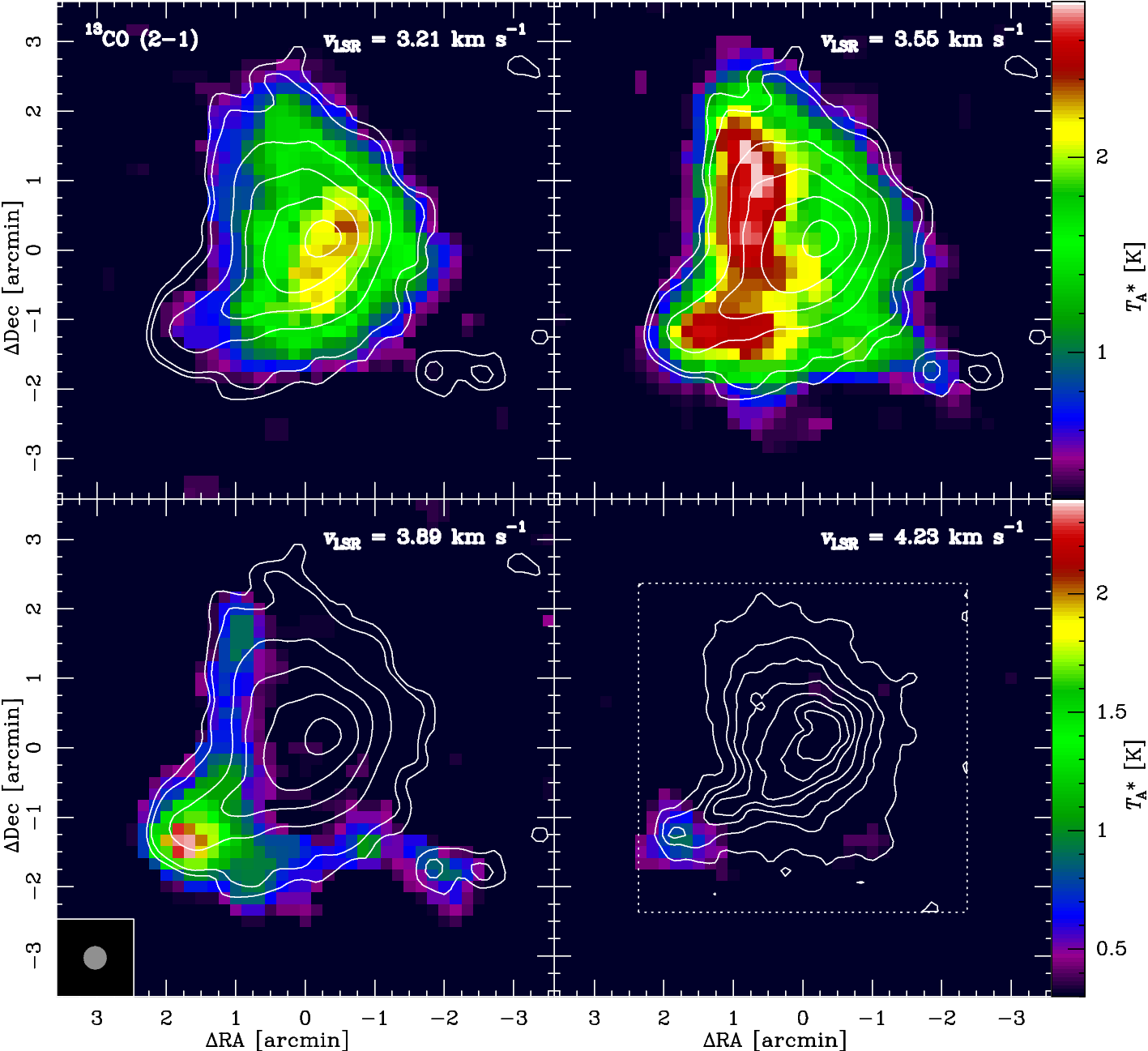}}
\caption{\label{f:13cochan}\element[][13]{CO}~(2--1) channel maps of the
molecular line observations of B68. The radial velocities are indicated. The
contours superimposed in the upper row and in the left panel of the lower row
represent the LABOCA data. The contours in the lower right panel show the
extinction map of \citet{alves01eso}. There is a general trend of the radial
velocities being somewhat higher towards the eastern edge of B68. The highest
velocities at $v_\mathrm{LSR}\approx4.1\pm0.2$~km\,s$^{-1}$ can be attributed to
a point source that coincides with a local density peak at the tip of
southeastern extension.}
\end{figure}

The results are shown in Fig.~\ref{f:b68aniso}. The upper left-hand image shows
the dust column number density $N$ of the two-clump model as an image. The
projection of the overlapping clump regions causes a local maximum. Likewise,
observed column density maxima or emission maxima at wavelengths where the core
is optically thin, can be an effect of projection without number density maxima
in the real structure. The middle and right upper images give the theoretical
intensity at 100 and 160~$\mu$m for the case of isotropic illumination. In both
images the central density peak of the main clump also shows an emission maximum
that is not seen in the observed images. For the 100~$\mu$m image, the global
emission peak appears in the overlap region of both clumps since the small clump
is not shielded but carries about a tenth of the gas mass of the main clump.
Even in the case of isotropic illumination, structures outside the main clump
will appear in emission at 100~$\mu$m. This wavelength range is sensitive to
temperature differences due to the exponentially rising Wien part of the thermal
emission. In the 160~$\mu$m image, the effect is less pronounced with the main
density peak also showing the strongest emission.

The lower panels show the observed PACS image at 100~$\mu$m, and the
simulated images at 100 and 160~$\mu$m for an illumination by an anisotropic
ISRF. The outer clump again appears in emission. The 100~$\mu$m image simulated
for anisotropic illumination, however, shows stronger emission in the southern
parts of the model core with a depression of flux in the central region as
seen in the PACS image. The 160~$\mu$m model image still shows more
emission in the southern part, but no depression near the column density centre,
again in qualitative agreement with the PACS image.

This means that a correct modelling of B68 requires the inclusion of the full
anisotropic radiation field and its wavelength dependency, including the
contributions from the nearby B2IV star \object{$\theta$~Oph} which in our
calculations accounted for 1\% of the full incident radiation field. This is
especially important for verifications of the balance between the thermal
pressure gradient and the gravitational force that is assumed when modelling B68
as a BES \citep{steinacker05ppv}. An investigation of nearby cores should reveal
one-sided heating in the same direction, if the cores of the snake nebula are
not shielding each other. This will be the focus of a subsequent publication.

\subsection{The core collision scenario}
\label{s:bullet}
Previous molecular line studies
\citep[e.g.~][]{bergin02,lada03,bergin06,redman06,maret07chem} have revealed a
high degree of gas depletion in the centre of B68, as well as a complex radial
velocity field that indicates oscillations with lifetimes that are typically
longer than the sound crossing time \citep{lada03,keto06,redman06}.
Nevertheless, other dynamical states like a global infall cannot be ruled out
\citep{keto06,broderick07}.

An intriguing scenario is suggested by \citet{burkert09}, who proposed an
explanation for the relative large lifetime of at least $3\times10^6$~yr that is
needed to explain the stable oscillations being driven by a subsonic velocity
field. They found this to be inconsistent with the suggestion of
\citet{alves01nature} that B68 is gravitationally unstable and collapses with a
free-fall timescale that is an order of magnitude smaller. In addition, the gas
density and temperature distribution can be best reproduced for an age of some
$10^5$~yr \citep{bergin06}. \citet{burkert09} interpret the local maxima of
column density in the southeastern trunk of B68 discovered in the NIR extinction
map of \citet{alves01eso} in terms of a small core of $0.2~M_{\sun}$ colliding
with the main body of B68. Based on this assumption, they were able to identify
such a collision as a possible trigger that recently pushed B68 beyond the
boundary of gravitational instability and initiated the gravitational collapse.
\citet{kitsionas07} have identified the collision of clumps of ISM as a possible
trigger for star formation in a more general way.

We have shown that a density peak appears in the \textit{Herschel} maps and can be
distinguished as a point source. In addition, we discovered at that very
location a global maximum in the integrated \element[][13]{CO} distribution (see
Sect.~\ref{s:co}), whose spectral line is slightly redshifted and broadened
relative to the main core of B68. Besides a global velocity gradient from west
to east, the individual channel maps of the \element[][13]{CO}~(2--1)
observations in Fig.~\ref{f:13cochan} identify a point source that is
kinematically separated from the main core. The highest radial velocities are
only present in this source. The contours of the LABOCA 870~$\mu$m and the
NIR extinction map of \citet{alves01eso} unambiguously identify this object with
the density peak at the tip of the southeastern extension of B68. Interestingly,
the second peak, which is located more to the centre of B68 and which appears to
have a higher column density \citep{alves01eso}, remains inconspicuous in our
continuum and molecular line maps, but may produce a weak signal in the
C$^{18}$O~(1--0) map of \citet{bergin02} and the CS~~(2--1) map of
\citet{lada03}.

It is noteworthy that previous molecular line investigations see the global
radial velocity gradient \citep[e.g.~][]{lada03}, but not the kinematically
isolated point source. The reason for this appears to be that either that
position was not covered \citep{lada03,redman06} or the velocities at $v_{\rm
LSR}\approx4.1\pm0.2$~km\,s$^{-1}$ were not included. Nonetheless, we have
identified this very object in the C$^{18}$O data of \citet{bergin02}
after our own re-analysis of the underlying observations. This new picture is
consistent with the scenario of a low-mass core, also referred to as the
``bullet'', colliding with a more massive core as portrayed by
\citet{burkert09}. They also argue that the chain of cores that include B68
and, to the southeast, B~69 to B~71, may be the remainder of a fragmented
filament in which the individual cores move along its direction and eventually
merge with each other.

\begin{figure}
\centering
\resizebox{0.789\hsize}{!}{\includegraphics{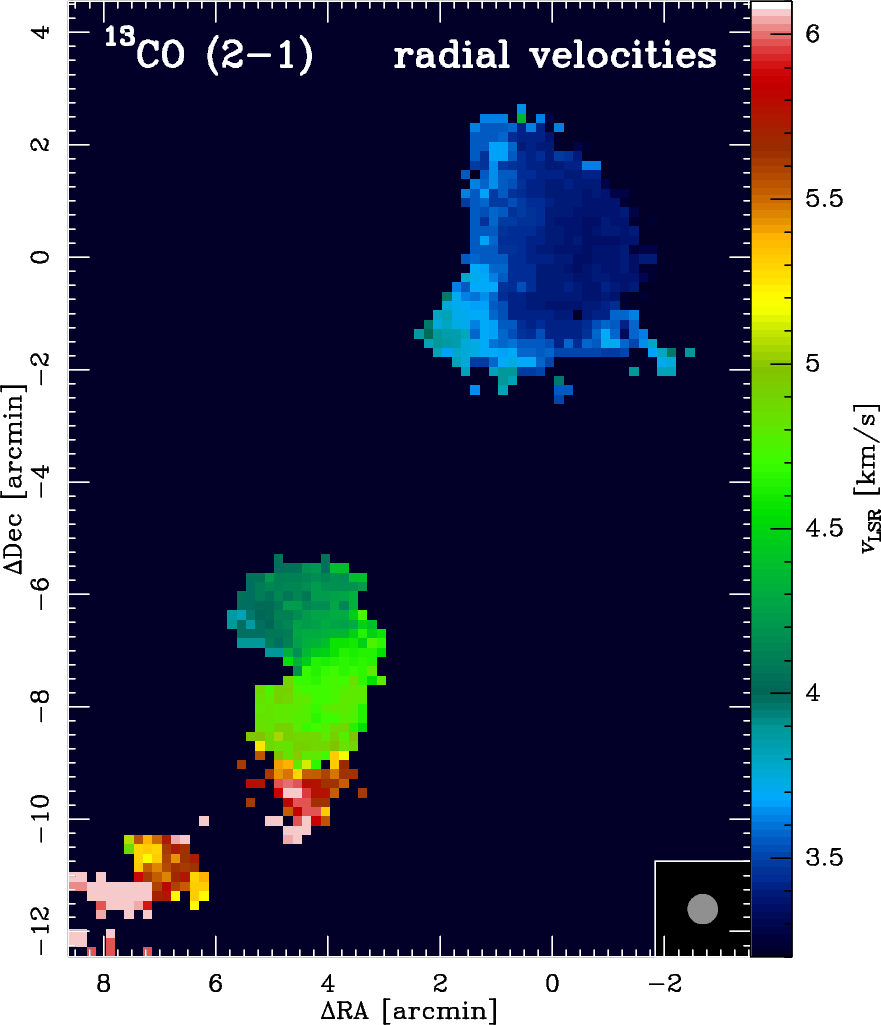}}
\caption{\label{f:13corvel}Radial velocity field of the larger area around B68
seen in the \element[][13]{CO}~(2--1) line.
There appears to be a gradient with increasing radial velocities from north to
south, which is consistent with a gas flow of constant velocity that is inclined
towards the LoS.}
\end{figure}

While most molecular line studies were restricted to B68, we have also included
the neighbouring southeastern cores of \object{Barnard~69} and
\object{Barnard~71} (see Fig.~\ref{f:b68obs}) in our CO-observing campaign.
Figure~\ref{f:13corvel} shows a radial velocity map of the
\mbox{\element[][13]{CO}~(2--1)} line of this area that reveals a large radial velocity
gradient from the north to the south, in which B68 is only a part of the larger
distribution of velocities. The greater picture therefore indicates that at
least some of the radial velocity gradient seen in B68 is not related to a
rotation of B68. The \element[][12]{CO}~(2--1) data show a similar picture.  If
we assume a constant stream velocity across the FoV of approximately $20'$, the
measured radial velocities can be explained by a small inclination from the PoS,
with the northern end tilted towards the observer. The stream velocity would
then amount to $\approx 500$~km~s$^{-1}$.  Such an unrealistically high value
suggests that other causes may contribute to the observed radial velocity
gradient.

Altogether, we may have found evidence that qualitatively supports the scenario
of colliding cores as suggested by \citet{burkert09}.

\section{Conclusions}
\label{s:conclusions}
We have presented FIR continuum maps of the isolated low-mass prestellar core,
B68, obtained with \textit{Herschel}. Using ancillary sub-mm continuum maps, all
convolved to the resolution of our SPIRE $500~\mu$m data, we presented two
approaches to modelling the dust temperature and column density distribution:
line-of-sight-averaged SED fitting and modelling by ray-tracing. The latter
modelling approach permitted us to also calculate the volume density
distribution.

We found the dust temperature rises from 8.2~K in the centre of 16.0~K at the
edge. The tenuous medium to the southeast, facing the galactic plane, reaches
almost 20~K. The column density peaks at $N_\mathrm{H} = 4.3 \times
10^{22}~\mathrm{cm}^{-2}$ and declines to $8 \times 10^{20}~\mathrm{cm}^{-2}$.
B68 has a total mass of $4.2~M_{\sun}$, or $3.1~M_{\sun}$ when considering only
material with $A_K > 0.2$~mag. We detected a new compact source in the
southeastern trunk of B68 at 160 and $250~\mu$m, which has a mass of about
$0.08~M_{\sun}$.

The mean radial column density profile can be fitted by a power-law with
$N_\mathrm{H} \propto r^{-2.5}$ between the flat inner core to the edge. The
corresponding volume density profile follows $n_\mathrm{H} \propto r^{-3.5}$,
ranging from $3.4 \times 10^5~\mathrm{cm}^{-3}$ (centre) to $4 \times
10^2~\mathrm{cm}^{-3}$ (edge), which is consistent with a "Plummer-like" profile
with $\eta = 4$ \citep{whitworth01}. However, we show that the mean radial
profile is not a good representation of the density distribution
beyond $40''$ (6000~AU), because deviations of up to a factor of 4 are seen in
the volume density profile beyond this radius. We will address this in detail
with 3D modelling in a forthcoming work.

We accounted for the effects of data-handling and differing angular
resolutions in our analysis of the core profiles. We also explored two physical
scenarios that can, either individually or in combination, explain the density
profile: (i) an isolated core in thermal and mechanical equilibrium
irradiated by an external ISRF with a significant UV component
\citep[e.g.][]{andre00,whitworth01}, and (ii) the remnants of a non-magnetic
isothermal cylinder \citep{ostriker64}. Both scenarios can produce density
profiles that are in better agreement with our results, which is steeper than
the commonly assumed singular isothermal sphere \citep[$n_\mathrm{H} \propto
r^{-2}$,][]{shu87}.

We observed a unique crescent-shaped asymmetry in the 100 and $160~\mu$m
emission profiles, intriguingly similar to the morphology shown in C$^{18}$O by
\citet{bergin02}. Using 3D radiative transfer modelling, such a morphology can
be reproduced in the continuum by an anisotropically irradiated core, including
contributions from the nearby B2IV star $\theta$~Oph and from the Galactic
centre and galactic plane ISRF. Put together, this demonstrates the importance
of accounting for an anisotropic ISRF when modelling a globule in detail.

We presented observations of $^{12}$CO and $^{13}$CO of a larger area
surrounding B68 than in previous molecular line studies. The new compact source
in the southeastern trunk is the peak in the integrated intensity, and it also
exhibits enhanced linewidths ($\sim 1.1$ and $0.65~\mathrm{km~s}^{-1}$, for
$^{12}$CO and $^{13}$CO, respectively) and a slightly redshifted centroid
velocity. Our observations lend qualitative credence to the ``bullet'' scenario
posited by \citet{burkert09}, in which B68 collides with a small core (our new,
kinematically distinct compact source), triggering gravitational collapse.

In addition, neighbouring globules, B69 and B71, show a smooth velocity
gradient, suggesting that this system of globules were once part of a single
filament that dispersed. This agrees with the predictions of the fragmenting
isothermal cylinder scenario mentioned above.

\begin{acknowledgements}
The authors would like to thank J.~F.~Alves and C.~G.~Rom\'an-Z\'u\~niga for
kindly providing their near-IR extinction data. In addition, we thank
G.~J.~Aniano \citep{aniano11} for supplying us with improved convolution kernels
for the \textit{Herschel} data that are based on empirically derived PSFs as
provided by the PACS ICC. We are particularly grateful to H.~Roussel, who helped
a lot with understanding and configuring the Scanamorphos software.

We greatly benefited from discussions on the data analysis
with Yaroslav Pavlyuchenkov (Russian Academy of Sciences, Moscow).

This work is based in part on observations made with the \textit{Spitzer} Space
Telescope, which is operated by the Jet Propulsion Laboratory, California
Institute of Technology under a contract with NASA.

This publication makes use of data products from the Two Micron All Sky Survey,
which is a joint project of the University of Massachusetts and the Infrared
Processing and Analysis Center/California Institute of Technology, funded by the
National Aeronautics and Space Administration and the National Science
Foundation.

This publication also makes use of data products from the Wide-field Infrared
Survey Explorer, which is a joint project of the University of California, Los
Angeles, and the Jet Propulsion Laboratory/California Institute of Technology,
funded by the National Aeronautics and Space Administration.

M.N., Z.B., and H.L. are funded by the Deutsches Zentrum f\"ur Luft- und
Raumfahrt (DLR). 

A.M.S., J.K., and S.R are supported by the Deutsche Forschungsgemeinschaft (DFG)
priority program 1573 (``Physics of the Interstellar Medium'').

We would like to thank the anonymous referee for the constructive comments.
\end{acknowledgements}

\bibliographystyle{aa}
\bibliography{references}

\begin{thebibliography}{142}
\expandafter\ifx\csname natexlab\endcsname\relax\def\natexlab#1{#1}\fi

\bibitem[{{Abergel} {et~al.}(1996){Abergel}, {Bernard}, {Boulanger},
  {Cesarsky}, {Desert}, {Falgarone}, {Lagache}, {Perault}, {Puget}, {Reach},
  {Nordh}, {Olofsson}, {Huldtgren}, {Kaas}, {Andre}, {Bontemps}, {Burgdorf},
  {Copet}, {Davies}, {Montmerle}, {Persi}, \& {Sibille}}]{abergel96}
{Abergel}, A., {Bernard}, J.~P., {Boulanger}, F., {et~al.} 1996, \aap, 315,
  L329

\bibitem[{{Adams}(1991)}]{adams91}
{Adams}, F.~C. 1991, \apj, 382, 544

\bibitem[{{Allende Prieto} {et~al.}(2001){Allende Prieto}, {Lambert}, \&
  {Asplund}}]{allende01}
{Allende Prieto}, C., {Lambert}, D.~L., \& {Asplund}, M. 2001, \apjl, 556, L63

\bibitem[{{Allende Prieto} {et~al.}(2002){Allende Prieto}, {Lambert}, \&
  {Asplund}}]{allende02}
{Allende Prieto}, C., {Lambert}, D.~L., \& {Asplund}, M. 2002, \apjl, 573, L137

\bibitem[{{Alves} \& {Franco}(2007)}]{alves07}
{Alves}, F.~O. \& {Franco}, G.~A.~P. 2007, \aap, 470, 597

\bibitem[{{Alves} {et~al.}(2001{\natexlab{a}}){Alves}, {Lada}, \&
  {Lada}}]{alves01eso}
{Alves}, J., {Lada}, C., \& {Lada}, E. 2001{\natexlab{a}}, The Messenger, 103,
  1

\bibitem[{{Alves} {et~al.}(2001{\natexlab{b}}){Alves}, {Lada}, \&
  {Lada}}]{alves01nature}
{Alves}, J.~F., {Lada}, C.~J., \& {Lada}, E.~A. 2001{\natexlab{b}}, \nat, 409,
  159

\bibitem[{{Andr{\'e}} {et~al.}(2004){Andr{\'e}}, {Bouwman}, {Belloche}, \&
  {Hennebelle}}]{andre04}
{Andr{\'e}}, P., {Bouwman}, J., {Belloche}, A., \& {Hennebelle}, P. 2004,
  \apss, 292, 325

\bibitem[{Andr{\'{e}} {et~al.}(1993)Andr{\'{e}}, Ward-Thompson, \&
  Barsony}]{andre93}
Andr{\'{e}}, P., Ward-Thompson, D., \& Barsony, M. 1993, ApJ, 406, 122

\bibitem[{{Andr{\'{e}}} {et~al.}(2000){Andr{\'{e}}}, {Ward-Thompson}, \&
  {Barsony}}]{andre00}
{Andr{\'{e}}}, P., {Ward-Thompson}, D., \& {Barsony}, M. 2000, Protostars and
  Planets IV, 59

\bibitem[{{Andr{\'{e}}} {et~al.}(1996){Andr{\'{e}}}, {Ward-Thompson}, \&
  {Motte}}]{andre96}
{Andr{\'{e}}}, P., {Ward-Thompson}, D., \& {Motte}, F. 1996, \aap, 314, 625

\bibitem[{{Aniano} {et~al.}(2011){Aniano}, {Draine}, {Gordon}, \&
  {Sandstrom}}]{aniano11}
{Aniano}, G., {Draine}, B.~T., {Gordon}, K.~D., \& {Sandstrom}, K. 2011, \pasp,
  123, 1218

\bibitem[{Bacmann {et~al.}(2000)Bacmann, Andr\'{e}, Puget, Abergel, Bontemps,
  \& Ward-Thompson}]{bacmann00}
Bacmann, A., Andr\'{e}, P., Puget, J.-L., {et~al.} 2000, A\&A, 361, 555

\bibitem[{{Barnard} {et~al.}(1927){Barnard}, {Frost}, \& {Calvert}}]{barnard27}
{Barnard}, E.~E., {Frost}, E.~B., \& {Calvert}, M.~R. 1927, {A photographic
  atlas of selected regions of the Milky way}

\bibitem[{{Bergin} {et~al.}(2002){Bergin}, {Alves}, {Huard}, \&
  {Lada}}]{bergin02}
{Bergin}, E.~A., {Alves}, J., {Huard}, T., \& {Lada}, C.~J. 2002, \apjl, 570,
  L101

\bibitem[{{Bergin} {et~al.}(2006){Bergin}, {Maret}, {van der Tak}, {Alves},
  {Carmody}, \& {Lada}}]{bergin06}
{Bergin}, E.~A., {Maret}, S., {van der Tak}, F.~F.~S., {et~al.} 2006, \apj,
  645, 369

\bibitem[{{Beuther} {et~al.}(2010){Beuther}, {Henning}, {Linz}, {Krause},
  {Nielbock}, \& {Steinacker}}]{beuther10}
{Beuther}, H., {Henning}, T., {Linz}, H., {et~al.} 2010, \aap, 518, L78

\bibitem[{{Beuther} {et~al.}(2012){Beuther}, {Tackenberg}, {Linz}, {Henning},
  {Krause}, {Ragan}, {Nielbock}, {Launhardt}, {Schmiedeke}, {Schuller},
  {Carlhoff}, {Nguyen-Luong}, \& {Sakai}}]{beuther12}
{Beuther}, H., {Tackenberg}, J., {Linz}, H., {et~al.} 2012, \aap, 538, A11

\bibitem[{{Bianchi} {et~al.}(2003){Bianchi}, {Gon{\c c}alves}, {Albrecht},
  {Caselli}, {Chini}, {Galli}, \& {Walmsley}}]{bianchi03}
{Bianchi}, S., {Gon{\c c}alves}, J., {Albrecht}, M., {et~al.} 2003, \aap, 399,
  L43

\bibitem[{{Black}(1994)}]{black94}
{Black}, J.~H. 1994, in Astronomical Society of the Pacific Conference Series,
  Vol.~58, The First Symposium on the Infrared Cirrus and Diffuse Interstellar
  Clouds, ed. {R.~M.~Cutri \& W.~B.~Latter}, 355

\bibitem[{{Bohlin} {et~al.}(1978){Bohlin}, {Savage}, \& {Drake}}]{bohlin78}
{Bohlin}, R.~C., {Savage}, B.~D., \& {Drake}, J.~F. 1978, ApJ, 224, 132

\bibitem[{{Bok}(1977)}]{bok77}
{Bok}, B.~J. 1977, \pasp, 89, 597

\bibitem[{{Bok} \& {Reilly}(1947)}]{bok47}
{Bok}, B.~J. \& {Reilly}, E.~F. 1947, \apj, 105, 255

\bibitem[{{Bonnor}(1956)}]{bonnor56}
{Bonnor}, W.~B. 1956, \mnras, 116, 351

\bibitem[{{Bourke} {et~al.}(1995){Bourke}, {Hyland}, \& {Robinson}}]{bourke95}
{Bourke}, T.~L., {Hyland}, A.~R., \& {Robinson}, G. 1995, \mnras, 276, 1052

\bibitem[{{Bouwman} {et~al.}(2008){Bouwman}, {Henning}, {Hillenbrand}, {Meyer},
  {Pascucci}, {Carpenter}, {Hines}, {Kim}, {Silverstone}, {Hollenbach}, \&
  {Wolf}}]{bouwman08}
{Bouwman}, J., {Henning}, T., {Hillenbrand}, L.~A., {et~al.} 2008, \apj, 683,
  479

\bibitem[{{Broderick} {et~al.}(2007){Broderick}, {Keto}, {Lada}, \&
  {Narayan}}]{broderick07}
{Broderick}, A.~E., {Keto}, E., {Lada}, C.~J., \& {Narayan}, R. 2007, \apj,
  671, 1832

\bibitem[{{Burkert} \& {Alves}(2009)}]{burkert09}
{Burkert}, A. \& {Alves}, J. 2009, \apj, 695, 1308

\bibitem[{{Cardelli} {et~al.}(1989){Cardelli}, {Clayton}, \&
  {Mathis}}]{cardelli89}
{Cardelli}, J.~A., {Clayton}, G.~C., \& {Mathis}, J.~S. 1989, \apj, 345, 245

\bibitem[{{Chapman} \& {Mundy}(2009)}]{chapman09b}
{Chapman}, N.~L. \& {Mundy}, L.~G. 2009, \apj, 699, 1866

\bibitem[{{Chapman} {et~al.}(2009){Chapman}, {Mundy}, {Lai}, \&
  {Evans}}]{chapman09a}
{Chapman}, N.~L., {Mundy}, L.~G., {Lai}, S.-P., \& {Evans}, II, N.~J. 2009,
  \apj, 690, 496

\bibitem[{{Chen} {et~al.}(2007){Chen}, {Launhardt}, \& {Henning}}]{chen07}
{Chen}, X., {Launhardt}, R., \& {Henning}, T. 2007, \apj, 669, 1058

\bibitem[{{Chiar} {et~al.}(2007){Chiar}, {Ennico}, {Pendleton}, {Boogert},
  {Greene}, {Knez}, {Lada}, {Roellig}, {Tielens}, {Werner}, \&
  {Whittet}}]{chiar07}
{Chiar}, J.~E., {Ennico}, K., {Pendleton}, Y.~J., {et~al.} 2007, \apjl, 666,
  L73

\bibitem[{{Chi\`{e}ze} \& {Pineau Des For\^{e}ts}(1987)}]{chieze87}
{Chi\`{e}ze}, J.-P. \& {Pineau Des For\^{e}ts}, G. 1987, \aap, 183, 98

\bibitem[{Chini {et~al.}(2003)Chini, K\"ampgen, Reipurth, Albrecht, Kreysa,
  Lemke, Nielbock, Reichertz, Sievers, \& Zylka}]{chini03}
Chini, R., K\"ampgen, K., Reipurth, B., {et~al.} 2003, A\&A, 409, 235

\bibitem[{{Chini} \& {Kr\"ugel}(1983)}]{chini83}
{Chini}, R. \& {Kr\"ugel}, E. 1983, \aap, 117, 289

\bibitem[{{Crapsi} {et~al.}(2007){Crapsi}, {Caselli}, {Walmsley}, \&
  {Tafalla}}]{crapsi07}
{Crapsi}, A., {Caselli}, P., {Walmsley}, M.~C., \& {Tafalla}, M. 2007, \aap,
  470, 221

\bibitem[{{Dapp} \& {Basu}(2009)}]{dapp09}
{Dapp}, W.~B. \& {Basu}, S. 2009, \mnras, 395, 1092

\bibitem[{{de Geus} {et~al.}(1989){de Geus}, {de Zeeuw}, \& {Lub}}]{degeus89}
{de Geus}, E.~J., {de Zeeuw}, P.~T., \& {Lub}, J. 1989, \aap, 216, 44

\bibitem[{{di Francesco} {et~al.}(2007){di Francesco}, {Evans}, {Caselli},
  {Myers}, {Shirley}, {Aikawa}, \& {Tafalla}}]{difrancesco07}
{di Francesco}, J., {Evans}, II, N.~J., {Caselli}, P., {et~al.} 2007,
  Protostars and Planets V, 17

\bibitem[{{di Francesco} {et~al.}(2008){di Francesco}, {Johnstone}, {Kirk},
  {MacKenzie}, \& {Ledwosinska}}]{difrancesco08}
{di Francesco}, J., {Johnstone}, D., {Kirk}, H., {MacKenzie}, T., \&
  {Ledwosinska}, E. 2008, \apjs, 175, 277

\bibitem[{{Doty} {et~al.}(2005){Doty}, {Everett}, {Shirley}, {Evans}, \&
  {Palotti}}]{doty05}
{Doty}, S.~D., {Everett}, S.~E., {Shirley}, Y.~L., {Evans}, N.~J., \&
  {Palotti}, M.~L. 2005, \mnras, 359, 228

\bibitem[{{Draine} \& {Lee}(1984)}]{draine84}
{Draine}, B.~T. \& {Lee}, H.~M. 1984, \apj, 285, 89

\bibitem[{{Ebert}(1955)}]{ebert55}
{Ebert}, R. 1955, \zap, 37, 217

\bibitem[{{Evans} {et~al.}(2001){Evans}, {Rawlings}, {Shirley}, \&
  {Mundy}}]{evans01}
{Evans}, II, N.~J., {Rawlings}, J.~M.~C., {Shirley}, Y.~L., \& {Mundy}, L.~G.
  2001, \apj, 557, 193

\bibitem[{{Falgarone} \& {Puget}(1985)}]{falgarone85}
{Falgarone}, E. \& {Puget}, J.~L. 1985, \aap, 142, 157

\bibitem[{{Fazio} {et~al.}(2004){Fazio}, {Hora}, {Allen}, {Ashby}, {Barmby},
  {Deutsch}, {Huang}, {Kleiner}, {Marengo}, {Megeath}, {Melnick}, {Pahre},
  {Patten}, {Polizotti}, {Smith}, {Taylor}, {Wang}, {Willner}, {Hoffmann},
  {Pipher}, {Forrest}, {McMurty}, {McCreight}, {McKelvey}, {McMurray}, {Koch},
  {Moseley}, {Arendt}, {Mentzell}, {Marx}, {Losch}, {Mayman}, {Eichhorn},
  {Krebs}, {Jhabvala}, {Gezari}, {Fixsen}, {Flores}, {Shakoorzadeh}, {Jungo},
  {Hakun}, {Workman}, {Karpati}, {Kichak}, {Whitley}, {Mann}, {Tollestrup},
  {Eisenhardt}, {Stern}, {Gorjian}, {Bhattacharya}, {Carey}, {Nelson},
  {Glaccum}, {Lacy}, {Lowrance}, {Laine}, {Reach}, {Stauffer}, {Surace},
  {Wilson}, {Wright}, {Hoffman}, {Domingo}, \& {Cohen}}]{spitzer-irac}
{Fazio}, G.~G., {Hora}, J.~L., {Allen}, L.~E., {et~al.} 2004, ApJS, 154, 10

\bibitem[{{Finger} {et~al.}(1998){Finger}, {Biereichel}, {Mehrgan}, {Meyer},
  {Moorwood}, {Nicolini}, \& {Stegmeier}}]{sofi}
{Finger}, G., {Biereichel}, P., {Mehrgan}, H., {et~al.} 1998, in Proc. SPIE,
  Vol. 3354, Infrared Astronomical Instrumentation, ed. A.~M. Fowler (The
  International Society for Optical Engineering), 87

\bibitem[{{Fitzpatrick} \& {Massa}(2009)}]{fitzpatrick09}
{Fitzpatrick}, E.~L. \& {Massa}, D. 2009, \apj, 699, 1209

\bibitem[{{Flaherty} {et~al.}(2007){Flaherty}, {Pipher}, {Megeath}, {Winston},
  {Gutermuth}, {Muzerolle}, {Allen}, \& {Fazio}}]{flaherty07}
{Flaherty}, K.~M., {Pipher}, J.~L., {Megeath}, S.~T., {et~al.} 2007, ApJ, 663,
  1069

\bibitem[{{Fritz} {et~al.}(2011){Fritz}, {Gillessen}, {Dodds-Eden}, {Lutz},
  {Genzel}, {Raab}, {Ott}, {Pfuhl}, {Eisenhauer}, \& {Yusef-Zadeh}}]{fritz11}
{Fritz}, T.~K., {Gillessen}, S., {Dodds-Eden}, K., {et~al.} 2011, \apj, 737, 73

\bibitem[{{Goldsmith}(2001)}]{goldsmith01}
{Goldsmith}, P.~F. 2001, \apj, 557, 736

\bibitem[{{Gordon} {et~al.}(2005){Gordon}, {Rieke}, {Engelbracht}, {Muzerolle},
  {Stansberry}, {Misselt}, {Morrison}, {Cadien}, {Young}, {Dole}, {Kelly},
  {Alonso-Herrero}, {Egami}, {Su}, {Papovich}, {Smith}, {Hines}, {Rieke},
  {Blaylock}, {P{\'e}rez-Gonz{\'a}lez}, {Le Floc'h}, {Hinz}, {Latter},
  {Hesselroth}, {Frayer}, {Noriega-Crespo}, {Masci}, {Padgett}, {Smylie}, \&
  {Haegel}}]{gordon05}
{Gordon}, K.~D., {Rieke}, G.~H., {Engelbracht}, C.~W., {et~al.} 2005, \pasp,
  117, 503

\bibitem[{{Griffin} {et~al.}(2010){Griffin}, {Abergel}, {Abreu}, {Ade},
  {Andr{\'e}}, {Augueres}, {Babbedge}, {Bae}, {Baillie}, {Baluteau}, {Barlow},
  {Bendo}, {Benielli}, {Bock}, {Bonhomme}, {Brisbin}, {Brockley-Blatt},
  {Caldwell}, {Cara}, {Castro-Rodriguez}, {Cerulli}, {Chanial}, {Chen},
  {Clark}, {Clements}, {Clerc}, {Coker}, {Communal}, {Conversi}, {Cox},
  {Crumb}, {Cunningham}, {Daly}, {Davis}, {de Antoni}, {Delderfield}, {Devin},
  {di Giorgio}, {Didschuns}, {Dohlen}, {Donati}, {Dowell}, {Dowell}, {Duband},
  {Dumaye}, {Emery}, {Ferlet}, {Ferrand}, {Fontignie}, {Fox}, {Franceschini},
  {Frerking}, {Fulton}, {Garcia}, {Gastaud}, {Gear}, {Glenn}, {Goizel},
  {Griffin}, {Grundy}, {Guest}, {Guillemet}, {Hargrave}, {Harwit}, {Hastings},
  {Hatziminaoglou}, {Herman}, {Hinde}, {Hristov}, {Huang}, {Imhof}, {Isaak},
  {Israelsson}, {Ivison}, {Jennings}, {Kiernan}, {King}, {Lange}, {Latter},
  {Laurent}, {Laurent}, {Leeks}, {Lellouch}, {Levenson}, {Li}, {Li},
  {Lilienthal}, {Lim}, {Liu}, {Lu}, {Madden}, {Mainetti}, {Marliani}, {McKay},
  {Mercier}, {Molinari}, {Morris}, {Moseley}, {Mulder}, {Mur}, {Naylor},
  {Nguyen}, {O'Halloran}, {Oliver}, {Olofsson}, {Olofsson}, {Orfei}, {Page},
  {Pain}, {Panuzzo}, {Papageorgiou}, {Parks}, {Parr-Burman}, {Pearce},
  {Pearson}, {P{\'e}rez-Fournon}, {Pinsard}, {Pisano}, {Podosek}, {Pohlen},
  {Polehampton}, {Pouliquen}, {Rigopoulou}, {Rizzo}, {Roseboom}, {Roussel},
  {Rowan-Robinson}, {Rownd}, {Saraceno}, {Sauvage}, {Savage}, {Savini},
  {Sawyer}, {Scharmberg}, {Schmitt}, {Schneider}, {Schulz}, {Schwartz},
  {Shafer}, {Shupe}, {Sibthorpe}, {Sidher}, {Smith}, {Smith}, {Smith},
  {Spencer}, {Stobie}, {Sudiwala}, {Sukhatme}, {Surace}, {Stevens}, {Swinyard},
  {Trichas}, {Tourette}, {Triou}, {Tseng}, {Tucker}, {Turner}, {Vaccari},
  {Valtchanov}, {Vigroux}, {Virique}, {Voellmer}, {Walker}, {Ward}, {Waskett},
  {Weilert}, {Wesson}, {White}, {Whitehouse}, {Wilson}, {Winter}, {Woodcraft},
  {Wright}, {Xu}, {Zavagno}, {Zemcov}, {Zhang}, \& {Zonca}}]{spire}
{Griffin}, M.~J., {Abergel}, A., {Abreu}, A., {et~al.} 2010, \aap, 518, L3

\bibitem[{{G{\"u}sten} {et~al.}(2006){G{\"u}sten}, {Booth}, {Cesarsky},
  {Menten}, {Agurto}, {Anciaux}, {Azagra}, {Belitsky}, {Belloche}, {Bergman},
  {De Breuck}, {Comito}, {Dumke}, {Duran}, {Esch}, {Fluxa}, {Greve}, {Hafok},
  {H{\"a}upl}, {Helldner}, {Henseler}, {Heyminck}, {Johansson}, {Kasemann},
  {Klein}, {Korn}, {Kreysa}, {Kurz}, {Lapkin}, {Leurini}, {Lis}, {Lundgren},
  {Mac-Auliffe}, {Martinez}, {Melnick}, {Morris}, {Muders}, {Nyman}, {Olberg},
  {Olivares}, {Pantaleev}, {Patel}, {Pausch}, {Philipp}, {Philipps},
  {Sridharan}, {Polehampton}, {Reveret}, {Risacher}, {Roa}, {Sauer}, {Schilke},
  {Santana}, {Schneider}, {Sepulveda}, {Siringo}, {Spyromilio}, {Stenvers},
  {van der Tak}, {Torres}, {Vanzi}, {Vassilev}, {Weiss}, {Willmeroth},
  {Wunsch}, \& {Wyrowski}}]{apex}
{G{\"u}sten}, R., {Booth}, R.~S., {Cesarsky}, C., {et~al.} 2006, in Proc. SPIE,
  Vol. 6267, Ground-based and Airborne Telescopes, ed. L.~M. Stepp (The
  International Society for Optical Engineering), 626714

\bibitem[{{Gutermuth} {et~al.}(2008){Gutermuth}, {Myers}, {Megeath}, {Allen},
  {Pipher}, {Muzerolle}, {Porras}, {Winston}, \& {Fazio}}]{gutermuth08}
{Gutermuth}, R.~A., {Myers}, P.~C., {Megeath}, S.~T., {et~al.} 2008, \apj, 674,
  336

\bibitem[{{G{\"u}ver} \& {{\"O}zel}(2009)}]{guever09}
{G{\"u}ver}, T. \& {{\"O}zel}, F. 2009, \mnras, 400, 2050

\bibitem[{{Henning} {et~al.}(2010){Henning}, {Linz}, {Krause}, {Ragan},
  {Beuther}, {Launhardt}, {Nielbock}, \& {Vasyunina}}]{henning10}
{Henning}, T., {Linz}, H., {Krause}, O., {et~al.} 2010, \aap, 518, L95

\bibitem[{{Higdon} {et~al.}(2004){Higdon}, {Devost}, {Higdon}, {Brandl},
  {Houck}, {Hall}, {Barry}, {Charmandaris}, {Smith}, {Sloan}, \&
  {Green}}]{higdon04smart}
{Higdon}, S.~J.~U., {Devost}, D., {Higdon}, J.~L., {et~al.} 2004, \pasp, 116,
  975

\bibitem[{{Holweger}(2001)}]{holweger01}
{Holweger}, H. 2001, in American Institute of Physics Conference Series, Vol.
  598, Joint SOHO/ACE workshop ''Solar and Galactic Composition'', ed. R.~F.
  {Wimmer-Schweingruber}, 23--30

\bibitem[{{Hotzel} {et~al.}(2002{\natexlab{a}}){Hotzel}, {Harju}, \&
  {Juvela}}]{hotzel02temp}
{Hotzel}, S., {Harju}, J., \& {Juvela}, M. 2002{\natexlab{a}}, \aap, 395, L5

\bibitem[{{Hotzel} {et~al.}(2002{\natexlab{b}}){Hotzel}, {Harju}, {Juvela},
  {Mattila}, \& {Haikala}}]{hotzel02co}
{Hotzel}, S., {Harju}, J., {Juvela}, M., {Mattila}, K., \& {Haikala}, L.~K.
  2002{\natexlab{b}}, \aap, 391, 275

\bibitem[{{Houck} {et~al.}(2004){Houck}, {Roellig}, {van Cleve}, {Forrest},
  {Herter}, {Lawrence}, {Matthews}, {Reitsema}, {Soifer}, {Watson}, {Weedman},
  {Huisjen}, {Troeltzsch}, {Barry}, {Bernard-Salas}, {Blacken}, {Brandl},
  {Charmandaris}, {Devost}, {Gull}, {Hall}, {Henderson}, {Higdon}, {Pirger},
  {Schoenwald}, {Sloan}, {Uchida}, {Appleton}, {Armus}, {Burgdorf},
  {Fajardo-Acosta}, {Grillmair}, {Ingalls}, {Morris}, \&
  {Teplitz}}]{spitzer-irs}
{Houck}, J.~R., {Roellig}, T.~L., {van Cleve}, J., {et~al.} 2004, \apjs, 154,
  18

\bibitem[{{Humphreys} \& {Larsen}(1995)}]{humphreys95}
{Humphreys}, R.~M. \& {Larsen}, J.~A. 1995, \aj, 110, 2183

\bibitem[{{Johnstone} {et~al.}(2003){Johnstone}, {Fiege}, {Redman}, {Feldman},
  \& {Carey}}]{johnstone03g11}
{Johnstone}, D., {Fiege}, J.~D., {Redman}, R.~O., {Feldman}, P.~A., \& {Carey},
  S.~J. 2003, \apjl, 588, L37

\bibitem[{{Juh{\'a}sz} {et~al.}(2010){Juh{\'a}sz}, {Bouwman}, {Henning},
  {Acke}, {van den Ancker}, {Meeus}, {Dominik}, {Min}, {Tielens}, \&
  {Waters}}]{juhasz10}
{Juh{\'a}sz}, A., {Bouwman}, J., {Henning}, T., {et~al.} 2010, \apj, 721, 431

\bibitem[{{Juvela} \& {Ysard}(2011)}]{juvela11}
{Juvela}, M. \& {Ysard}, N. 2011, \apj, 739, 63

\bibitem[{{Kainulainen} {et~al.}(2009){Kainulainen}, {Beuther}, {Henning}, \&
  {Plume}}]{kainulainen09}
{Kainulainen}, J., {Beuther}, H., {Henning}, T., \& {Plume}, R. 2009, \aap,
  508, L35

\bibitem[{{Keene} {et~al.}(1983){Keene}, {Davidson}, {Harper}, {Hildebrand},
  {Jaffe}, {Loewenstein}, {Low}, \& {Pernic}}]{keene83}
{Keene}, J., {Davidson}, J.~A., {Harper}, D.~A., {et~al.} 1983, \apjl, 274, L43

\bibitem[{{Keto} {et~al.}(2006){Keto}, {Broderick}, {Lada}, \&
  {Narayan}}]{keto06}
{Keto}, E., {Broderick}, A.~E., {Lada}, C.~J., \& {Narayan}, R. 2006, \apj,
  652, 1366

\bibitem[{{Kirk} {et~al.}(2007){Kirk}, {Ward-Thompson}, \&
  {Andr{\'e}}}]{kirk07}
{Kirk}, J.~M., {Ward-Thompson}, D., \& {Andr{\'e}}, P. 2007, \mnras, 375, 843

\bibitem[{{Kitsionas} \& {Whitworth}(2007)}]{kitsionas07}
{Kitsionas}, S. \& {Whitworth}, A.~P. 2007, \mnras, 378, 507

\bibitem[{{Kr\"ugel}(2003)}]{kruegel03}
{Kr\"ugel}, E. 2003, {The physics of interstellar dust}, IoP Series in
  astronomy and astrophysics (Bristol, UK: The Institute of Physics)

\bibitem[{{Lada} {et~al.}(2003){Lada}, {Bergin}, {Alves}, \& {Huard}}]{lada03}
{Lada}, C.~J., {Bergin}, E.~A., {Alves}, J.~F., \& {Huard}, T.~L. 2003, \apj,
  586, 286

\bibitem[{{Langer} \& {Willacy}(2001)}]{langer01}
{Langer}, W.~D. \& {Willacy}, K. 2001, \apj, 557, 714

\bibitem[{{Larson}(1972)}]{larson72}
{Larson}, R.~B. 1972, \mnras, 157, 121

\bibitem[{{Launhardt} {et~al.}(2010){Launhardt}, {Nutter}, {Ward-Thompson},
  {Bourke}, {Henning}, {Khanzadyan}, {Schmalzl}, {Wolf}, \&
  {Zylka}}]{launhardt10}
{Launhardt}, R., {Nutter}, D., {Ward-Thompson}, D., {et~al.} 2010, \apjs, 188,
  139

\bibitem[{{Lesaffre} {et~al.}(2005){Lesaffre}, {Belloche}, {Chi{\`e}ze}, \&
  {Andr{\'e}}}]{lesaffre05}
{Lesaffre}, P., {Belloche}, A., {Chi{\`e}ze}, J.-P., \& {Andr{\'e}}, P. 2005,
  \aap, 443, 961

\bibitem[{{Linz} {et~al.}(2010){Linz}, {Krause}, {Beuther}, {Henning}, {Klein},
  {Nielbock}, {Stecklum}, {Steinacker}, \& {Stutz}}]{linz10}
{Linz}, H., {Krause}, O., {Beuther}, H., {et~al.} 2010, \aap, 518, L123

\bibitem[{{Lombardi}(2005)}]{lombardi05}
{Lombardi}, M. 2005, \aap, 438, 169

\bibitem[{{Lombardi}(2009)}]{lombardi09}
{Lombardi}, M. 2009, \aap, 493, 735

\bibitem[{{Lombardi} {et~al.}(2006){Lombardi}, {Alves}, \& {Lada}}]{lombardi06}
{Lombardi}, M., {Alves}, J., \& {Lada}, C.~J. 2006, \aap, 454, 781

\bibitem[{{Maret} {et~al.}(2007){Maret}, {Bergin}, \& {Lada}}]{maret07chem}
{Maret}, S., {Bergin}, E.~A., \& {Lada}, C.~J. 2007, \apjl, 670, L25

\bibitem[{{Martin} {et~al.}(2012){Martin}, {Roy}, {Bontemps},
  {Miville-Desch{\^e}nes}, {Ade}, {Bock}, {Chapin}, {Devlin}, {Dicker},
  {Griffin}, {Gundersen}, {Halpern}, {Hargrave}, {Hughes}, {Klein}, {Marsden},
  {Mauskopf}, {Netterfield}, {Olmi}, {Patanchon}, {Rex}, {Scott}, {Semisch},
  {Truch}, {Tucker}, {Tucker}, {Viero}, \& {Wiebe}}]{martin12}
{Martin}, P.~G., {Roy}, A., {Bontemps}, S., {et~al.} 2012, \apj, 751, 28

\bibitem[{{Mathis} {et~al.}(1983){Mathis}, {Mezger}, \& {Panagia}}]{mathis83}
{Mathis}, J.~S., {Mezger}, P.~G., \& {Panagia}, N. 1983, \aap, 128, 212

\bibitem[{{Mazzei} \& {Barbaro}(2008)}]{mazzei08}
{Mazzei}, P. \& {Barbaro}, G. 2008, \mnras, 390, 706

\bibitem[{{Moffat}(1969)}]{moffat69}
{Moffat}, A.~F.~J. 1969, \aap, 3, 455

\bibitem[{{Moorwood} {et~al.}(1998){Moorwood}, {Cuby}, \& {Lidman}}]{sofi2}
{Moorwood}, A., {Cuby}, J.-G., \& {Lidman}, C. 1998, The Messenger, 91, 9

\bibitem[{{Motte} \& {Andr{\'e}}(2001)}]{motte01}
{Motte}, F. \& {Andr{\'e}}, P. 2001, \aap, 365, 440

\bibitem[{{Nutter} {et~al.}(2009){Nutter}, {Stamatellos}, \&
  {Ward-Thompson}}]{nutter09}
{Nutter}, D., {Stamatellos}, D., \& {Ward-Thompson}, D. 2009, \mnras, 396, 1851

\bibitem[{{Nyman} {et~al.}(2001){Nyman}, {Lerner}, {Nielbock}, {Anciaux},
  {Brooks}, {Chini}, {Albrecht}, {Lemke}, {Kreysa}, {Zylka}, {Johansson},
  {Bronfman}, {Kontinen}, {Linz}, \& {Stecklum}}]{nyman01}
{Nyman}, L., {Lerner}, M., {Nielbock}, M., {et~al.} 2001, The Messenger, 106,
  40

\bibitem[{{Olofsson} \& {Olofsson}(2010)}]{olofsson10}
{Olofsson}, S. \& {Olofsson}, G. 2010, \aap, 522, A84

\bibitem[{{Ossenkopf} \& {Henning}(1994)}]{ossenkopf94}
{Ossenkopf}, V. \& {Henning}, T. 1994, \aap, 291, 943

\bibitem[{{Ostriker}(1964)}]{ostriker64}
{Ostriker}, J. 1964, \apj, 140, 1056

\bibitem[{{Ott}(2010)}]{hipe}
{Ott}, S. 2010, in Astronomical Society of the Pacific Conference Series, Vol.
  434, Astronomical Data Analysis Software and Systems XIX, ed. {Y.~Mizumoto,
  K.-I.~Morita, \& M.~Ohishi}, 139

\bibitem[{{Pagani} {et~al.}(2007){Pagani}, {Bacmann}, {Cabrit}, \&
  {Vastel}}]{pagani07}
{Pagani}, L., {Bacmann}, A., {Cabrit}, S., \& {Vastel}, C. 2007, \aap, 467, 179

\bibitem[{{Pagani} {et~al.}(2004){Pagani}, {Bacmann}, {Motte}, {Cambr{\'e}sy},
  {Fich}, {Lagache}, {Miville-Desch{\^e}nes}, {Pardo}, \& {Apponi}}]{pagani04}
{Pagani}, L., {Bacmann}, A., {Motte}, F., {et~al.} 2004, \aap, 417, 605

\bibitem[{{Pagani} {et~al.}(2010){Pagani}, {Steinacker}, {Bacmann}, {Stutz}, \&
  {Henning}}]{pagani10}
{Pagani}, L., {Steinacker}, J., {Bacmann}, A., {Stutz}, A., \& {Henning}, T.
  2010, Science, 329, 1622

\bibitem[{{Pavlyuchenkov} {et~al.}(2007){Pavlyuchenkov}, {Henning}, \&
  {Wiebe}}]{pavlyuchenkov07}
{Pavlyuchenkov}, Y., {Henning}, T., \& {Wiebe}, D. 2007, \apjl, 669, L101

\bibitem[{{Pilbratt} {et~al.}(2010){Pilbratt}, {Riedinger}, {Passvogel},
  {Crone}, {Doyle}, {Gageur}, {Heras}, {Jewell}, {Metcalfe}, {Ott}, \&
  {Schmidt}}]{herschel}
{Pilbratt}, G.~L., {Riedinger}, J.~R., {Passvogel}, T., {et~al.} 2010, \aap,
  518, L1

\bibitem[{{Plummer}(1911)}]{plummer11}
{Plummer}, H.~C. 1911, \mnras, 71, 460

\bibitem[{{Poglitsch} {et~al.}(2010){Poglitsch}, {Waelkens}, {Geis},
  {Feuchtgruber}, {Vandenbussche}, {Rodriguez}, {Krause}, {Renotte}, {van
  Hoof}, {Saraceno}, {Cepa}, {Kerschbaum}, {Agn{\`e}se}, {Ali}, {Altieri},
  {Andreani}, {Augueres}, {Balog}, {Barl}, {Bauer}, {Belbachir}, {Benedettini},
  {Billot}, {Boulade}, {Bischof}, {Blommaert}, {Callut}, {Cara}, {Cerulli},
  {Cesarsky}, {Contursi}, {Creten}, {De Meester}, {Doublier}, {Doumayrou},
  {Duband}, {Exter}, {Genzel}, {Gillis}, {Gr{\"o}zinger}, {Henning},
  {Herreros}, {Huygen}, {Inguscio}, {Jakob}, {Jamar}, {Jean}, {de Jong},
  {Katterloher}, {Kiss}, {Klaas}, {Lemke}, {Lutz}, {Madden}, {Marquet},
  {Martignac}, {Mazy}, {Merken}, {Montfort}, {Morbidelli}, {M{\"u}ller},
  {Nielbock}, {Okumura}, {Orfei}, {Ottensamer}, {Pezzuto}, {Popesso},
  {Putzeys}, {Regibo}, {Reveret}, {Royer}, {Sauvage}, {Schreiber}, {Stegmaier},
  {Schmitt}, {Schubert}, {Sturm}, {Thiel}, {Tofani}, {Vavrek}, {Wetzstein},
  {Wieprecht}, \& {Wiezorrek}}]{pacs}
{Poglitsch}, A., {Waelkens}, C., {Geis}, N., {et~al.} 2010, \aap, 518, L2

\bibitem[{{Racca} {et~al.}(2009){Racca}, {Vilas-Boas}, \& {de la
  Reza}}]{racca09}
{Racca}, G.~A., {Vilas-Boas}, J.~W.~S., \& {de la Reza}, R. 2009, \apj, 703,
  1444

\bibitem[{{Ragan} {et~al.}(2012){Ragan}, {Henning}, {Krause}, {Pitann},
  {Beuther}, {Linz}, {Tackenberg}, {Balog}, {Hennemann}, {Launhardt}, {Lippok},
  {Nielbock}, {Schmiedeke}, {Schuller}, {Steinacker}, {Stutz}, \&
  {Vasyunina}}]{ragan12}
{Ragan}, S., {Henning}, T., {Krause}, O., {et~al.} 2012, \aap, in press

\bibitem[{{Redman} {et~al.}(2006){Redman}, {Keto}, \& {Rawlings}}]{redman06}
{Redman}, M.~P., {Keto}, E., \& {Rawlings}, J.~M.~C. 2006, \mnras, 370, L1

\bibitem[{{Riedinger}(2009)}]{hcss}
{Riedinger}, J.~R. 2009, ESA Bulletin, 139, 14

\bibitem[{{Rieke} {et~al.}(2004){Rieke}, {Young}, {Engelbracht}, {Kelly},
  {Low}, {Haller}, {Beeman}, {Gordon}, {Stansberry}, {Misselt}, {Cadien},
  {Morrison}, {Rivlis}, {Latter}, {Noriega-Crespo}, {Padgett}, {Stapelfeldt},
  {Hines}, {Egami}, {Muzerolle}, {Alonso-Herrero}, {Blaylock}, {Dole}, {Hinz},
  {Le Floc'h}, {Papovich}, {P{\'e}rez-Gonz{\'a}lez}, {Smith}, {Su}, {Bennett},
  {Frayer}, {Henderson}, {Lu}, {Masci}, {Pesenson}, {Rebull}, {Rho}, {Keene},
  {Stolovy}, {Wachter}, {Wheaton}, {Werner}, \& {Richards}}]{spitzer-mips}
{Rieke}, G.~H., {Young}, E.~T., {Engelbracht}, C.~W., {et~al.} 2004, \apjs,
  154, 25

\bibitem[{{Rom{\'a}n-Z{\'u}{\~n}iga} {et~al.}(2010){Rom{\'a}n-Z{\'u}{\~n}iga},
  {Alves}, {Lada}, \& {Lombardi}}]{roman10}
{Rom{\'a}n-Z{\'u}{\~n}iga}, C.~G., {Alves}, J.~F., {Lada}, C.~J., \&
  {Lombardi}, M. 2010, \apj, 725, 2232

\bibitem[{{Rom{\'a}n-Z{\'u}{\~n}iga} {et~al.}(2007){Rom{\'a}n-Z{\'u}{\~n}iga},
  {Lada}, {Muench}, \& {Alves}}]{roman07}
{Rom{\'a}n-Z{\'u}{\~n}iga}, C.~G., {Lada}, C.~J., {Muench}, A., \& {Alves},
  J.~F. 2007, ApJ, 664, 357

\bibitem[{{Roussel}(2012)}]{scanamorphos_arxiv}
{Roussel}, H. 2012, ArXiv e-prints, 1205.2576

\bibitem[{{Ryter}(1996)}]{ryter96}
{Ryter}, C.~E. 1996, \apss, 236, 285

\bibitem[{{Schuller} {et~al.}(2009){Schuller}, {Menten}, {Contreras},
  {Wyrowski}, {Schilke}, {Bronfman}, {Henning}, {Walmsley}, {Beuther},
  {Bontemps}, {Cesaroni}, {Deharveng}, {Garay}, {Herpin}, {Lefloch}, {Linz},
  {Mardones}, {Minier}, {Molinari}, {Motte}, {Nyman}, {Reveret}, {Risacher},
  {Russeil}, {Schneider}, {Testi}, {Troost}, {Vasyunina}, {Wienen}, {Zavagno},
  {Kovacs}, {Kreysa}, {Siringo}, \& {Wei{\ss}}}]{schuller09}
{Schuller}, F., {Menten}, K.~M., {Contreras}, Y., {et~al.} 2009, \aap, 504, 415

\bibitem[{{Shu}(1977)}]{shu77}
{Shu}, F.~H. 1977, \apj, 214, 488

\bibitem[{{Shu} {et~al.}(1987){Shu}, {Adams}, \& {Lizano}}]{shu87}
{Shu}, F.~H., {Adams}, F.~C., \& {Lizano}, S. 1987, \araa, 25, 23

\bibitem[{{Sipil{\"a}} {et~al.}(2011){Sipil{\"a}}, {Harju}, \&
  {Juvela}}]{sipilae11}
{Sipil{\"a}}, O., {Harju}, J., \& {Juvela}, M. 2011, \aap, 535, A49

\bibitem[{{Siringo} {et~al.}(2009){Siringo}, {Kreysa}, {Kov{\'a}cs},
  {Schuller}, {Wei{\ss}}, {Esch}, {Gem{\"u}nd}, {Jethava}, {Lundershausen},
  {Colin}, {G{\"u}sten}, {Menten}, {Beelen}, {Bertoldi}, {Beeman}, \&
  {Haller}}]{laboca}
{Siringo}, G., {Kreysa}, E., {Kov{\'a}cs}, A., {et~al.} 2009, \aap, 497, 945

\bibitem[{{Skrutskie} {et~al.}(2006){Skrutskie}, {Cutri}, {Stiening},
  {Weinberg}, {Schneider}, {Carpenter}, {Beichman}, {Capps}, {Chester},
  {Elias}, {Huchra}, {Liebert}, {Lonsdale}, {Monet}, {Price}, {Seitzer},
  {Jarrett}, {Kirkpatrick}, {Gizis}, {Howard}, {Evans}, {Fowler}, {Fullmer},
  {Hurt}, {Light}, {Kopan}, {Marsh}, {McCallon}, {Tam}, {Van Dyk}, \&
  {Wheelock}}]{2mass}
{Skrutskie}, M.~F., {Cutri}, R.~M., {Stiening}, R., {et~al.} 2006, \aj, 131,
  1163

\bibitem[{{Sodroski} {et~al.}(1997){Sodroski}, {Odegard}, {Arendt}, {Dwek},
  {Weiland}, {Hauser}, \& {Kelsall}}]{sodroski97}
{Sodroski}, T.~J., {Odegard}, N., {Arendt}, R.~G., {et~al.} 1997, \apj, 480,
  173

\bibitem[{{Steinacker} {et~al.}(2005{\natexlab{a}}){Steinacker}, {Bacmann},
  {Henning}, \& {Klessen}}]{steinacker05ppv}
{Steinacker}, J., {Bacmann}, A., {Henning}, T., \& {Klessen}, R.
  2005{\natexlab{a}}, in Protostars and Planets V, 8107--+

\bibitem[{{Steinacker} {et~al.}(2005{\natexlab{b}}){Steinacker}, {Bacmann},
  {Henning}, {Klessen}, \& {Stickel}}]{steinacker05}
{Steinacker}, J., {Bacmann}, A., {Henning}, T., {Klessen}, R., \& {Stickel}, M.
  2005{\natexlab{b}}, \aap, 434, 167

\bibitem[{{Steinacker} {et~al.}(2003){Steinacker}, {Henning}, {Bacmann}, \&
  {Semenov}}]{steinacker03}
{Steinacker}, J., {Henning}, T., {Bacmann}, A., \& {Semenov}, D. 2003, \aap,
  401, 405

\bibitem[{{Steinacker} {et~al.}(2010){Steinacker}, {Pagani}, {Bacmann}, \&
  {Guieu}}]{steinacker10}
{Steinacker}, J., {Pagani}, L., {Bacmann}, A., \& {Guieu}, S. 2010, \aap, 511,
  A9+

\bibitem[{{Steinacker} {et~al.}(1996){Steinacker}, {Thamm}, \&
  {Maier}}]{steinacker96}
{Steinacker}, J., {Thamm}, E., \& {Maier}, U. 1996, \jqsrt, 56, 97

\bibitem[{{Stutz} {et~al.}(2010){Stutz}, {Launhardt}, {Linz}, {Krause},
  {Henning}, {Kainulainen}, {Nielbock}, {Steinacker}, \& {Andr{\'e}}}]{stutz10}
{Stutz}, A., {Launhardt}, R., {Linz}, H., {et~al.} 2010, \aap, 518, L87

\bibitem[{{Stutz} {et~al.}(2007){Stutz}, {Bieging}, {Rieke}, {Shirley},
  {Balog}, {Gordon}, {Green}, {Keene}, {Kelly}, {Rubin}, \& {Werner}}]{stutz07}
{Stutz}, A.~M., {Bieging}, J.~H., {Rieke}, G.~H., {et~al.} 2007, \apj, 665, 466

\bibitem[{{Stutz} {et~al.}(2009){Stutz}, {Rieke}, {Bieging}, {Balog},
  {Heitsch}, {Kang}, {Peters}, {Shirley}, \& {Werner}}]{stutz09}
{Stutz}, A.~M., {Rieke}, G.~H., {Bieging}, J.~H., {et~al.} 2009, \apj, 707, 137

\bibitem[{{Swain} {et~al.}(2008){Swain}, {Bouwman}, {Akeson}, {Lawler}, \&
  {Beichman}}]{swain08}
{Swain}, M.~R., {Bouwman}, J., {Akeson}, R.~L., {Lawler}, S., \& {Beichman},
  C.~A. 2008, \apj, 674, 482

\bibitem[{{Tafalla} {et~al.}(1998){Tafalla}, {Mardones}, {Myers}, {Caselli},
  {Bachiller}, \& {Benson}}]{tafalla98}
{Tafalla}, M., {Mardones}, D., {Myers}, P.~C., {et~al.} 1998, \apj, 504, 900

\bibitem[{{Tassis} \& {Yorke}(2011)}]{tassis11}
{Tassis}, K. \& {Yorke}, H.~W. 2011, \apjl, 735, L32+

\bibitem[{{van Leeuwen}(2007)}]{hipparcos07}
{van Leeuwen}, F. 2007, \aap, 474, 653

\bibitem[{{Vrba} {et~al.}(1993){Vrba}, {Coyne}, \& {Tapia}}]{vrba93}
{Vrba}, F.~J., {Coyne}, G.~V., \& {Tapia}, S. 1993, \aj, 105, 1010

\bibitem[{{Vuong} {et~al.}(2003){Vuong}, {Montmerle}, {Grosso}, {Feigelson},
  {Verstraete}, \& {Ozawa}}]{vuong03}
{Vuong}, M.~H., {Montmerle}, T., {Grosso}, N., {et~al.} 2003, \aap, 408, 581

\bibitem[{{Ward-Thompson}(2002)}]{ward-thompson02}
{Ward-Thompson}, D. 2002, Science, 295, 76

\bibitem[{{Watson}(2011)}]{watson11}
{Watson}, D. 2011, \aap, 533, A16

\bibitem[{{Werner} {et~al.}(2004){Werner}, {Roellig}, {Low}, {Rieke}, {Rieke},
  {Hoffmann}, {Young}, {Houck}, {Brandl}, {Fazio}, {Hora}, {Gehrz}, {Helou},
  {Soifer}, {Stauffer}, {Keene}, {Eisenhardt}, {Gallagher}, {Gautier}, {Irace},
  {Lawrence}, {Simmons}, {Van Cleve}, {Jura}, {Wright}, \&
  {Cruikshank}}]{spitzer}
{Werner}, M.~W., {Roellig}, T.~L., {Low}, F.~J., {et~al.} 2004, ApJS, 154, 1

\bibitem[{{Whittet}(1992)}]{whittet92}
{Whittet}, D.~C.~B. 1992, {Dust in the galactic environment} (Institute of
  Physics Publishing)

\bibitem[{{Whitworth} \& {Ward-Thompson}(2001)}]{whitworth01}
{Whitworth}, A.~P. \& {Ward-Thompson}, D. 2001, \apj, 547, 317

\bibitem[{{Wright} {et~al.}(2010){Wright}, {Eisenhardt}, {Mainzer}, {Ressler},
  {Cutri}, {Jarrett}, {Kirkpatrick}, {Padgett}, {McMillan}, {Skrutskie},
  {Stanford}, {Cohen}, {Walker}, {Mather}, {Leisawitz}, {Gautier}, {McLean},
  {Benford}, {Lonsdale}, {Blain}, {Mendez}, {Irace}, {Duval}, {Liu}, {Royer},
  {Heinrichsen}, {Howard}, {Shannon}, {Kendall}, {Walsh}, {Larsen}, {Cardon},
  {Schick}, {Schwalm}, {Abid}, {Fabinsky}, {Naes}, \& {Tsai}}]{wise}
{Wright}, E.~L., {Eisenhardt}, P.~R.~M., {Mainzer}, A.~K., {et~al.} 2010, \aj,
  140, 1868

\bibitem[{{Young} \& {Evans}(2005)}]{young05}
{Young}, C.~H. \& {Evans}, II, N.~J. 2005, \apj, 627, 293

\bibitem[{{Yun} \& {Clemens}(1991)}]{yun91}
{Yun}, J.~L. \& {Clemens}, D.~P. 1991, \apj, 381, 474

\bibitem[{{Zhilkin} {et~al.}(2009){Zhilkin}, {Pavlyuchenkov}, \&
  {Zamozdra}}]{zhilkin09}
{Zhilkin}, A.~G., {Pavlyuchenkov}, Y.~N., \& {Zamozdra}, S.~N. 2009, Astronomy
  Reports, 53, 590

\bibitem[{{Zucconi} {et~al.}(2001){Zucconi}, {Walmsley}, \&
  {Galli}}]{zucconi01}
{Zucconi}, A., {Walmsley}, C.~M., \& {Galli}, D. 2001, \aap, 376, 650

\end{thebibliography}

\Online
\clearpage
\onecolumn
\begin{appendix}
\section{Conversion between dust opacity and extinction}
\label{s:kappa}

In Eq.~\ref{e:kappa}, we have provided a conversion between the frequency
dependent dust opacity $\kappa_\mathrm{d}(\nu)$ and the NIR extinction law
\citep{cardelli89} in order to be able to compare the behaviour of the
scatter-free dust opacities of \citet{ossenkopf94} and the extinction law at NIR
wavelengths in Fig.~\ref{f:opacities}. Below, we derive this equation in detail
by starting from the relation between the extinction value and the optical depth
at a given wavelength:
\begin{eqnarray*}
A_\lambda &=& 1.086\,\tau_\lambda = 1.086\,\kappa_\lambda\,\rho_\mathrm{tot}\,x =
1.086\,\kappa_\lambda\,\Sigma_\mathrm{g,d} \quad .
\end{eqnarray*}
Here, $\rho$ is the density of the material that causes the extinction
$A_\lambda$ with its opacity $\kappa_\lambda$ along the LoS path $x$.
This can be written in terms of a surface density $\Sigma_\mathrm{g,d}$ that
includes both dust and gas. Their properties are being disentangled as
follows, where $X_\mathrm{g,d}$ is the gas-to-dust ratio, which is assumed to be
150 in this paper \citep{sodroski97}:
\begin{eqnarray*}
A_\lambda &=& 1.086\,\kappa_\mathrm{d}(\nu)\,X^{-1}_\mathrm{g,d}\,\left(\Sigma_{\rm
g}+\Sigma_\mathrm{d}\right) \\
&=& 1.086\,\kappa_\mathrm{d}(\nu)\,X^{-1}_\mathrm{g,d}\,\left(\Sigma_\mathrm{g}+X^{-1}_\mathrm{g,d}\Sigma_\mathrm{g}\right) \\
&=& 1.086\,\kappa_\mathrm{d}(\nu)\,X^{-1}_\mathrm{g,d}\,\left(1+X^{-1}_\mathrm{g,d}\right)\Sigma_\mathrm{g} \\
&=& 1.086\,\kappa_\mathrm{d}(\nu)\,\left(X^{-1}_\mathrm{g,d}+X^{-2}_{\rm
g,d}\right)\Sigma_\mathrm{g} \\
&=& 1.086\,\kappa_\mathrm{d}(\nu)\,\left(X^{-1}_\mathrm{g,d}+X^{-2}_{\rm
g,d}\right)\Sigma_\mathrm{H} \frac{m_\mathrm{g}}{m_\mathrm{H}} \\
&=& 1.086\,\kappa_\mathrm{d}(\nu)\,\left(X^{-1}_\mathrm{g,d}+X^{-2}_{\rm
g,d}\right)N_\mathrm{H}\,m_\mathrm{H} \frac{m_\mathrm{g}}{m_\mathrm{H}}
\quad .
\end{eqnarray*}
At this point, we have come to a relation between the extinction $A_\lambda$ and
the dust opacity $\kappa_\mathrm{d}(\nu)$ that depends on the column density of
hydrogen atoms $N_\mathrm{H}$. The conversion between the hydrogen and the mean
gas mass is $m_\mathrm{g}/m_\mathrm{H} = 1.36$. When replacing the constants
with the corresponding numbers, we get
\begin{eqnarray*}
A_\lambda &=& 1.086\,\kappa_\mathrm{d}(\nu)\,\left(150^{-1}+150^{-2}
\right)N_\mathrm{H}\cdot 1.675\times10^{-24}\ \mathrm{g}\cdot 1.36 \\
&=& 1.66\times10^{-26}\ \mathrm{g}\cdot\kappa_\mathrm{d}(\nu)\,N_\mathrm{H}
\quad .
\end{eqnarray*}
Expressed as a relation for $\kappa_\mathrm{d}(\nu)$, we find
\begin{eqnarray*}
\kappa_\mathrm{d}(\nu) &=& 6.02\times10^{25}\ \mathrm{g}^{-1}\,\left(\frac{N_{\rm
H}}{A_\lambda}\right)^{-1}
\quad .
\end{eqnarray*}
In the NIR, the extinction law is insensitive to variations in
$R_V=A_V/E_{B-V}$. Therefore, we can use \citet{cardelli89} to transform the
ratios $N_H/A_J$ \citep{vuong03} and $N_H/E_{J-K}$ \citep{martin12} to hydrogen
column density vs. NIR extinction ratios. If the $N_H/A_V$ or $N_H/E_{B-V}$
ratios are used instead, they implicitly introduce a dependence of $R_V$ as
follows:
\begin{eqnarray*}
\kappa_\mathrm{d}(\nu) &=& 6.02\times10^{25}\ \mathrm{g}^{-1}\cdot
\left(\frac{A_\lambda}{A_V}\right)(R_V)\cdot \left(\frac{N_{\rm
H}}{A_V}\right)^{-1}\\
&=& 6.02\times10^{25}\ \mathrm{g}^{-1}\cdot
\left(\frac{A_\lambda}{A_V}\right)(R_V)\cdot R_V \left(\frac{N_\mathrm{H}}{E_{B-V}}\right)^{-1}
\quad .
\end{eqnarray*}
The ratio $A_\lambda/A_V$ as a function of $R_V$ represents the extinction law
of \citet{cardelli89}. Several values for the ratio between the hydrogen column
density and the colour excess $E_{B-V}$ can be applied. If, for instance, the
ratio of \citet{bohlin78} is used, we get
\begin{eqnarray*}
\kappa_\mathrm{d}(\nu) &=& 10400\ \mathrm{cm}^{2}\ \mathrm{g}^{-1}\cdot R_V\,
\left(\frac{A_\lambda}{A_V}\right)(R_V)
\quad .
\end{eqnarray*}
For a more recent characterisation of that ratio
\citep{ryter96,guever09,watson11}, the numerical value in this conversion is
modified by about 18\%. To be able to convert the quoted ratio
$N_\mathrm{H}/A_V$ to $N_\mathrm{H}/E_{B-V}$, we have assumed $R_V = 3.1$ as the
mean value for the diffuse ISM, and replaced the corresponding ratio in the
equation above:
\begin{eqnarray*}
\kappa_\mathrm{d}(\nu) &=& 8800\ \mathrm{cm}^{2}\ \mathrm{g}^{-1}\cdot R_V\,
\left(\frac{A_\lambda}{A_V}\right)(R_V)
\quad .
\end{eqnarray*}

\twocolumn

\section{Spitzer IRS spectra}
\begin{figure}[!h]
\centering
\resizebox{\hsize}{!}{\includegraphics{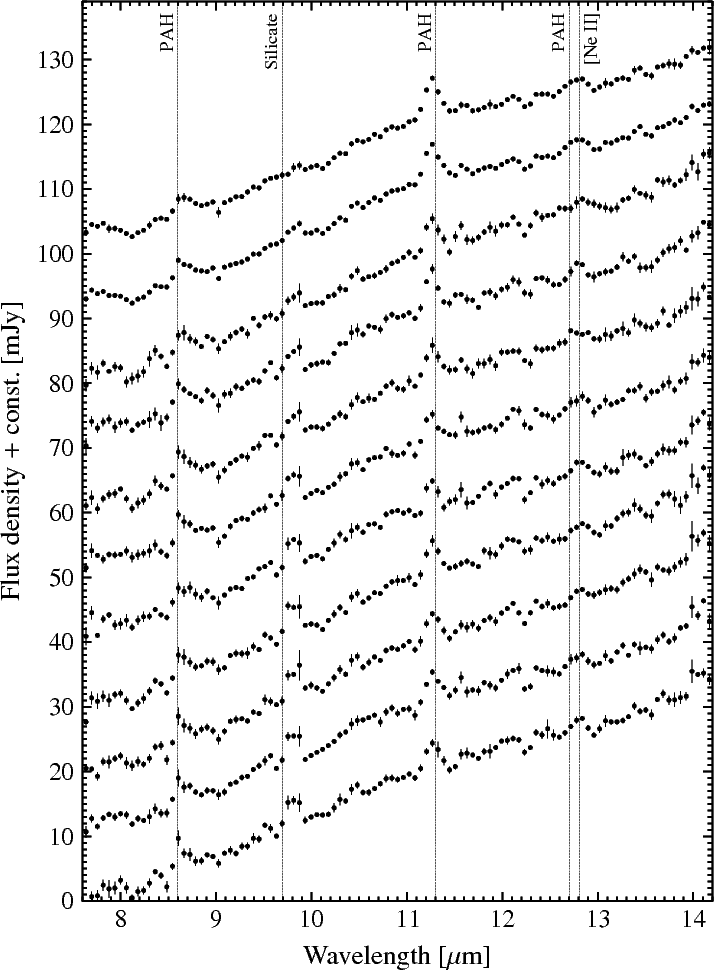}}
\caption{\label{f:irs}Mid-infrared spectra at 11 positions across B68 obtained
with the \textit{Spitzer} IRS instrument. The most prominent spectral features
are indicated by dashed lines. The flux density is given in mJy. The individual
spectra are offset by 10~mJy each. There is no significant variation in the
strength of the PAH features.}
\end{figure}

\begin{table}[!h]
\caption[]{\label{t:irs}Coordinates of the \textit{Spitzer} IRS observations used
to extract the background spectra. Apertures of 5 detector pixels in size
corresponding to $9''$ were placed $20''$ left and right (in the detector frame)
from the nominal target position to avoid contamination from the stellar object
a given observation was aimed at. The slit width was $3\farcs6$. The coordinates
are given in the order of Fig.~\ref{f:irs}.}
\begin{tabular}{c@{$^{\rm h}$}c@{$^{\rm m}$}c@{\fs}ccc@{$^\circ$}c@{$'$}c@{\farcs}c}
\hline
\hline
\multicolumn{4}{c}{RA (J2000)} & & \multicolumn{4}{c}{Dec (J2000)} \\
\hline
17&22&37&94&&$-$23&48&51&6\\
17&22&45&15&&$-$23&48&39&6\\
17&22&40&31&&$-$23&48&55&7\\
17&22&41&62&&$-$23&50&26&4\\
17&22&39&04&&$-$23&50&36&4\\
17&22&37&93&&$-$23&49&24&7\\
17&22&38&80&&$-$23&49&14&5\\
17&22&39&05&&$-$23&50&00&8\\
17&22&39&05&&$-$23&50&36&6\\
17&22&37&94&&$-$23&49&24&9\\
17&22&45&03&&$-$23&48&53&3\\
17&22&44&87&&$-$23&49&04&7\\
17&22&44&14&&$-$23&49&16&8\\
\hline
\end{tabular}
\end{table}
\clearpage

\section{CO spectra}
\begin{figure}[!h]
\centering
\resizebox{0.9\hsize}{!}{\includegraphics{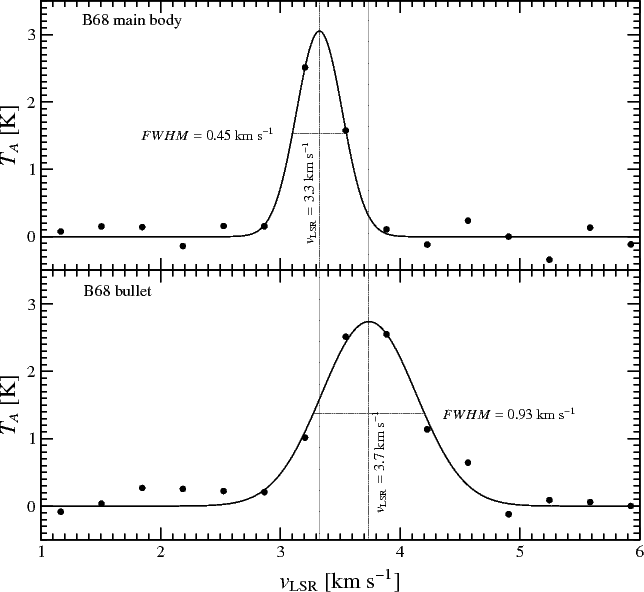}}
\caption{\label{f:13coline}\element[][13]{CO}~(2--1) spectra extracted at two
positions in the corresponding map. The upper spectrum shows a line on the
western rim of B68 (RA~(J2000) = $17^{\rm h}22^{\rm m}34\fs1$, Dec~(J2000) =
$-23^\circ49'53''$), while the lower spectrum is taken at
the location of the point source mentioned in Sect.~\ref{s:morphology}
A Gaussian profile fit was applied to both spectra.}
\end{figure}
\clearpage

\section{Radial NIR extinction profiles}
\begin{figure}[!h]
\centering
\resizebox{0.9\hsize}{!}{\includegraphics{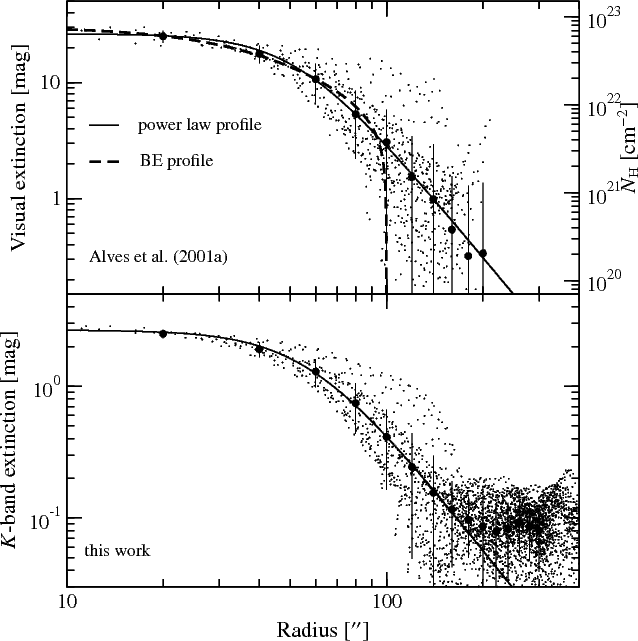}}
\caption{\label{f:extprofiles}Radial profiles extracted from the extinction map
of \citet{alves01eso} (upper panel) and the newly calculated extinction map
shown in Fig.~\ref{f:b68ext} (lower panel). The small dots represent the data in
the maps after regridding them to the same scale; the large dots are the mean
values in bins of $20\arcsec$ each with the error bars indicating the
corresponding r.m.s. scatter. The mean distributions follow a profile fit
similar to Eq.~\ref{e:N_powerlaw} (solid line). The power-law indices are
$\alpha=3.3$ in the upper panel and $\alpha=3.1$ in the lower panel. For
comparison, we added the profile of the BES (dashed line) given by
\citet{alves01nature}.}
\end{figure}

\end{appendix}

\end{document}